\def\published#1{Pulished~#1\par}
\documentclass{ws-ijmpa}

\usepackage[super,compress]{cite}

\usepackage{url}
\usepackage{hyperref} 
\usepackage{graphicx}
\usepackage{tabularx}
\usepackage{amsmath}
\usepackage{slashed}
\usepackage[usenames]{color}
\usepackage{longtable}

\catchline{}{}{}{}{}

\begin{document} 
%%%%%%%%%%%%%%%%%%%%%%
\title{Recent Developments  on the CKM Matrix  }

\author{ Wei Wang}

\address{INPAC, Department of Physics and Astronomy, Shanghai Jiao Tong University, Shanghai, 200240, P. R. China \\
and \\
Helmholtz-Institut f\"ur Strahlen- und Kernphysik, Universit\"at Bonn, Bonn 53115, Germany\\
weiwang@hiskp.uni-bonn.de}

\maketitle

\begin{history}
\received{xxx}
\accepted{xxx}
\published{xxx}
\end{history}.pdf

\begin{abstract}
In   Standard Model,  CP violation arises from an irreducible complex phase in the quark mixing
matrix,  now under the name Cabibbo-Kobayashi-Maskawa matrix.  This description has  shown remarkably overall  agreement  with various experimental  measurements.  In this review, 
we  discuss   recent experimental data and  theoretical  developments on three quantities of CKM matrix that are most uncertain:  the $V_{ub}$,  including its magnitude and the phase  $\gamma$ in   standard parametrization,  and the $B_s-\bar B_s$ mixing phase $\beta_s$.

\end{abstract}
 
\ccode{PACS numbers: 12.15.Hh, 13.20.He, 13.25.Hw} 

\tableofcontents
%%%%%%%%%%%%%%%%%%%%%%
\section{Introduction}
\label{sec:Introduction}
%%%%%%%%%%%%%%%%%%%%%%

In the Standard Model (SM) of particle physics the three observed generations
of quarks   show a remarkable feature that their weakly interacting eigenstates do  not coincide with their mass eigenstates. This
``misalignment" gives  rise to flavor changing transitions that   can be represented by a
$3\times 3$ mixing matrix. The  mixing scheme was formulated first
by Cabibbo~\cite{Cabibbo:1963yz}  for two generations, and   extended later to   three   generations by Kobayashi and Maskawa~\cite{Kobayashi:1973fv}, now referred to as  the
Cabibbo-Kobayashi-Maskawa (CKM) matrix. In this unitary  $3 \times 3$   matrix, there is an   irreducible complex phase, which can   give  rise to  a difference in the decays of a particle and its antiparticle, namely CP violation.

In the SM, all flavour-changing interactions of   quarks, from the lowest energies (such as nuclear transitions and pion decays) to the highest energies that can be   reached  at high energy  accelerators for instance the Large Hadron Collider (LHC),  are described by  four parameters of the CKM matrix, which makes it a remarkably predictive paradigm. However  new physics (NP) degrees of freedom are  generally believed to exist  at the TeV scale or not too much higher, and may show patterns deviating from the CKM mechanism.  Thus a precise  determination of CKM parameters is not only useful for   the test of SM but can also serve as an indirect   probes     for the NP.

The CKM  matrix~\cite{Cabibbo:1963yz,Kobayashi:1973fv} describes the mixing between the three   families of quarks.
Thereby it is  a $3\times 3$ unitary matrix, and can be parametrized  in terms of four real parameters.
For example, in   Wolfenstein parametrisation~\cite{Wolfenstein:1983yz} it can be expressed as
\begin{eqnarray}
  \label{eq:ckm}
  V_{\rm CKM} & = & 
  \Bigg(
  \begin{array}{ccc}
    V_{ud} & V_{us} & V_{ub} \\
    V_{cd} & V_{cs} & V_{cb} \\
    V_{td} & V_{ts} & V_{tb} \\
  \end{array}
  \Bigg)  \nonumber\\
  &= &
  \left(
  \begin{array}{ccc}
    1 -  \frac{\lambda^2}{2}  & \lambda & A \lambda^3 ( \rho - i \eta ) \\
    - \lambda & 1 - \frac{\lambda^2}{2} & A \lambda^2 \\
    A \lambda^3 ( 1 - \rho - i \eta ) & - A \lambda^2 & 1 \\
  \end{array}
  \right) + {\cal O}\left( \lambda^4 \right) \, ,
\end{eqnarray}
where the expansion parameter $\lambda$ is the sine of the Cabibbo angle.  The diagonal elements are of unity, but  an empirical relation exists for the off-diagonal  elements~\footnote{Due to the smallness of $|V_{ub}|$, it has been proposed  for instance in  Ref.\cite{Ahn:2011fg}  that the $V_{ub}\sim\lambda^{4}$.  I thank Prof. Zhi-Zhong Xing for an interesting  discussion on this relation.}: 
\begin{eqnarray}
|V_{ij}|\sim \lambda^{ max(i, j)+ |i-j| -2},
\end{eqnarray}
where $i,j$ is the family index. 
With four independent parameters, a $3\times 3$ unitary matrix cannot be forced to be real-valued, and hence $CP$ violation arises as a consequence of the fact that the couplings for quarks and antiquarks have different phases, {\it i.e.} $V_{\rm CKM} \neq V_{\rm CKM}^*$.
In the SM, all $CP$ violation in the quark sector indeed arises from this fact, which is encoded in the Wolfenstein parameter $\eta$.

In this review, we will  concentrate on three   quantities in the CKM matrix:  the $|V_{ub}|$,  the phase   $\gamma$,  and   $\beta_{s}$, whose determinations  are most uncertain nowadays  but  have been greatly  improved in the past a few years.  Most precise results  on the $|V_{ub}|$ arise   from exclusive and inclusive semi-leptonic $b\to u$ decays, and see for instance  Ref.~\cite{Ricciardi:2013cda,Ricciardi:2014iga,Xiao:2014ana,Beringer:1900zz,Lenz:2014nka} for recent reviews. 
These determinations   rely on different theoretical calculations and on
different experimental measurements which very likely  have  uncorrelated
statistical and systematic uncertainties. The independence in these determinations  from inclusive and exclusive decays makes
the comparison of  $|V_{ub}|$   a
powerful test of the CKM mechanism.

\begin{figure}[ht]
  \begin{center}
    \includegraphics[width=0.6\columnwidth]{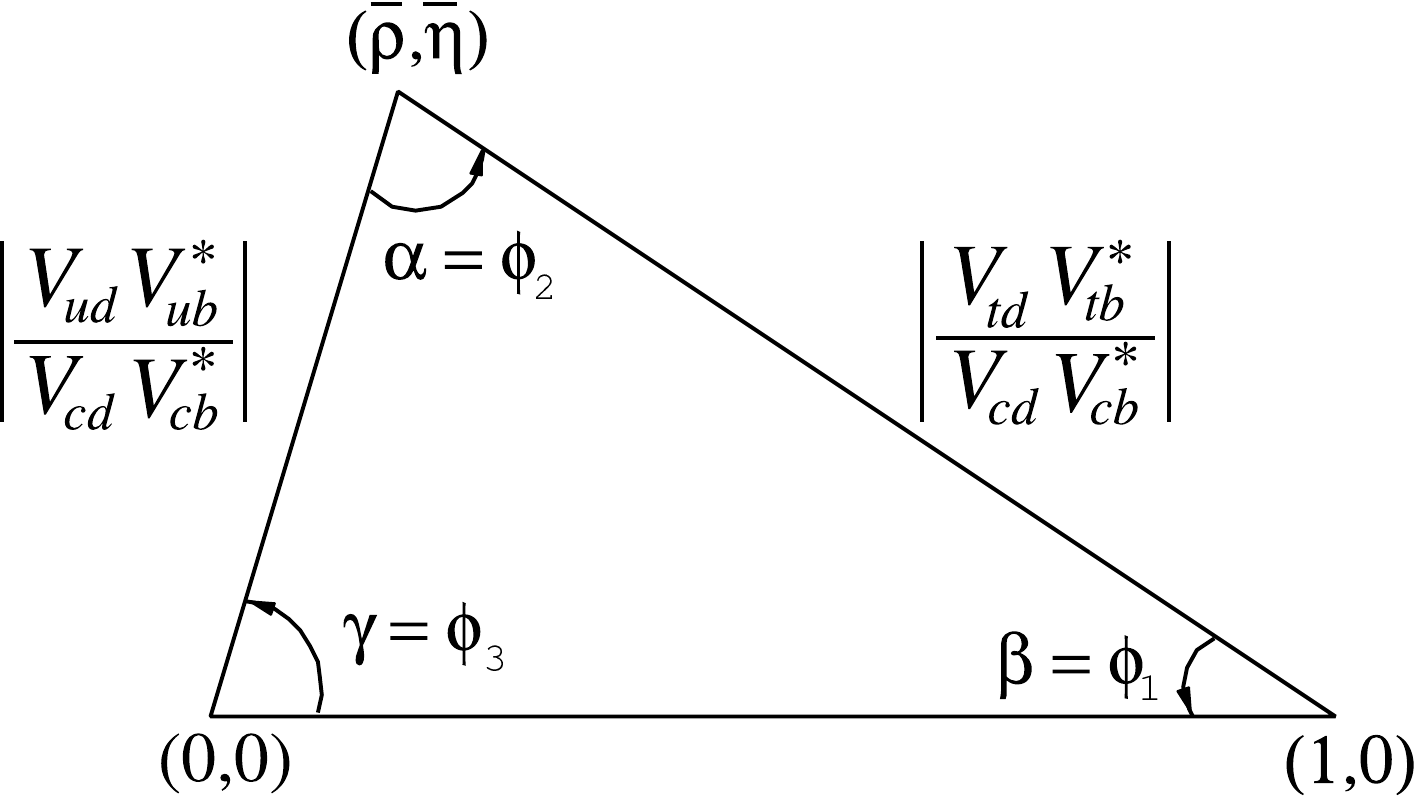} 
    \caption{ A sketch of the $bd$ unitary triangle  formed by $V_{ud}V_{ub}^* +V_{cd}V_{cb}^*+ V_{td}V_{tb}^*=0$.   }
    \label{fig:sketchTriangle}
  \end{center}
\end{figure}

\begin{figure}[ht]
  \begin{center}
    \includegraphics[width=0.32\columnwidth]{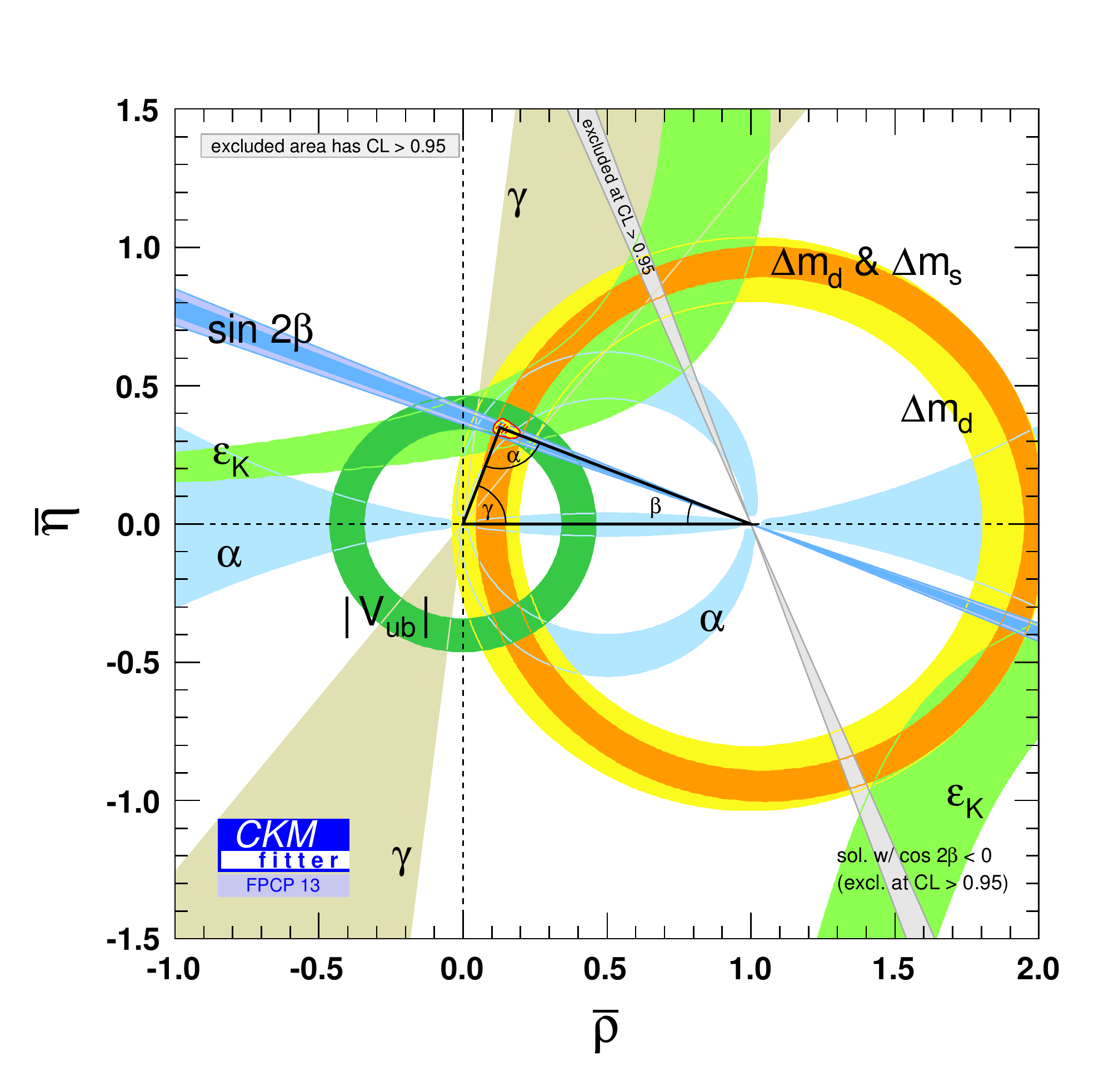} 
    \includegraphics[width=0.32\columnwidth]{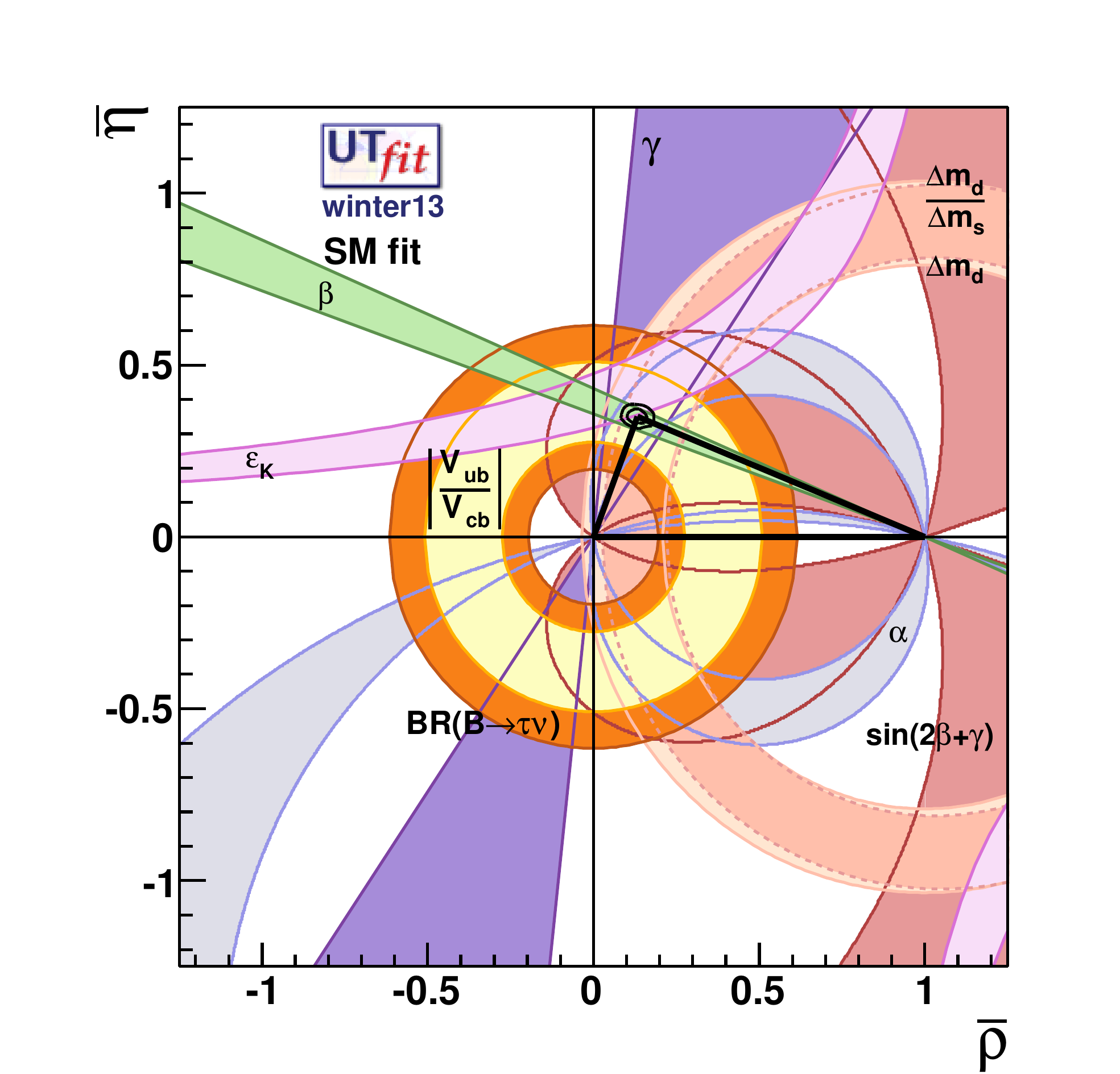} 
    \includegraphics[width=0.32\columnwidth]{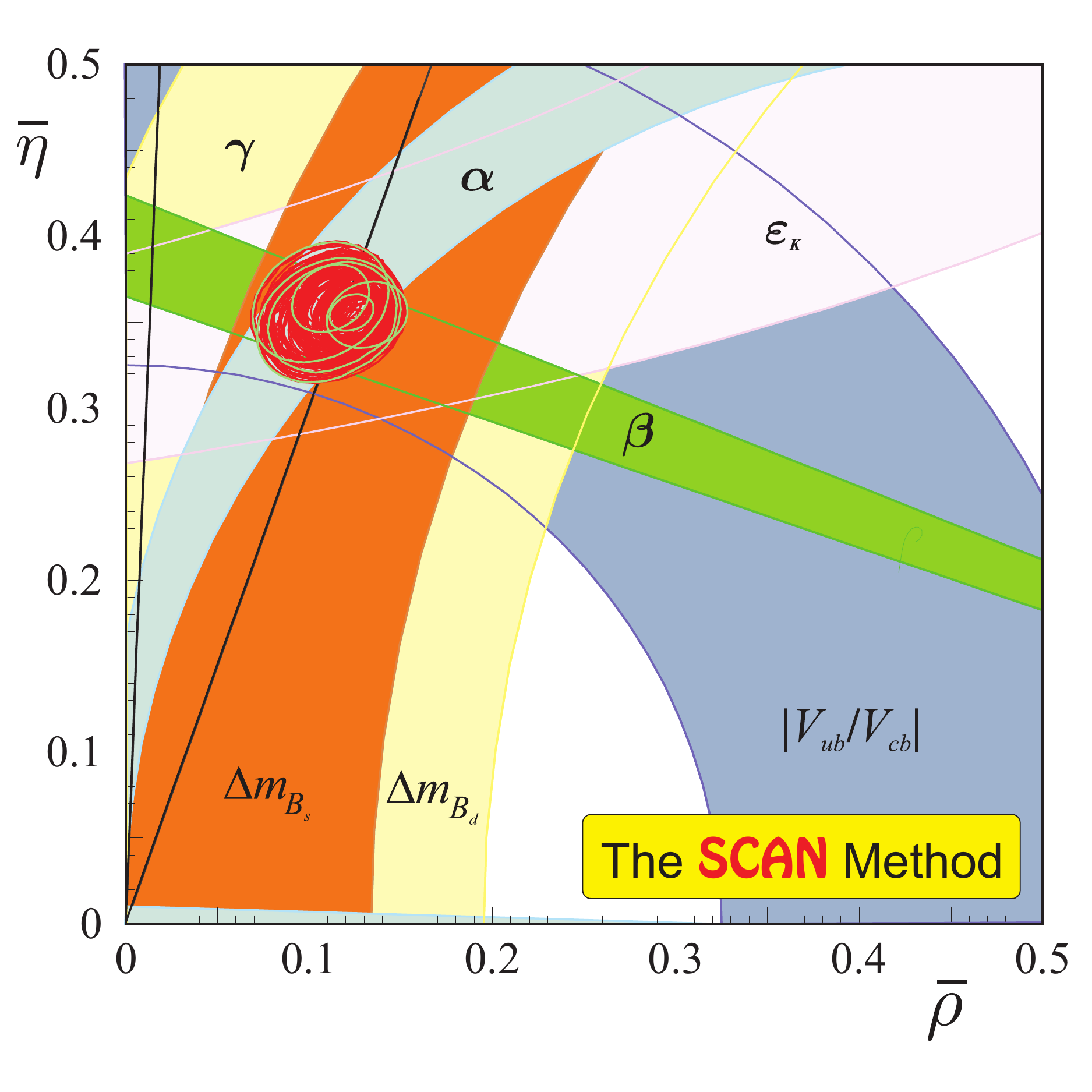} 
    \caption{Global fitting  constraints on the $(bd)$ unitary triangle  from the CKMfitter Group~\cite{Charles:2004jd},  UTfit  Group~\cite{Ciuchini:2000de}, and the scan method~\cite{Eigen:2013cv}.  }
    \label{fig:rhoetaCKM}
  \end{center}
\end{figure}

The unitarity constraints on the CKM matrix can be represented as triangles in the complex plane:
the lengths of whose sides are the moduli of  CKM matrix element products, while the angles
are constructed from the relative phases.  For instance the orthogonality 
\begin{eqnarray}
 V_{ud}V_{ub}^*  + V_{cd}V_{cb}^* + V_{td}V_{tb}^*  = 0, \label{eq:bdUT}
\end{eqnarray} 
form the commonly-studied ($bd$) unitarity triangle as shown  in Fig.~\ref{fig:sketchTriangle}.  The current global fitting  results on this triangle  by the CKMfitter~\cite{Charles:2004jd}, and UTfit Group~\cite{Ciuchini:2000de}, and in the scan method~\cite{Eigen:2013cv}  can be summarised in Fig.~\ref{fig:rhoetaCKM}.

The three angles  $(\alpha,\beta,\gamma)$ as shown  in Fig.~\ref{fig:sketchTriangle} satisfy   the constraint, $\alpha+\beta+\gamma=180^\circ$, that can be tested via experimental  measurements.   The world averages from   Particle Data Group (PDG) in 2012 are given as~\cite{Beringer:1900zz}
\begin{eqnarray} 
 \alpha= (89.0^{+4.4}_{-4.2})^{\circ},\;\;\;{\rm PDG 2012} \\
 \gamma= (68^{+10}_{-11})^{\circ}, \;\;\;{\rm PDG 2012} 
\end{eqnarray} 
while the result 
\begin{eqnarray}
 \sin(2\beta)= (0.679\pm 0.020) \;\;\;{\rm PDG 2012} 
\end{eqnarray}
has a four-fold ambiguity for the $\beta$ angle. 
The error in  $\gamma$ is  about $10^{\circ}$~\cite{Amhis:2012bh,Beringer:1900zz,Charles:2004jd,Ciuchini:2000de,Eigen:2013cv}. Though it is one of  the main sources of   current uncertainties on the apex of the unitary triangle,   recent measurements and theoretical investigations are progressing very fast  and will be covered later in this review. With a large amount of data accumulated in the future,  the LHCb  would be  able to diminish the errors in $\gamma$ to about $4^\circ$ from the  $B\to DK$ until 2018, and to  $1^\circ$ after the upgrade~\cite{Bediaga:2012py}.  On the SuperB factories the error can be   reduced to $2^\circ$ \cite{Bona:2007qt,Aushev:2010bq}.

In contrast with the $(bd)$ triangle,  the $(bs)$ triangle, $V_{tb}V_{ts}^*+V_{cb}V_{cs}^*+V_{ub}V_{us}^*=0$, has a much  smaller complex phase:
\begin{eqnarray}
\phi_s= (-0.036 \pm 0.002) \quad { \rm rad},\label{eq:phisSM}
\end{eqnarray}
with $ \phi_s=-2\beta_s = -2 {\rm arg} [- V_{ts}V_{tb}^*/ (V_{cs}V_{cb}^*)]$~\cite{Charles:2011va}.
The smallness of the $\beta_{s}$ can provide a null test of the SM, and  the observation of a large  non-zero  value   would probably  indicate  a signal for NP  beyond the SM.  Previous data obtained by the CDF~\cite{Aaltonen:2007he} and D0~\cite{Abazov:2008af} collaborations, based on  the angular analysis of  $B_s\to J/\psi \phi$,  indicate much larger values with sizeable uncertainties. This deviation from the SM value has been treated as signals of NP but is softened by new measurements  in the physics programs at the LHC.

%%%%%%%%%%%%%%%%%%%%%%
%%%%%%%%%%%%%%%%%%%%%%
%%%%%%%%%%%%%%%%%%%%%%
%%%%%%%%%%%%%%%%%%%%%%

%%%%%%%%%%%%%%%%%%%%%%
%%%%%%%%%%%%%%%%%%%%%%
%%%%%%%%%%%%%%%%%%%%%%
%%%%%%%%%%%%%%%%%%%%%%

The rest of this review is organised as follows.  Sec.~\ref{sec:Vub} will discuss the extraction of the $|V_{ub}|$ from semi-leptonic and leptonic $B$ decays.  In this section, we will summarise the latest results from the experimental data on various channels, and give a look at the future prospect. At the same time, considerable focus will be spent on the recent  developments of theoretical techniques that can be applied to multi-body semileptonic $B$ decays like the $B\to \pi\pi\ell\bar\nu_{\ell}$. In Sec.~\ref{sec:gamma}, we will review the status on the   angle $\gamma$, including the newly released  experimental data and new theoretical insights. In Sec.~\ref{sec:betas}, we will discuss the recent  progress in the extraction of $\beta_{s}$ through the $B_{s}\to J/\psi\phi$ and $B_{s}\to J/\psi f_{0}(980)$. We conclude in Sec.~\ref{sec:conclusions}.

%%%%%%%%%%%%%%%%%%%%%%%%%%%%%%%%%%%%%%%%
%%%%%%%%%%%%%%%%%%%%%%%%%%%%%%%%%%%%%%%%
\section{$|V_{ub}|$}
\label{sec:Vub}
%%%%%%%%%%%%%%%%%%%%%%%%%%%%%%%%%%%%%%%%
%%%%%%%%%%%%%%%%%%%%%%%%%%%%%%%%%%%%%%%%

As one can see from Fig.~\ref{fig:sketchTriangle}, the length of the side opposite the $\beta$ angle is proportional to the $|V_{ub}|$ and thus its determination is of great importance. 
However since  the $|V_{ub}|$ is the smallest matrix element,  its determination  has a limited precision:~\cite{Beringer:1900zz}
\begin{eqnarray}
 |V_{ub}| =  
\bigg\{
\begin{array}{cc}
  &  (4.41\pm0.15^{+0.15}_{-0.17})\times 10^{-3}   \;\;\; {\rm inclusive}  \\
  &    (3.23\pm0.31)\times 10^{-3}\;\;\; \;\;\;  \;\;\; \;\;   {\rm exclusive} \\ 
\end{array}\;\;\;\;
 {\rm PDG 2012}, \label{eq:VubPDG}
\end{eqnarray}
where the errors are about  $10\%$.

%%%%%%%%%%%%%%%%%%%%%%
%%%%%%%%%%%%%%%%%%%%%%
\begin{figure}[ht]
  \begin{center}
    \includegraphics[width=0.3\columnwidth]{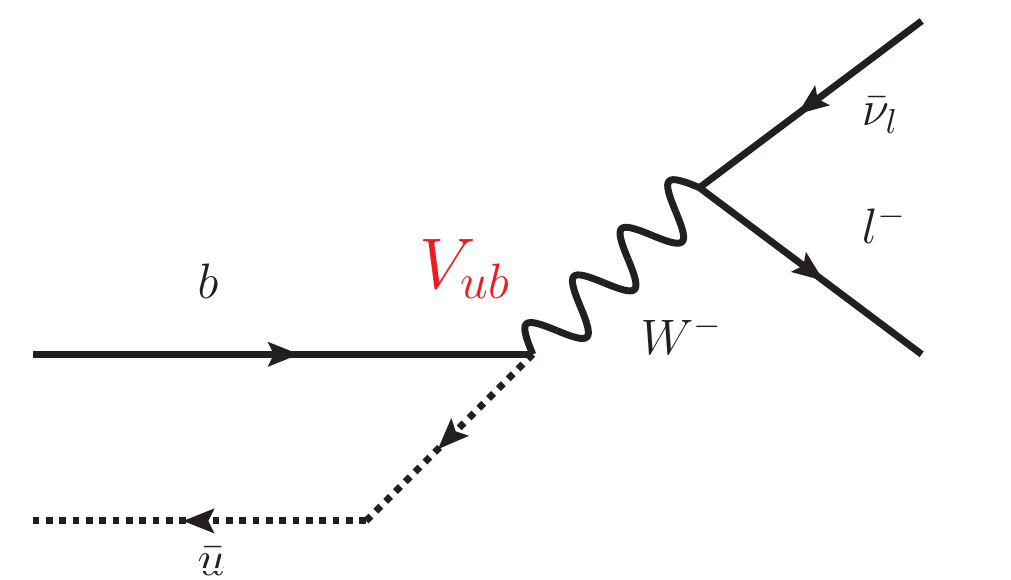} 
    \includegraphics[width=0.3\columnwidth]{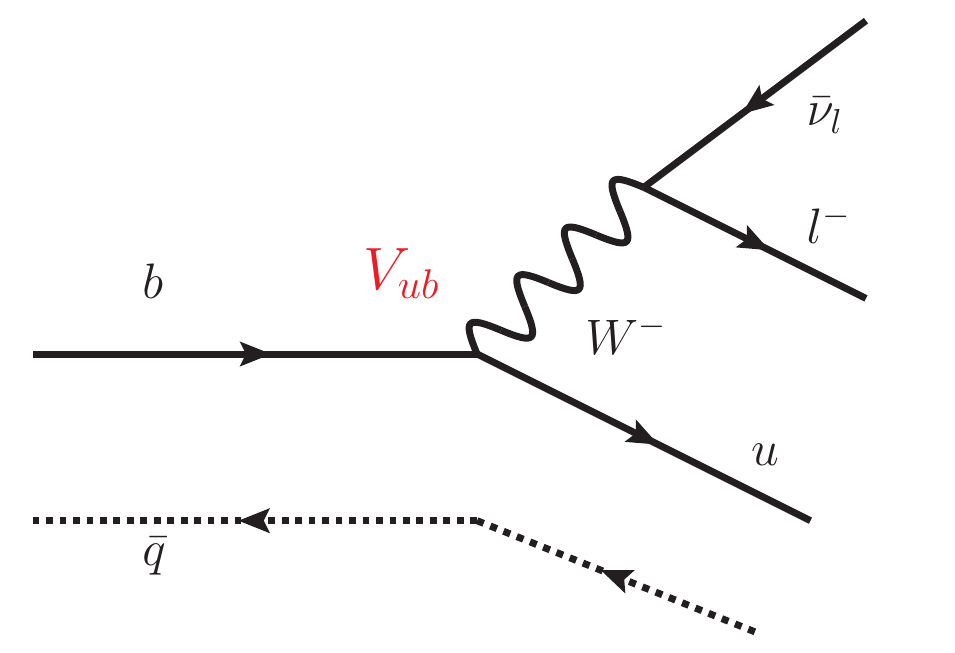} 
    \includegraphics[width=0.3\columnwidth]{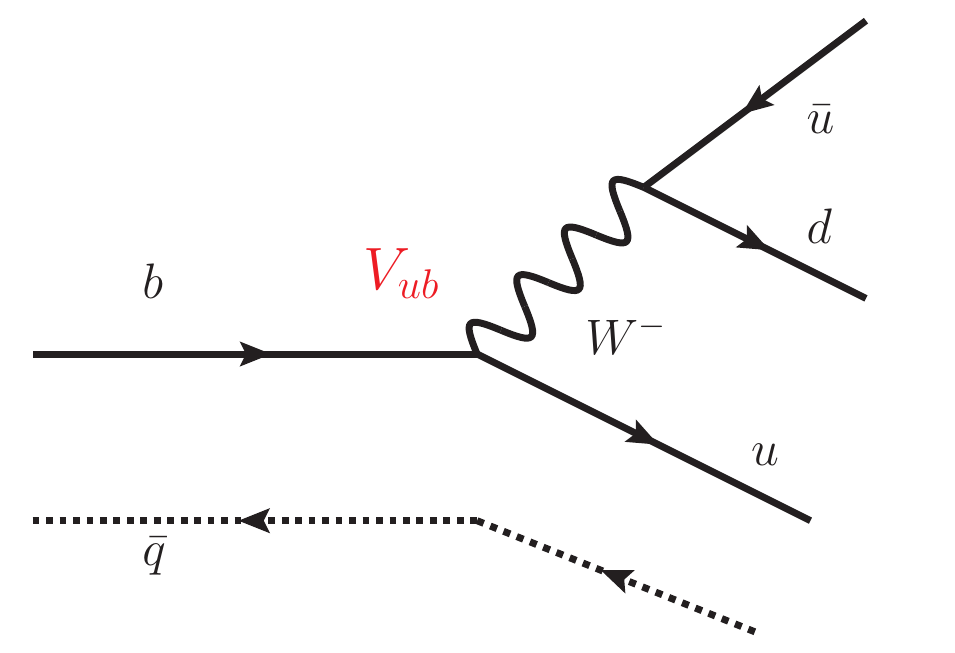} 
    \caption{Decay modes that can be used to extract the  $|V_{ub}|$: purely leptonic (first panel), semi-leptonic (second panel) and non-leptonic  $B$ decays (last panel).     }
    \label{fig:sketchVub}
  \end{center}
\end{figure} 
%%%%%%%%%%%%%%%%%%%%%%
%%%%%%%%%%%%%%%%%%%%%%

The magnitude $\left| V_{\text{ub}} \right|$ can be determined from a multitude of  weak $B$-decays
governed by the $b\to u$ transition which involve either inclusive or exclusive 
final states, whose Feynman diagrams are sketched   in Fig.~\ref{fig:sketchVub}.  These processes  exhibit different experimental and  theoretical challenges.  Compared to leptonic and semi-leptonic decay modes,    non-leptonic  processes    receive additional complexity due to the entanglement with the emitted hadron in final state, and thus its constraint on the $|V_{ub}|$ is quite uncertain (see Ref.~\cite{Kim:2010wr} for a recent discussion).

%%%%%%%%%%%%%%%%%%%%%%%%%%%%%%%%%%%%%%%%
%%%%%%%%%%%%%%%%%%%%%%%%%%%%%%%%%%%%%%%%
\begin{figure}[ht]
\begin{center}
\includegraphics[width=0.5\columnwidth]{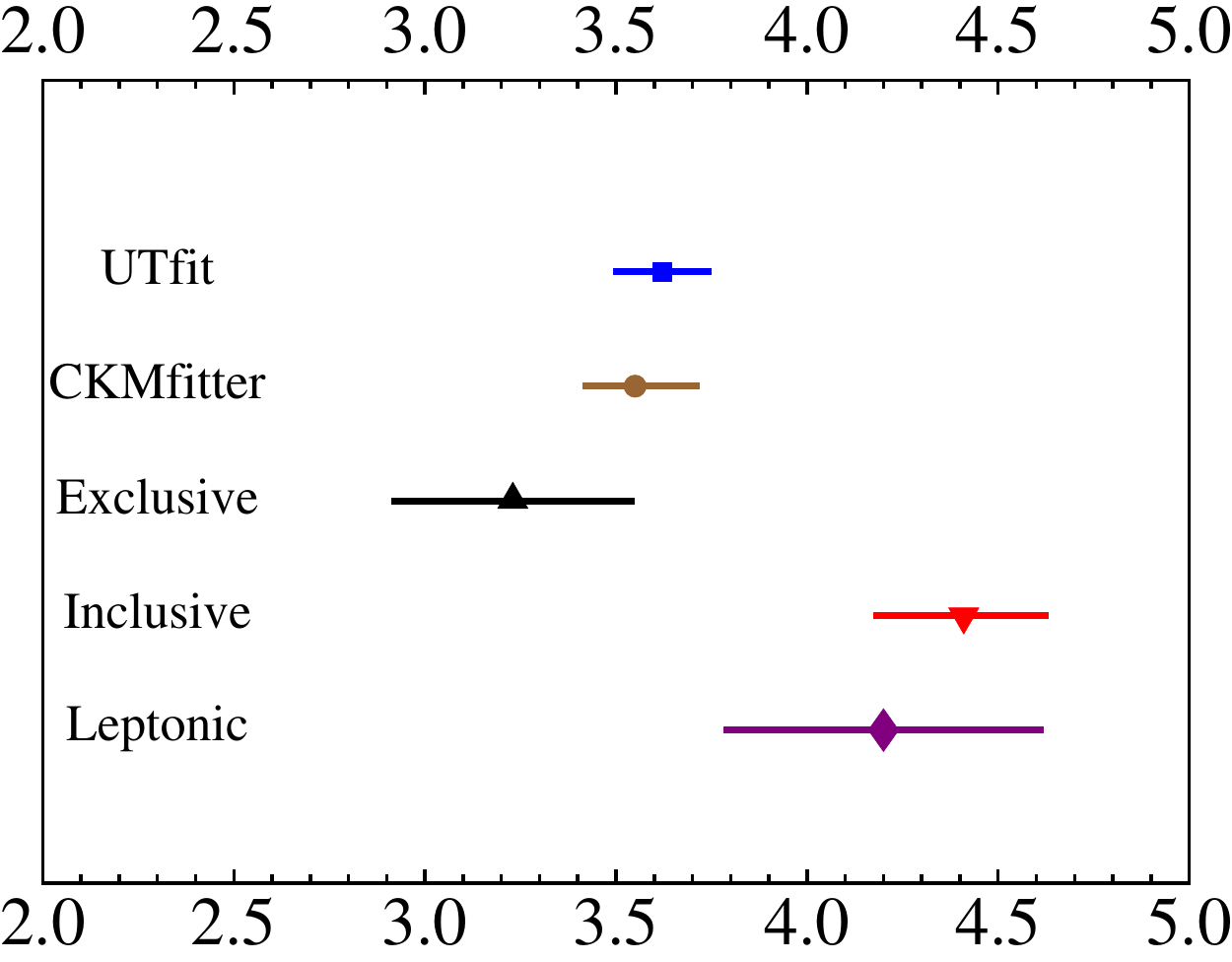}  
\caption{ $|V_{ub}|$ (in units of $10^{{-3}}$) obtained  from experimental data and  the global fitting approach. }
\label{fig:VubValues}
\end{center}
\end{figure} 
%%%%%%%%%%%%%%%%%%%%%%%%%%%%%%%%%%%%%%%%
%%%%%%%%%%%%%%%%%%%%%%%%%%%%%%%%%%%%%%%%

At the current stage,  the increased precision has made manifest
a tension between the values of $|V_{ub}|$ extracted from   exclusive and inclusive semileptonic decays. As shown in Eq.~\eqref{eq:VubPDG}, the inclusive determinations
mostly  yield a central value   larger than
$4  \times 10^{-3}$, while  exclusive analyses produce central values 
below this. 
In Fig.~\ref{fig:VubValues}, we   have collected   the results  from exclusive and inclusive processes as shown in Eq.~\eqref{eq:VubPDG} together with the indirect fits~\cite{Charles:2004jd,Ciuchini:2000de}
\begin{eqnarray}
 |V_{ub}|  &=& (3.65 \pm  0.13) \times 10^{-3}, \;\;\; {\rm UTfit},\\
|V_{ub}|  &=& (3.49^{+0.21}_{-0.10}) \times 10^{-3}, \;\;\; {\rm CKMfitter},
\end{eqnarray}
and the value  from leptonic process later shown in Eq.~\eqref{eq:leptonic}. 
The global fit approaches  prefer to a lower $|V_{ub}|$ that is closer to the exclusive  determination, while the leptonic result is more consistent with the inclusive determination. 
Although the tension in  $|V_{ub}|$ is only    approximately  
3$\sigma$, it has already created a significant amount of speculations
about possible NP effects.  See Ref.~\cite{Crivellin:2014zpa} for a recent discussion.

%%%%%%%%%%%%%%%%%%
\subsection{Inclusive decays}
\label{subsectionInclusive decays}
%%%%%%%%%%%%%%%%%%

By integrating  the off-shell $W$ boson out,  one can obtain 
the effective   Hamiltonian
for the $b \to u \ell^-\bar{\nu}_\ell$ transition  
\begin{equation}
\mathcal{H}_{\rm eff} = \frac{G_F}{\sqrt{2}}V_{ub} \bar{u} \gamma_\mu(1-\gamma_5)b \bar{\ell} \gamma^\mu (1-\gamma_5) \nu_{\ell} + h.c.,
\end{equation}
with the Fermi constant $G_F$.

In inclusive decays $B\to X_{u}\ell\bar\nu_{\ell}$, $X_u$ refers to the sum of all possible final states.
The theoretical description of   $B\to X_{u}\ell\bar\nu_{\ell}$ decays is based on   heavy quark expansion, which has been validated  in various studies. Two-loop ${\cal O}(\alpha_{s}^{2})$ corrections have also been recently  calculated, for instance, in Ref.~\cite{Brucherseifer:2013cu}. Unfortunately, the total decay rate is very difficult  to measure due to the large background from the $B\to X_{c}\ell\bar\nu_{\ell}$.

Theoretical calculation of the partial decay rate in the region where the $B\to X_{c}\ell\bar\nu_{\ell}$ is suppressed requests the knowledge of an unknown  non-perturbative distribution function.  The explicit realisations  differ significantly
in the treatment of perturbative corrections and the parameterization of non-perturbative effects.

The shape function approach~\cite{Lange:2005yw,Bosch:2004th,Gambino:2007rp} is based on the introduction of  shape function that at leading order is
universal, and can be constrained from the $B\to X_{s}\gamma$. The shape function takes care of singular terms in the theoretical
spectrum; it has the role of a momentum distribution function of the $b$-quark
in the $B$ meson. However, no prediction is available for
the shape function and an ansatz is needed for its functional form. The
subleading shape functions and are not process
independent and thus are difficult to constrain.

Predictions based on resummed perturbative QCD use resummed perturbation
theory  to provide a perturbative calculation of the
on-shell decay spectrum in the entire phase space. It can extend the standard
Sudakov resummation framework by adding non-perturbative corrections, whose structure is determined by renormalon
resumming  \cite{Andersen:2005mj} or by an effective QCD coupling \cite{ Aglietti:2004fz, Aglietti:2006yb,  Aglietti:2007ik}.

On the experimental side, efforts have been made   to enlarge
the  experimental range, so as to reduce  the weight of the endpoint region.
Latest results by Belle \cite{Urquijo:2009tp} can
access $\sim  90$\% of the $ \bar B \rightarrow X_u \ell\bar\nu_{\ell}$ phase space, claiming an overall uncertainty of 7\% on $|V_{ub}|$.
A similar portion of the phase space is also covered in recent BaBar analysis \cite{Lees:2011fv}.

Though conceptually   different, all the above approaches can 
lead to   consistent results when the same inputs are used, and this situation has been reviewed in Ref.~\cite{Ricciardi:2013cda,Ricciardi:2014iga}.  The averaged values  have been given in Eq.~\eqref{eq:VubPDG}.

%%%%%%%%%%%%%%%%%%%%%%%%%%%%%%%%%%%%%%
\subsection{Exclusive decays $B\to \pi \ell\bar\nu$}
\label{subsec:Exclusive decays}
%%%%%%%%%%%%%%%%%%%%%%%%%%%%%%%%%%%%%%

%%%%%%%%%%%%%%%%%%%%%%
%%%%%%%%%%%%%%%%%%%%%%
\begin{figure}[ht]
\begin{center}
\includegraphics[width=0.64\columnwidth]{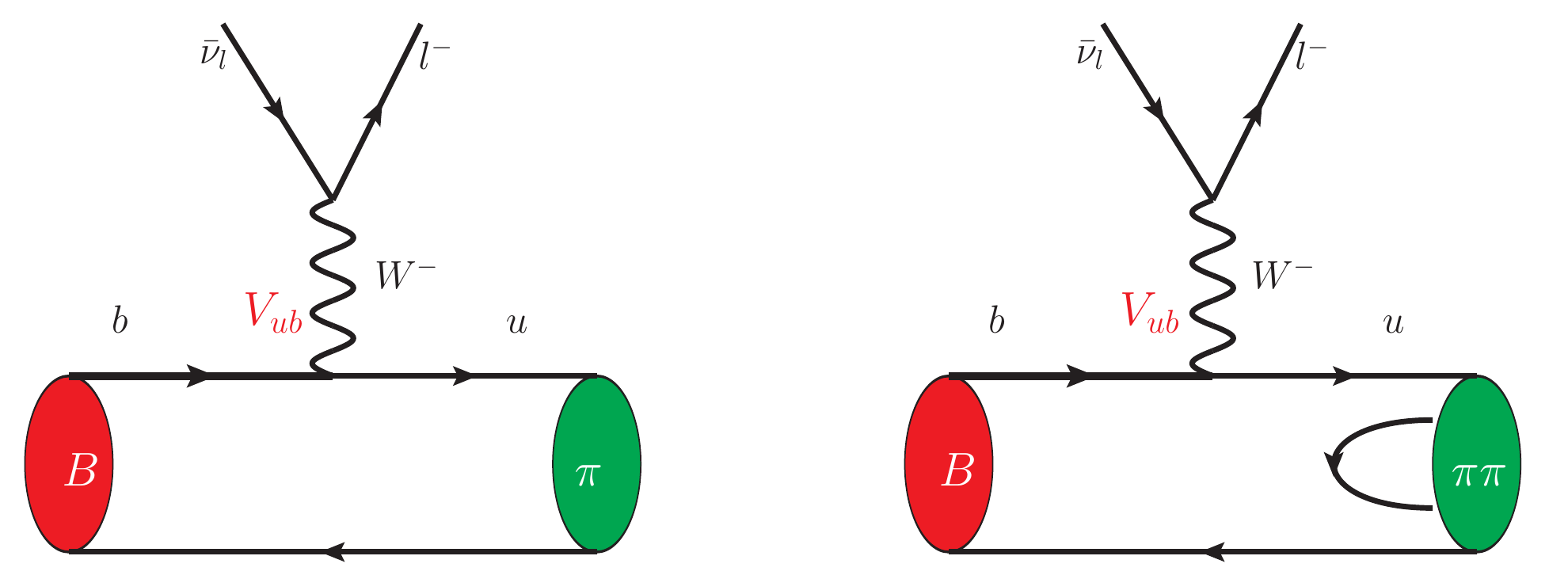} \hspace{5mm}
\includegraphics[width=0.3\columnwidth]{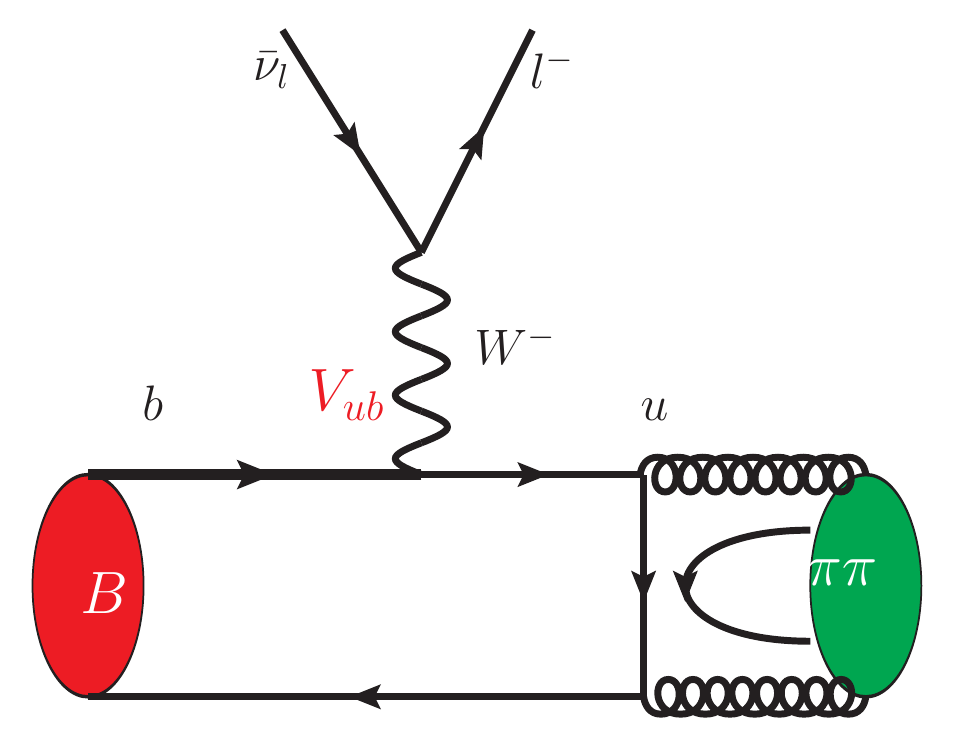}  
\caption{ Feynman diagrams   for the  $B\to \pi\ell\bar\nu$  (left panel) and $B\to \pi\pi\ell\bar\nu$ (second panel). Replacing the spectator quark, one may obtain the ones for the $B_{s}\to K\ell\bar\nu$ and $B_{s}\to K\pi\ell\bar\nu$. The last panel denotes  two gluon contributions to  $B\to \pi\pi\ell\bar\nu$. }
\label{fig:FeynmanBtopiVub}
\end{center}
\end{figure} 
%%%%%%%%%%%%%%%%%%%%%%
%%%%%%%%%%%%%%%%%%%%%%

Among all charmless $B$ decays observed  so far,
the $B\to \pi \ell\bar\nu_\ell(\ell=e,\mu)$ has been considered as the most 
reliable exclusive channel to extract the $|V_{ub}|$. Feynman diagrams for the  $B\to \pi\ell\bar\nu$   and $B\to \pi\pi\ell\bar\nu$ (for the convenience of later discussion) are shown in Fig.~\ref{fig:FeynmanBtopiVub}. There is a steady progress in
measuring the branching fractions and $q^2$-distribution on the experimental 
side~\cite{Lees:2012vv,Sibidanov:2013rkk}. On the theoretical side, at low $q^{2}$ (large recoil)   QCD light-cone sum rules (LCSR)  is applicable and the leading-twist  ${\cal O}(\alpha_{s}^{2}\beta_{0})$ corrections have been calculated in Ref.~\cite{Bharucha:2012wy} (see also Ref.~\cite{Li:2012gr,Khodjamirian:2011ub} for recent LCSR update of $B\to \pi$ form factors),   while the perturbative QCD calculation (pQCD) in $k_{T}$ factorisation~\cite{Keum:2000ph,Keum:2000wi,Lu:2000em,Lu:2000hj,Kurimoto:2001zj} is   rapidly developing~\cite{Li:2010nn,Li:2012nk,Wang:2012ab,Hu:2012cp,Cheng:2014fwa,Li:2012md,Li:2013xna}.  At high $q^{2}$ (low recoil) region, the Lattice QCD (LQCD) simulation has also achieved great progress~\cite{Bailey:2008wp,Dalgic:2006dt,AlHaydari:2009zr,Kawanai:2013qxa,Du:2013kea}.

For the sake of clarification,  let us consider a generic semi-leptonic decay $B\to P \ell \bar\nu_{\ell}$, where  $P$
stands for a light pseudoscalar meson.
The transition  is induced  by the vector current $V^\mu=\bar u\gamma^{\mu} b$ and the hadronic matrix element between the initial and final state can be decomposed as
\begin{eqnarray}
\langle P(p_P)| V^\mu |\overline B(p_B) \rangle = &&  F_{1}(q^2) \left (P^\mu-\frac{m^2_B-m_P^2}{q^2} q^\mu \right)  +
 F_0(q^2) \frac{m^2_B-m_P^2}{q^2} q^\mu,
 \label{eq:BtoPFormFactor}
\end{eqnarray}
where $P^\mu=p_B^\mu + p_P^\mu$. 
The   $F_{1}(q^2)$ and $F_0(q^2)$ depend only on $ q^\mu \equiv p_B^\mu - p_P^\mu$,   the momentum transferred to the lepton pair. In the
approximation where the leptons are massless, only $F_{1}(q^2)$
enters the partial decay rate: 
\begin{eqnarray}
\frac{d \Gamma( B \rightarrow \pi  \ell \bar\nu_{\ell})}{dq^2}= \frac{G_F^2  |\mathbf{p_\pi}|^3}{24 \pi^3} |V_{ub}|^2 |F_{1}(q^2)|^2,
\end{eqnarray}
where  $\mathbf{p_\pi}$ is the pion momentum  in the $B$ meson rest frame. For the results with massive leptons and various angular distributions, see Ref.~\cite{Meissner:2013pba}.

A great advantage in the study of $B$ decays is that the mass $m_b$ of  the $ b$-quark is large compared to the QCD  hadronic scale $\Lambda$
and therefore approximations and techniques of   heavy quark effective theory   can be used. Moreover in the large recoil region, the energy of the $\pi$ is also large compared to $\Lambda$ and thus simplifications of form factors can be achieved in soft-collinear effective  theory (SCET)~\cite{Bauer:2000ew,Bauer:2000yr,Bauer:2001yt,Beneke:2003pa}. Despite the factorisation property can be proved, non-perturbative  theoretical predictions for form factors are usually confined to limited regions of $q^2$.

Lattice  calculations have been performed in the kinematic region where the outgoing light hadron carries little energy. The  first lattice determinations of  $F_{1}(q^2)$  based on unquenched  simulations have been obtained by the Fermilab/MILC collaboration~\cite{Bailey:2008wp}  and the HPQCD collaboration
\cite{Dalgic:2006dt}, and they are  in substantial agreement.
In Ref. \cite{Bailey:2008wp},  the $b$-quark
is simulated by  using  the so-called Fermilab heavy-quark method, while
 the dependence of the form factor from $q^2$ is parameterized according to the
z-expansion  \cite{Becher:2005bg, Arnesen:2005ez, Bourrely:2008za}.
In Ref. \cite{Dalgic:2006dt},
the $b$-quark
is simulated by using nonrelativistic QCD and  the BK  parameterization \cite{Becirevic:1999kt}  is  extensively used for the $q^2$ dependence.
Recent results are also available on a fine lattice (lattice spacing $a \sim 0.04$ fm) in the quenched approximations by the QCDSF collaboration \cite{AlHaydari:2009zr}.  Preliminary results  from unquenched Lattice QCD simulation by the FNAL/MILC collaboration can be found in Ref.~\cite{Du:2013kea}. Based on the $2+1$ flavour domain-wall fermion and Iwasaki gauge-field ensembles generated by the RBC/UKQCD collaboration, Ref.~\cite{Kawanai:2013qxa} has also updated the $B\to \pi$ form factors.

As a reconciliation of the original QCD sum rule approach~\cite{Shifman:1978bx,Shifman:1978by}  and the
application of perturbation theory to hard processes,  LCSR
exhibit  several advantages in the calculation of quantities like   meson form 
factors~\cite{Craigie:1982ng,Braun:1988qv,Chernyak:1990ag,Belyaev:1994zk,Colangelo:2000dp}.   In the hard scattering region the light-cone operator product expansion (OPE) is applicable, based on which form 
factors are expressed as a convolution of  light-cone distribution amplitudes (LCDA) with a perturbatively  calculable hard kernel. Leading twist  and a few sub-leading twist LCDA are dominant. Contributions
corresponding to higher twist and/or higher multiplicity pion  distribution amplitudes  are
suppressed by  powers of $1/m_{b}$ allowing one to
truncate the expansion after a few low twist contributions.

Latest experimental  data  on   $ B \rightarrow \pi \ell\bar\nu_{\ell}$ decays come  from BaBar \cite{Lees:2012vv} and  Belle \cite{Sibidanov:2013rkk}.
The measured differential decay rates can be fit at low and high $q^2$  according to LCSR and lattice QCD approaches, respectively. A simultaneous
fit to  LQCD results    has been performed by the two collaborations, which lead to
\begin{eqnarray}
 |V_{ub}| =  
\bigg\{
\begin{array}{cc}
  &    (3.52\pm0.29)\times 10^{-3}   \;\;\; {\rm Belle}  \\
  &    (3.25\pm0.31)\times 10^{-3}\;\;\;    {\rm Babar} \\ 
\end{array},
\end{eqnarray} 
where  errors are the combined experimental and theoretical uncertainty. 
Both values  are consistent with   previous results, but the Belle result has a higher central value by about $1\sigma$.

%%%%%%%%%%%%%%%%% 
\subsection{Other  Semi-Leptonic  $B$ decay modes} 
%%%%%%%%%%%%%%%%%

If the hadronic  final state is a vector meson $V$,  both   vector and axial currents contribute to the  $B \to V \ell \bar\nu_{\ell}$
 \begin{eqnarray}
  \langle V(p_{V},\epsilon)|V^{\mu}|\overline B(p_B)\rangle
   &=&-\frac{2V(q^2)}{m_B+m_{V}}\epsilon^{\mu\nu\rho\sigma} \epsilon^*_{\nu}  p_{B\rho}p_{V\sigma}, 
   \end{eqnarray}
   %\\
   \begin{eqnarray}
  \langle  V(p_V,\epsilon)|A^{\mu}|\overline B(p_B)\rangle
   &=&2im_{V} A_0(q^2)\frac{\epsilon^* \cdot  q }{ q^2}q^{\mu}
    +i(m_B+m_{V})A_1(q^2)\left[ \epsilon^*_{\mu }
    -\frac{\epsilon^* \cdot  q }{q^2}q^{\mu} \right] \nonumber\\
    &&-iA_2(q^2)\frac{\epsilon^* \cdot  q }{  m_B+m_{V} }
     \left[ P^{\mu}-\frac{m_B^2-m_{V}^2}{q^2}q^{\mu} \right],
 \end{eqnarray}
where $ \varepsilon_{\mu\sigma \nu \rho}$ is the Levi-Civita  tensor with the convention $\epsilon^{0123}=1$, $\epsilon^\mu$ is
the vector polarization vector  and
$P=p_{B}^{\mu}+p_{V}^{\mu}$. Here the momentum transferred to the lepton pair is $q^\mu \equiv  p_B^\mu-p_V^\mu$. The differential decay width of $B\to P_{1}P_{2}\ell\bar\nu_{\ell}$ including the resonating contribution from $B\to V\ell\bar\nu_{\ell}$ can be found in   Ref.~\cite{Meissner:2013pba}.

Recently, BaBar and Belle collaborations have reported significantly improved branching ratios of other
 heavy-to-light semileptonic decays, that reflects on
increased precision
for $|V_{ub}|$ values inferred by these decays.
These channels include the $B\to \rho\ell\bar\nu_{\ell}$~\cite{Sibidanov:2013rkk}, $B\to \omega\ell\bar\nu_{\ell}$~\cite{Lees:2012vv,Lees:2013gja,Sibidanov:2013rkk} and $ B \rightarrow \eta^{(\prime)}\ell \nu_{\ell} $~\cite{Lees:2012vv}.  For  the $B\to \rho$ and $B\to\omega$ form factors, Belle~\cite{Sibidanov:2013rkk}  has used   LCSR~\cite{Ball:2004rg} and LQCD from   UKQCD collaboration~\cite{DelDebbio:1997kr},  and the extracted $|V_{ub}|$ is in agreement  with the ones from   $B\to \pi\ell\bar\nu_{\ell}$. Babar measurement of the $B\to\omega\ell\bar\nu_{\ell}$ used  LCSR form factors~\cite{Ball:2004rg} and obtained similar values~\cite{Lees:2012vv}. 
The experimental data on$ B \rightarrow \eta^{(\prime)} \ell \bar\nu_{\ell} $~\cite{Lees:2012vv} are consistent with theoretical predictions in Refs.~\cite{Chen:2009qk,Wang:2013ix}, but no result on $|V_{ub}|$ is extracted.

Apart from the observed processes, new channels  that are able to extract  $|V_{ub}|$ and thus can reduce statistical and systematic 
uncertainties  also deserve theoretical and experimental investigations in future.  
The $\overline B_s^0\to K^+\ell^-\bar \nu$ and $\overline B_s^0\to K^{*+}\ell^-\bar \nu$ decays  are of this type and 
have been studied  using the state-of-the-art knowledge 
of   form factors in Ref.~\cite{Meissner:2013pba}; those include not only the recent LQCD
calculation~\cite{Horgan:2013hoa} and  the  LCSR~\cite{Ball:2004rg},  
but also various sets of calculations  from the factorisation approach~\cite{Wang:2012ab} 
and QCD-inspired  models~\cite{Faustov:2013ima,Cheng:2003sm,Lu:2007sg,Verma:2011yw}.

The baryonic $\Lambda_b \to p$ matrix elements of the vector and axial vector $b\to u$ currents are parametrized in terms of six
independent form factors. At  leading-order in $1/m_{b}$,
which becomes exact in the limit $m_b\to \infty$ and is a good approximation at the physical value of $m_b$,
only two independent form factors remain, and the matrix element with arbitrary Dirac matrix $\Gamma$ in the current
can be written as \cite{Mannel:1990vg, Hussain:1990uu, Hussain:1992rb}
\begin{equation}
\langle p(p', s') | \:\bar{u} \Gamma b \: | \Lambda_b(v, s) \rangle =
\overline{u}_p(p',s')\left[ F_1 + \slashed{v}\:F_2 \right] \Gamma\: u_{\Lambda_b}(v, s). \label{eq:FFdef}
\end{equation}
Here, $v$ is the four-velocity of the $\Lambda_b$ baryon, and the form factors $F_1$, $F_2$ are functions of $p'\cdot v$,
the energy of the proton in the $\Lambda_b$ rest frame. Note that in leading-order SCET, which applies in the limit of large $p'\cdot v$,
the form factor $F_2$ vanishes \cite{Feldmann:2011xf, Mannel:2011xg, Wang:2011uv}.
Calculations of the $\Lambda_b \to p$ form factors have been performed using light-front quark model~\cite{Wei:2009np}, QCD sum rules \cite{Huang:1998rq,Carvalho:1999ia}
and LCSR~\cite{Huang:2004vf, Wang:2009hra, Azizi:2009wn, Khodjamirian:2011jp}, and LQCD~\cite{Detmold:2013nia}.  We shall wait for the future experimental measurements from  LHC and SuperB factories  which will make this decay mode also useful to extract the $|V_{ub}|$.

%%%%%%%%%%%%%%%%%%%%%%%%
\subsection{Purely leptonic decays}
\label{subsec:purelyLeptonic}
%%%%%%%%%%%%%%%%%%%%%%%%

In the absence of NP, $B^{-}  \to \ell^{-}\bar \nu_\ell$  decays  are simple tree-level decays,
where  the two quarks in the initial state, $b$ and $\bar u$, annihilate to a $W^-$ boson.
They are particularly sensitive to physics beyond the SM, since a new particle,
for example a charged Higgs boson, may lead the decay taking
the place of the $W^-$ boson. 
In the SM, the  ${\cal B}(B^-  \to \tau^- \bar \nu_\tau)$  is given as
\begin{eqnarray} 
{\cal{B}}(B^-  \to \tau^- \bar \nu_\tau) = \frac{G_F^2 m_B m_\tau^2}{8 \pi} \left(1- \frac{m_\tau^2}{m_B^2} \right)^2 f_B^2 |V_{ub}|^2 \tau_B,
\end{eqnarray}
 and its  measurement can provide a direct experimental determination of the
product  $f_B  |V_{ub}|$. See Ref.~\cite{Soffer:2014kxa} for a recent review on $B$ decays into a $\tau$-lepton.

Experimentally, it is challenging to identify the $B^- \to \tau^-  \bar \nu_\tau$   decay because it involves more than one neutrino
in the final state and therefore cannot be kinematically
constrained. This can be measured in $\Upsilon(4S)$ decays, where one of the $B$ mesons from the $\Upsilon(4S)$ can be tagged in hadronic and semileptonic final states.   One
then compares properties of the remaining particles to those expected for
signal and background.
The   $ B^{-} \to \tau^{-}\bar \nu_{\tau}$ was first  observed  by Belle in 2006 \cite{Ikado:2006un}, and the new average   has combined the results from BaBar~\cite{Aubert:2009wt,Lees:2012ju} and Belle~\cite{Hara:2010dk,Adachi:2012mm}:
\begin{eqnarray} 
{\cal{B}}( B^- \to \tau^- \bar\nu_\tau) =(1.14\pm0.23) \times 10^{-4}.\label{eq:BtoTauNuBR}
\end{eqnarray}

%%%%%%%%%%%%%%%%%%%%%%
%%%%%%%%%%%%%%%%%%%%%%
\begin{figure}[ht]
\begin{center}
\includegraphics[width=0.4\columnwidth]{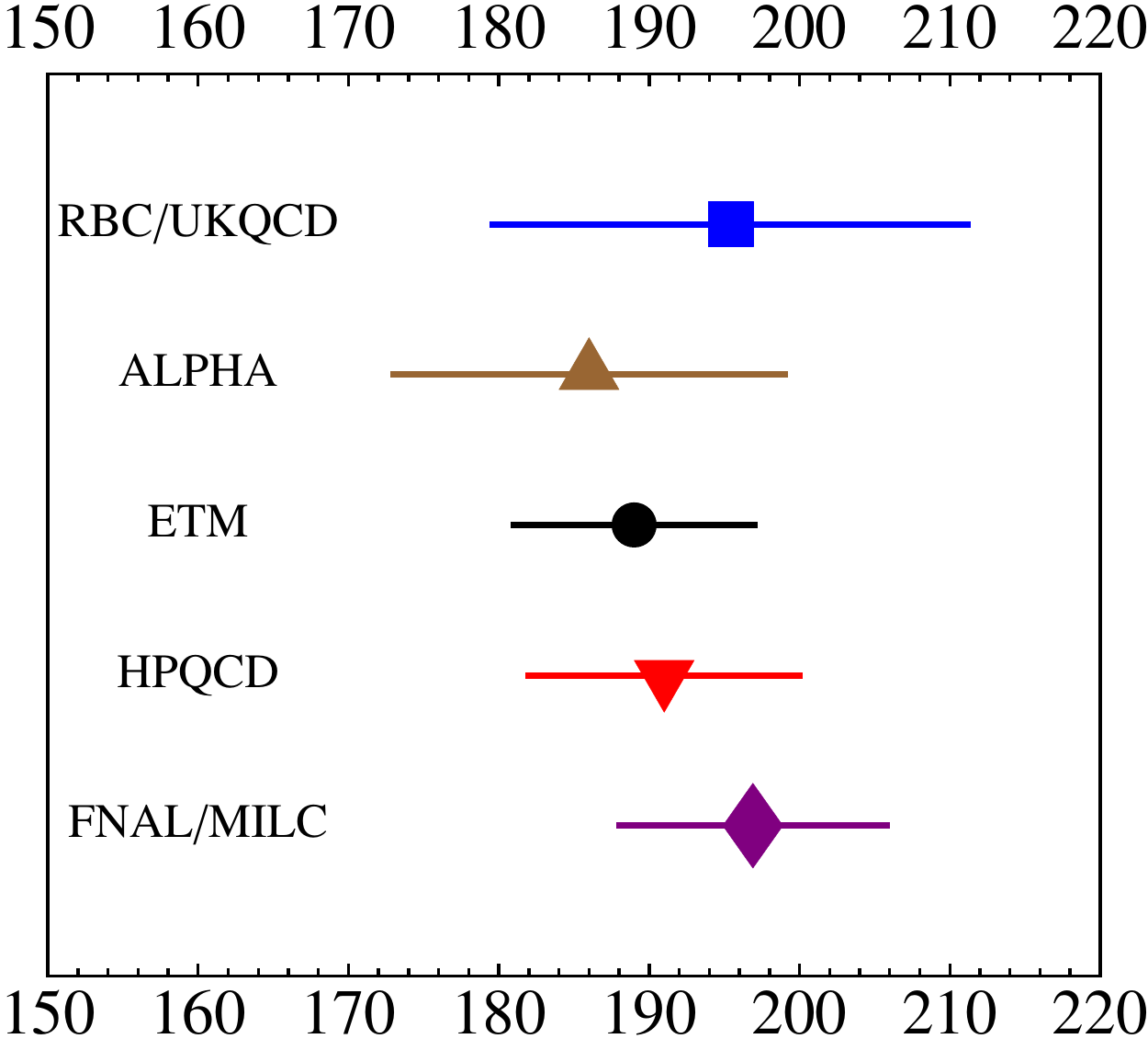}  
\caption{ Decay constant $f_{B}$ (in units of MeV) from recent  Lattice QCD simulations:   FNAL/MILC~\cite{Bazavov:2011aa}, HPQCD~\cite{Na:2012kp},  ETM with twisted -mass~\cite{Carrasco:2013zta}, ALPHA~\cite{Bernardoni:2014fva}, RBC/UKQCD~\cite{Christ:2014uea}. }
\label{fig:fBCompare}
\end{center}
\end{figure} 
%%%%%%%%%%%%%%%%%%%%%%
%%%%%%%%%%%%%%%%%%%%%%

The extraction of $|V_{ub}|$ relies on the decay constant $f_{B}$, and recent LQCD simulations include    FNAL/MILC~\cite{Bazavov:2011aa}, HPQCD~\cite{Na:2012kp},  ETM with twisted -mass~\cite{Carrasco:2013zta}, ALPHA~\cite{Bernardoni:2014fva}, RBC/UKQCD~\cite{Christ:2014uea}. Based on these results that are collected in Fig.~\ref{fig:fBCompare} and assuming the errors are independent, we obtain an average
\begin{eqnarray}
f_{B}=(191.6\pm 4.4) {\rm MeV},
\end{eqnarray}
which corresponds to
\begin{eqnarray}
 |V_{ub}|= (4.2\pm 0.4\pm 0.1) \times 10^{-3}.\label{eq:leptonic}
\end{eqnarray}
The first errors come from the ${\cal B}( B^- \to \tau^- \bar\nu_\tau) $ as shown in Eq.~\eqref{eq:BtoTauNuBR} and the second ones are from  $f_{B}$.  This value seems to be more consistent with the result from the inclusive  $b\to u\ell\bar\nu$ decay mode. 

It is worthwhile to point out that compared to the averaged branching fraction  in Eq.~\eqref{eq:BtoTauNuBR}, the recent Belle measurement has a lower central value
\begin{eqnarray}
{\cal{B}}( B^- \to \tau^- \bar\nu_\tau) =(0.72^{+0.27}_{-0.25}\pm0.11) \times 10^{-4}.\label{eq:BtoTauNuBRNew}
\end{eqnarray}
This corresponds to a smaller $|V_{ub}|$ with a larger uncertainty. Future measurements of ${\cal B}( B^- \to \tau^- \bar\nu_\tau) $ will be of great value to make clarifications.

\subsection{Theoretical developments on multi-body semileptonic  B decays}

The $B\to \rho \ell\bar \nu$ and $B_s\to K^* \ell\bar \nu$  reactions receive a 
complexity due to the large width of $\rho$ (about 150~MeV) and $K^*$ (about~50 MeV).  As both $\rho$ and $K^{*}$ decay into two pseudo-scalars,
these processes  are  quasi-four-body decays, and  in
principle   other    resonant and nonresonant states may    contribute
in the same final state and thereby   the $|V_{ub}|$  extraction is contaminated.

%%%%%%%%%%%%%%%%%%%%%%
%%%%%%%%%%%%%%%%%%%%%%
%%%%%%\begin{figure}[ht]
%%%%%%  \begin{center}
%%%%%%    \includegraphics[width=0.8\columnwidth]{plots/four-body.pdf} 
%%%%%%   \caption{ Kinematics in the $B\to \pi\pi l\bar\nu$  and $B_{s}\to K\pi l\bar\nu$      }
%%%%%%    \label{fig:kinematicsFourBody}
%%%%%%  \end{center}
%%%%%%\end{figure} 
%%%%%%%%%%%%%%%%%%%%%%
%%%%%%%%%%%%%%%%%%%%%%

%%%%%%We consider the kinematics  for the $B_s\to K\pi  \ell \bar \nu$ as shown in 
%%%%%%Fig.~\ref{fig:kinematicsFourBody}.  {The} $K\pi $ system  moves along the $z$-axis 
%%%%%%in the $\overline B_s$ rest-frame. 
%%%%%%$\theta_K(\theta_l)$ is defined in the $K\pi$ (lepton pair) rest frame as the angle
%%%%%%between $z$-axis and the direction of motion of the $K $ ($\ell^-$), respectively.  
%%%%%%The azimuth angle $\phi$ is the relative angle between the $K\pi$ decay and lepton pair
%%%%%%planes. 

%%%%%%%%%%%%%%%%%%%%%%
%%%%%%%%%%%%%%%%%%%%%%
\begin{figure}[ht]
\begin{center}
\includegraphics[width=0.5\columnwidth]{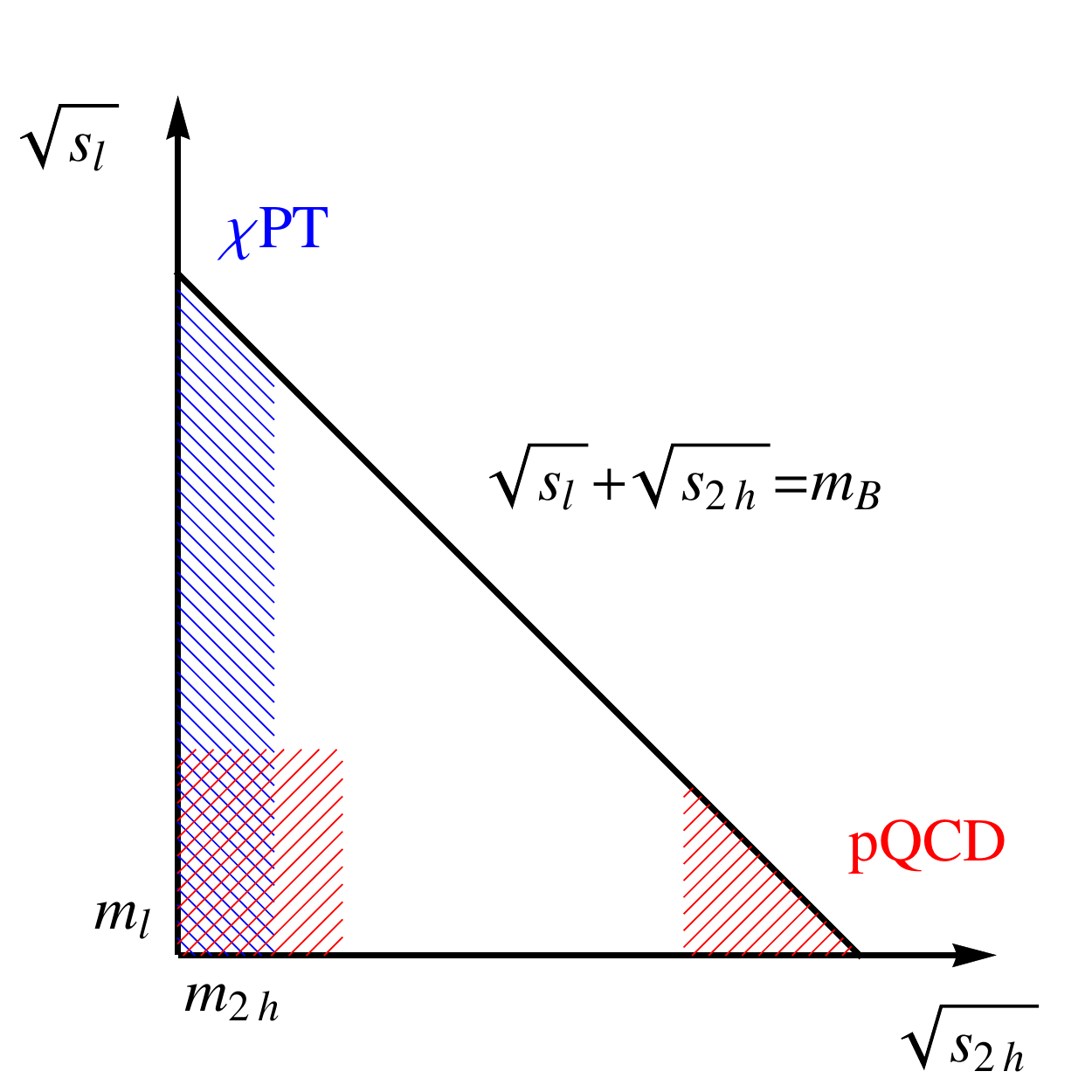} 
\caption{ A sketch of the phase space in the $B\to P_{1}P_{2} \ell\bar\nu$. $\sqrt{s_{l}}$ is the lepton pair invariant mass, while   $\sqrt{s_{2h}}$ is the invariant mass of the two hadrons. When the two hadron has a small invariant mass, the interaction is   strong and can be described by the chiral perturbation theory. If one or two hadrons in the final state move fast, the hard scattering amplitude can be calculated in QCD.  }
\label{fig:sls2hKinematicRegion}
\end{center}
\end{figure} 
%%%%%%%%%%%%%%%%%%%%%%
%%%%%%%%%%%%%%%%%%%%%%

A general formalism has been developed  to incorporate various partial-wave
contributions~\cite{Meissner:2013pba} (similar with the $B\to K^*_J(\to K\pi)\ell^+\ell^-$ 
case~\cite{Kruger:1999xa,Lu:2011jm,Doring:2013wka,Li:2010ra,Becirevic:2012dp,Matias:2012qz,Blake:2012mb,Bobeth:2012vn,Descotes-Genon:2013vna,Descotes-Genon:2013wba} and see also~\cite{Kopp:1990yx,Lee:1992ih,Ananthanarayan:2005us,Faller:2013dwa}),  through which   branching fractions, 
forward-backward asymmetries and
polarisations can be   projected out.   It is worthwhile to  stress  that   
the S-wave, whose effects are not negligible,   
can {\it not} be expressed    in terms of  a Breit-Wigner formula,  especially for the broad scalar
meson $\kappa\equiv K^*_0(800)$ and $\sigma\equiv f_{0}(600)$. This broad nature is also stressed  from  the Roy-Steiner representations of the $\pi K$ scattering~\cite{Buettiker:2003pp,DescotesGenon:2006uk}.

The kinematics of the $B\to P_{1}P_{2}\ell\bar\nu_{\ell}$ is shown in Fig.~\ref{fig:sls2hKinematicRegion}. In this figure the $\sqrt{s_{l}}$ is the lepton pair invariant mass, while the $\sqrt{s_{2h}}$ is the invariant mass of the two hadrons.  When the invariant mass of the two hadron is small for instance below 1GeV,   chiral perturbation theory ($\chi$PT) is applicable to handle their interactions. When one or two hadrons move fast in the final state, there is a large momentum transfer and thus QCD perturbation theory can be used to calculate the transition.

Using   $B\to K\pi$ as the explicit example, 
the  matrix elements 
\begin{eqnarray}
    \langle (K\pi)_S|\bar s \gamma_\mu\gamma_5 b|\overline B 
 \rangle  &=&   \frac{-i}{m_{K\pi}} \bigg\{ \bigg[P_{\mu}
 -\frac{m_{B }^2-m_{K\pi}^2}{q^2} q_\mu \bigg] {\cal F}_{1}^{B\to K\pi}(m_{K\pi}^2, q^2) \nonumber\\
 &&
 +\frac{m_{B }^2-m_{K\pi}^2}{q^2} q_\mu  {\cal F}_{0}^{B\to K\pi}(m_{K\pi}^2, q^2)  \bigg\},\nonumber\\ 
  \langle (K\pi)_S|\bar s \sigma_{\mu\nu} q^\nu \gamma_5 b|\overline B
 \rangle  &=& -   \frac{{\cal F}_T^{B\to K\pi}(m_{K\pi}^2, q^2)}{m_{K\pi}(m_B+m_{K\pi})}  \big[q^2 P_{\mu}- (m_{B }^2-m_{K\pi}^2) q_\mu \big],
 \label{eq:generalized_form_factors}
\end{eqnarray}
define the S-wave generalized form factors ${\cal F}_i$~\cite{Meissner:2013pba,Doring:2013wka,Meissner:2013hya} and project out the S-wave contributions. 
Here,  $P=p_B+ p_{K\pi}$ and $q=p_B- p_{K\pi}$.  

%%%%%%%%%%%%%%%%%%%%%%
%%%%%%%%%%%%%%%%%%%%%%
\begin{figure}[ht]
\begin{center}
\includegraphics[width=0.8\columnwidth]{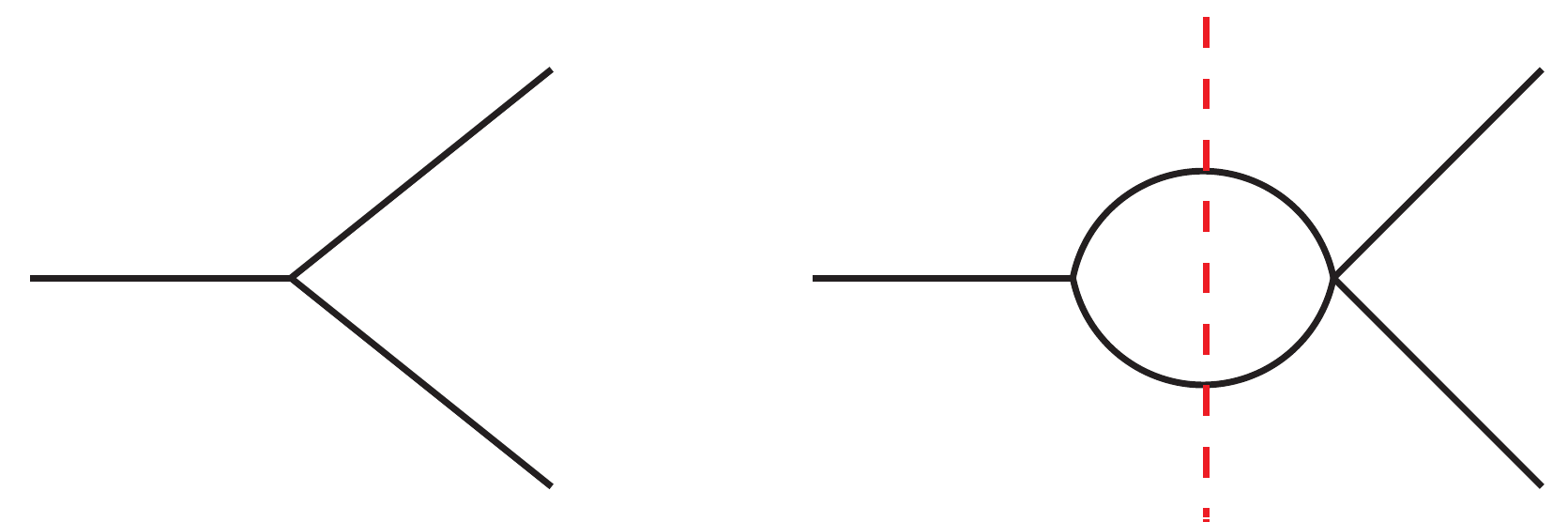}  
\caption{In the elastic region, the imaginary part (discontinuity) of a form factor (left panel) comes from the one loop diagram shown in the right panel, guaranteed by the Watson's theorem }
\label{fig:watson}
\end{center}
\end{figure} 
%%%%%%%%%%%%%%%%%%%%%%
%%%%%%%%%%%%%%%%%%%%%%

To avoid the finite-width problem, we will make  use of  Watson's
theorem~\cite{Watson:1952ji} which allows a reliable description in terms of   scalar form factors. 
As depicted in Fig.~\ref{fig:watson},
Watson's theorem  implies that  phases measured in the $K\pi$ elastic scattering and in a decay channel where the
$K\pi$ system decouple with other hadrons  are equal (modulo
$\pi$ radians). This leads to
\begin{eqnarray}
 \langle (K\pi)_S |\bar s \Gamma b|\overline B\rangle   
 \propto F_{K\pi}(m_{K\pi}^2),
\end{eqnarray}
where  the strangeness-changing scalar $K\pi$ form factors.

\begin{figure}[ht]
  \begin{center}
    \includegraphics[width=0.5\columnwidth]{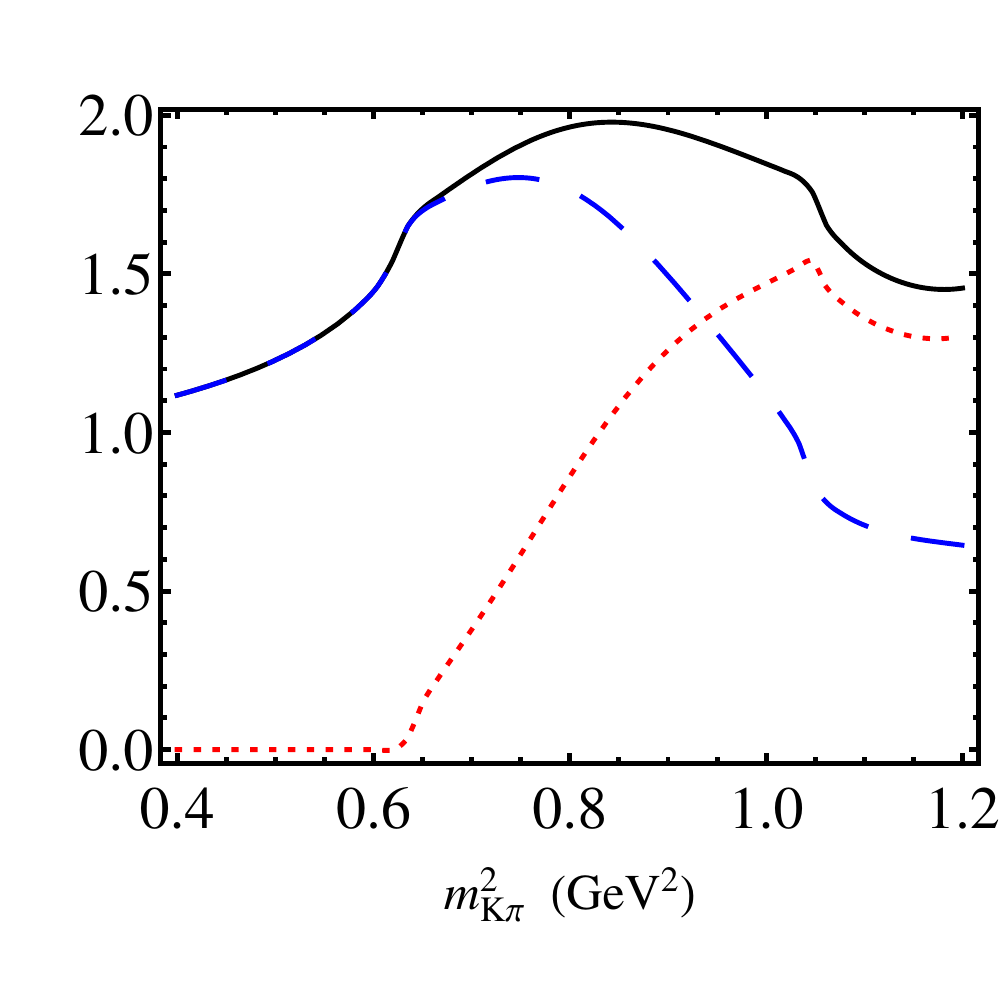}  
    \caption{ Scalar $K\pi$ form factors calculated in unitarized $\chi$PT approach. Solid, dashed and dotted lines correspond to the magnitude, the real and   imaginary part, in order.}
    \label{fig:scalarKpiFormFactor}
  \end{center}
\end{figure}

The $K \pi$ scattering  is strictly elastic below the $K+3\pi$ threshold, about 911 MeV. Inelastic contributions in the $K \pi$ scattering comes from the $K+3\pi$ or $K\eta^{(')}$.  In the region from 911 MeV to 1 GeV, the $K+3\pi$ channel has a limited phase space, and thus is generically suppressed.  Moreover, as a case-dependent study, it has been demonstrated   states with two additional pions will  not give sizeable contributions to physical observables~\cite{Bar:2012ce}. 
The $K\eta^{(')}$ coupled-channel effects can be included in the unitarized approach of $\chi$PT~\cite{Gasser:1990bv,Oller:2000ug,Gardner:2001gc,Meissner:2000bc,Frink:2002ht,Bijnens:2003uy,Lahde:2006wr,Guo:2012yt,Donoghue:1990xh,Jamin:2000wn,Jamin:2001zq,Jamin:2006tj,Bernard:2007tk,Bernard:2009ds}.  
We quote   recently updated  results   from Ref.~\cite{Doring:2013wka}  in Fig.~\ref{fig:scalarKpiFormFactor}. Solid, dashed and dotted lines correspond to the magnitude, the real and   imaginary part, in order.

%%%%%%%%%%%%%%%%%%%%%%
%%%%%%%%%%%%%%%%%%%%%%
%%%%%%%%%%%%%%%%%%%%%%
%%%%%%%%%%%%%%%%%%%%%%

\begin{figure}[ht]
\begin{center}
\includegraphics[width=1.0\columnwidth]{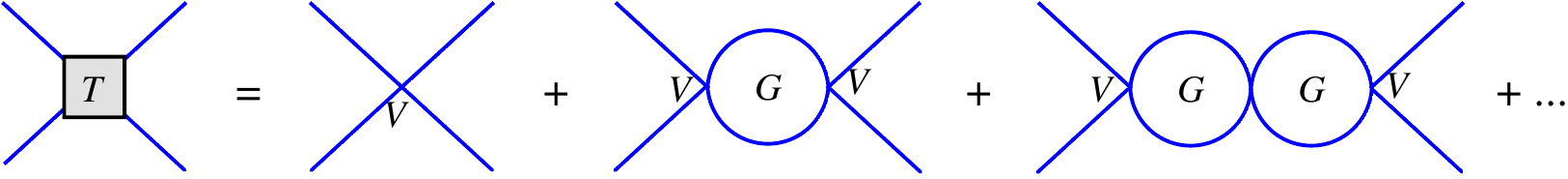}  
\caption{Sketch of the resummation scheme in the unitarized approach. Here, T denotes the total
scattering amplitude, V is the leading order amplitude and G is the loop integral.  }
\label{fig:uchpt}
\end{center}
\end{figure}

For the $\pi\pi$ and $K\bar K$ channel, the coupled channel effects should be taken into account, and moreover the standard $\chi$PT may fail to describe the $K\bar K$ system as the involved invariant mass is close to 1 GeV. It has been proposed that the unitarized approach which can sum  higher order corrections and extend the applicability to energy around 1 GeV. A sketch of the resummation scheme  is shown in Fig.~\ref{fig:uchpt}.  Here, T denotes the total
scattering amplitude, V is the leading order amplitude and G is the loop integral. For detailed discussions on these form factors, we refer the reader to Ref.~\cite{Meissner:2000bc,Lahde:2006wr}. The generalisation to the P-wave case is under progress~\cite{twoHadronFormFactor}. Once these quantities are available, they can be used in the study of charmless three-body $B$ decays~\cite{Chen:2002th,ElBennich:2009da,Bediaga:2013ela,Zhang:2013oqa,Cheng:2013dua,Xu:2013rua,Xu:2013dta,Cheng:2014uga,Li:2014fla,Wang:2014ira,Li:2014oca,Bhattacharya:2013cvn}.

%%%%%%%%%%%%%%%%%%%%%%
%%%%%%%%%%%%%%%%%%%%%%
%%%%%%%%%%%%%%%%%%%%%%
%%%%%%%%%%%%%%%%%%%%%%

In the large recoil region, the $K\pi$ system  with invariant mass below 1 GeV  moves very fast and therefore can be treated as a light hadron.  As shown later this  system has similar LCDA  with the ones for  a light hadron.  The transition matrix elements for $B\to K\pi$ may   be factorized in the same way as the ordinary  $B$-to-light ones. 
At the leading power, the form factors obey   factorization~\cite{Meissner:2013hya}: 
\begin{eqnarray}
{\cal F}_i = C_i \xi(q^2) + \Delta {\cal F}_i, 
\end{eqnarray}
where $C_i$ are the short-distance and calculable functions, and $\xi$ is a universal soft form factor derived  from  the heavy quark $m_b\to\infty$  and large energy  $E\to\infty$ limit~\cite{Charles:1998dr}. Symmetry breaking terms, starting at order $\alpha_s$,  can be encoded into $\Delta {\cal F}_i$, and   expressed as a convolution  in terms of the LCDA~\cite{Beneke:2000wa,Bauer:2002aj,Beneke:2003pa,Beneke:2004rc,Beneke:2005gs}. 

%As for the soft form factors, dispute exists about the expansion in $\alpha_s$. If this form factor is at $\alpha_s^0$, we can say at leading order in $\alpha_s$, there is only one form factor parametrizing the matrix elements. If $\xi(q^2)$ starts at $\alpha_s$, the reduction  does not exists.  

%In the heavy quark $m_b\to\infty$  and large energy  $E_{K\pi}\to\infty$ limit, interactions of the heavy and light system can be  expanded in  $1/m_b$.  At the leading-order in $\alpha_s$ and $1/m_b$, one has the so-called  large
%recoil symmetries, which to a large extent simplifies the dynamics:  
%\begin{eqnarray}
 %{\cal F}_1= \frac{m_{B}^2-m_{K\pi}^2}{m_{B}^2-m_{K\pi}^2-q^2 } 
 %{\cal F}_0=  \frac{m_B}{m_B+m_{K\pi}}  {\cal F}_T \equiv {\cal \xi} (m_{K\pi}^2, q^2)~. 
%\end{eqnarray}
%Once the distribution amplitudes of the two-hadron state are available as we will discuss in the following, symmetry breaking effects  can systematically be explored  in an analogous way with the analysis of the $B$ to light pseudoscalar form factors~\cite{Beneke:2000wa,Beneke:2003pa,Beneke:2004rc,Beneke:2005gs}. 

In Ref.~\cite{Meissner:2013hya},    LCSR has been chosen  to calculate the  ${\cal F}_i$.  
The calculation  is based on the expansion of the
T-product in the correlation function  near the light-cone, which  produces
matrix elements of non-local quark-gluon operators. 
These quantities  are in terms of the generalized LCDA  of
increasing twist~\cite{Diehl:1998dk,Polyakov:1998ze,Kivel:1999sd,Diehl:2003ny}: 
\begin{eqnarray}
 \langle {(K\pi)_S}|\bar s (x)\Gamma d(0)|0\rangle,
\end{eqnarray}
with $\Gamma$ being a Dirac matrix.  Higher-order calculation will  request    the gluonic LCDA,  whose  contribution  is shown in Fig.~\ref{fig:FeynmanBtopiVub}.  Based on the perturbative calculation in Ref.~\cite{Charng:2006zj,Wang:2009rc,Wang:2013ix}, there is no endpoint singularity and one may directly adopt the collinear  factorisation scheme.

%%%%%%%%%%%%%%%%%%%%%%%%%%%%%%%%%%%%%%%%%%%%%%%%%%%%%%%%%%%%%%%%%%%%%%%%%%%%%%%%
\begin{figure}[ht]
\begin{center}
\includegraphics[width=0.45\columnwidth]{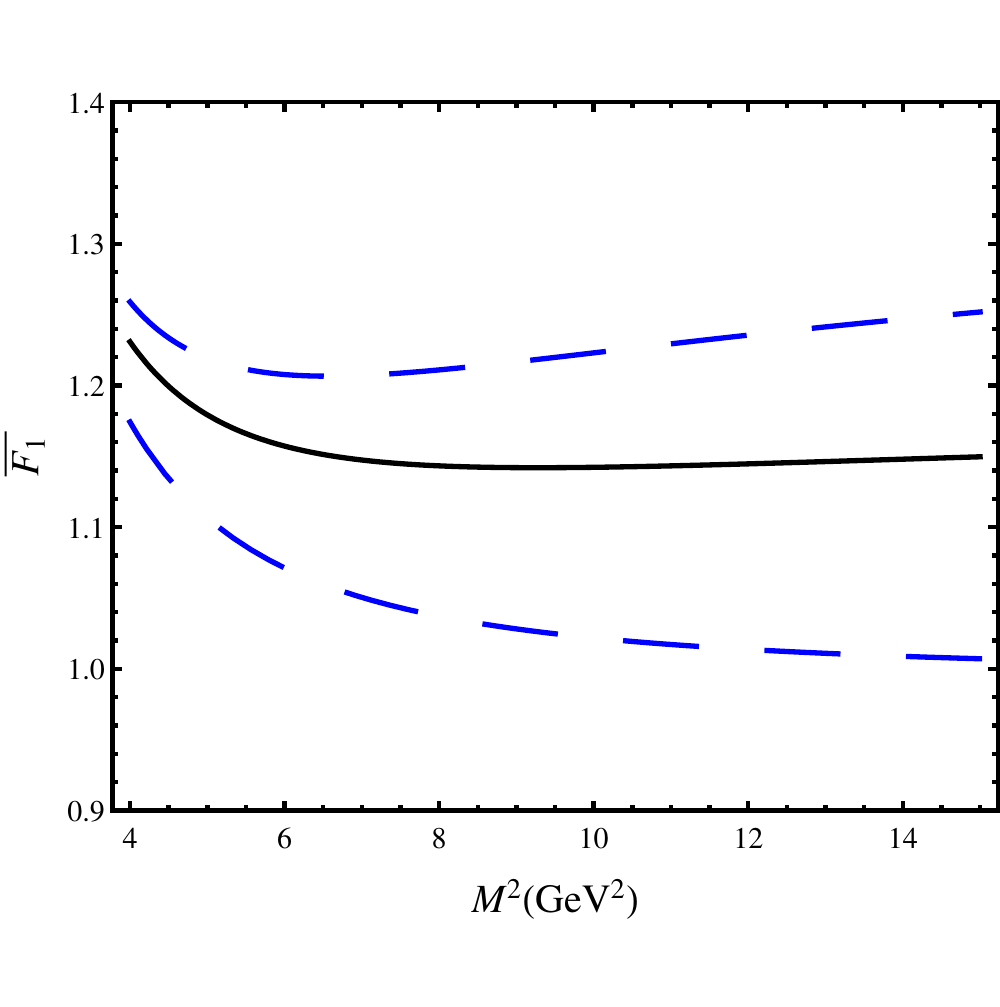}  
\includegraphics[width=0.45\columnwidth]{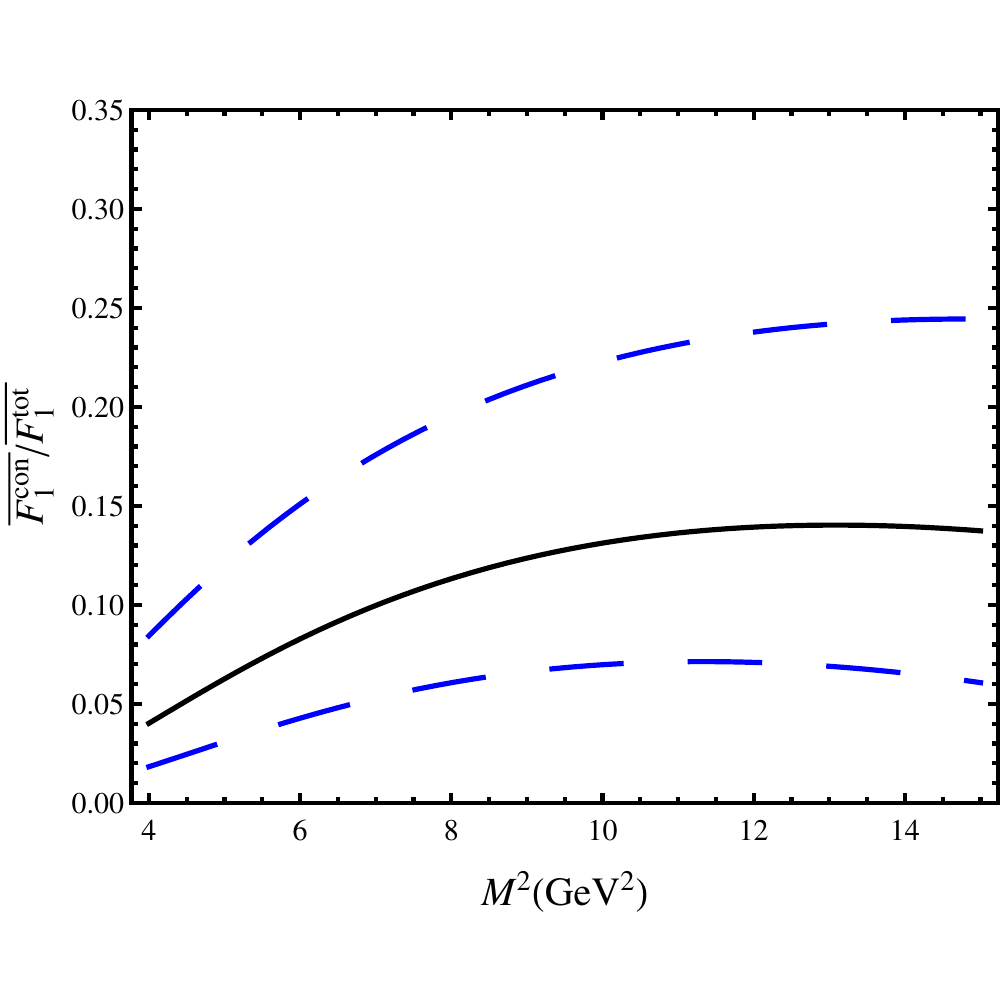}  
\caption{The dependence of   $\bar F_1$ (left panel) and the ratio of   continuum and total contributions (right panel) on the Borel parameter. Solid lines denote the central value while   dashed curves correspond to variations of threshold parameter: $s_0= (34\pm 2) {\rm GeV}^2$.  Results for $\bar F_1$  are   stable when 
$M^2>  6\,  {\rm GeV}^2$, while the continuum contribution is mostly  smaller than  $30\%$.  } 
\label{fig:BorelDepen}
\end{center}
\end{figure}
%%%%%%%%%%%%%%%%%%%%%%%%%%%%%%%%%%%%%%%%%%%%%%%%%%%%%%%%%%%%%%%%%%%%%%%%%%%%%%%% 

%%%%%%%%%%%%%%%%%%%%%%%%%%%%%%%%%%%%%%%%%%%%%%%%%%%%%%%%%%%%%%%%%%%%%%%%%%%%%%%%
\begin{figure}[ht]
\begin{center}
\includegraphics[width=0.7\columnwidth]{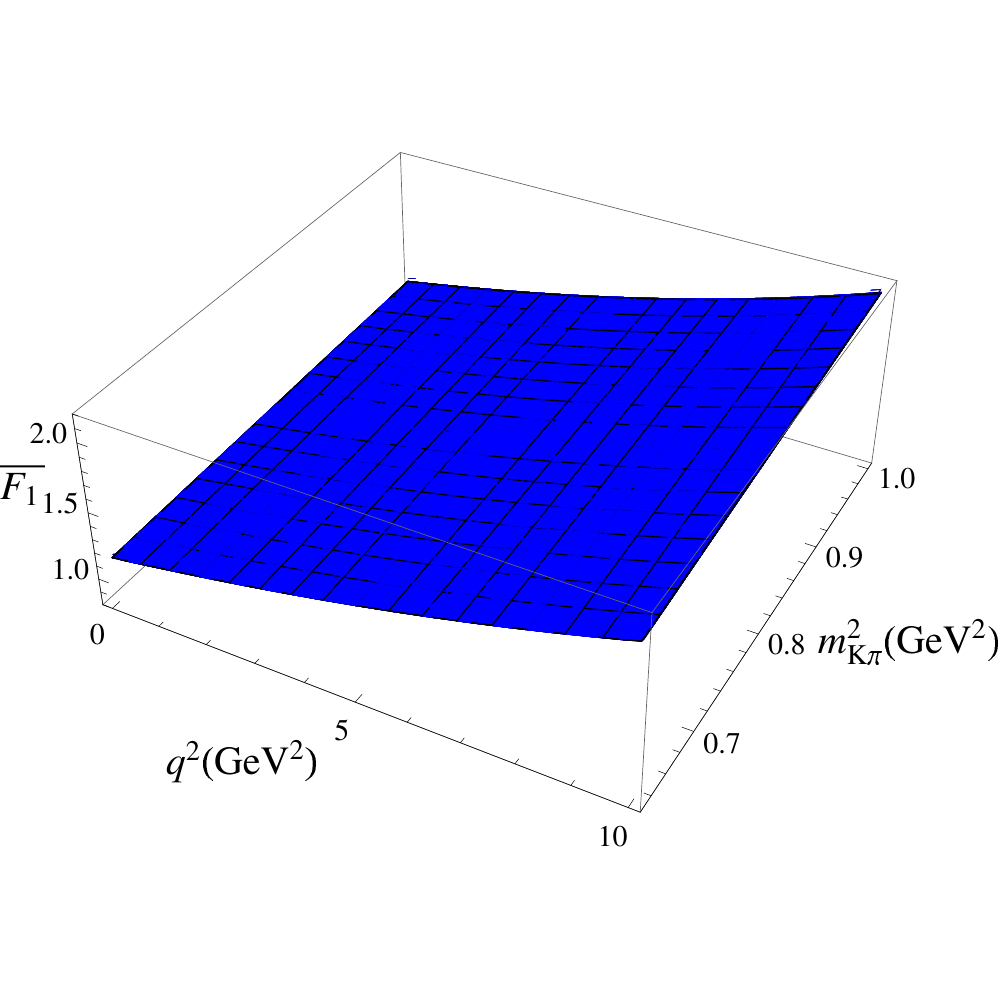} 
\caption{The dependence of $\bar F_1$ on the squared momentum transfer $q^2$ and the two-hadron invariant mass square $m_{K\pi}^2$.} 
\label{fig:MKpiQ2}
\end{center}
\end{figure}
%%%%%%%%%%%%%%%%%%%%%%%%%%%%%%%%%%%%%%%%%%%%%%%%%%%%%%%%%%%%%%%%%%%%%%%%%%%%%%%% 

For presentation, one can introduce~\cite{Meissner:2013hya}
\begin{eqnarray}
 {\cal F}_{i}(q^2, m_{K\pi}^2) = C_X\frac{m_{K}^{2}-m_{\pi}^{2}}{m_{s}-m_{u}}m_{K\pi} F_{K\pi}(m_{K\pi}^2)  \overline   F_i (m_{K\pi}^2,q^2).
\end{eqnarray}     
 
The criteria in LCSR  to find sets of parameters $M^2$ (the Borel
parameter) and $s_0$ (the continuum threshold) is  that the resulting form factor does not
depend   much on the precise values of these parameters;  additionally both  the continuum contribution and the higher power corrections, arising from the neglected  higher twist  LCDA,  should not be  significant. The $s_0$ is to separate the ground state  from higher mass contributions,
and thus  shall be below the next known   resonance, in this case, $B_1$ with $J^P=1^+$. Thus approximately this parameter should be close to $33~{\rm GeV}^2$~\cite{Beringer:1900zz}. 
Studies  of ordinary heavy-to-light  form factors in LCSR, see for instance Ref.~\cite{Ball:2004rg},     suggested a similar result, ranging from $33~{\rm GeV}^2$ to $36~{\rm GeV}^2$, while some bigger values are derived  in the recent update of $B\to \pi$ form factor in LCSR~\cite{Khodjamirian:2011ub}.

Numerical results for the auxiliary function  $\overline  F_1$ at the  $K\pi$ threshold $m_{K\pi}= m_{K}+m_{\pi}$  are given in 
Fig.~\ref{fig:BorelDepen}, where the dependence of the form factor $\bar F_1$ (left panel) and the continuum/total ratio (right panel) on the Borel parameter are shown. The continuum contribution to the form factors is  obtained by invoking the quark-hadron duality above the threshold $s_0$ and calculating the correlation function on   QCD side.  Solid lines denote the central value while   dashed curves correspond to variations of threshold parameter: $s_0= (34\pm 2) {\rm GeV}^2$. From this figure, we can see that results for $\bar F_1$  are   stable  when $M^2> 6 \, {\rm GeV}^2$, and meanwhile the continuum contribution is typically  smaller than  $30\%$.   Unfortunately, due to the lack of knowledge on the 3-particle twist-3  and higher twist  generalized LCDA,    their contributions have been neglected. 

The dependence on the squared invariant mass of the $K\pi$ system and the squared momentum transfer $q^2$ is shown in Fig.~\ref{fig:MKpiQ2}  with the value $M^2=8 {\rm GeV}^2$.

%%%%%%%%%%%%%%%%%%%%%%
%%%%%%%%%%%%%%%%%%%%%%
\begin{figure}[ht]
\begin{center}
\includegraphics[width=0.6\columnwidth]{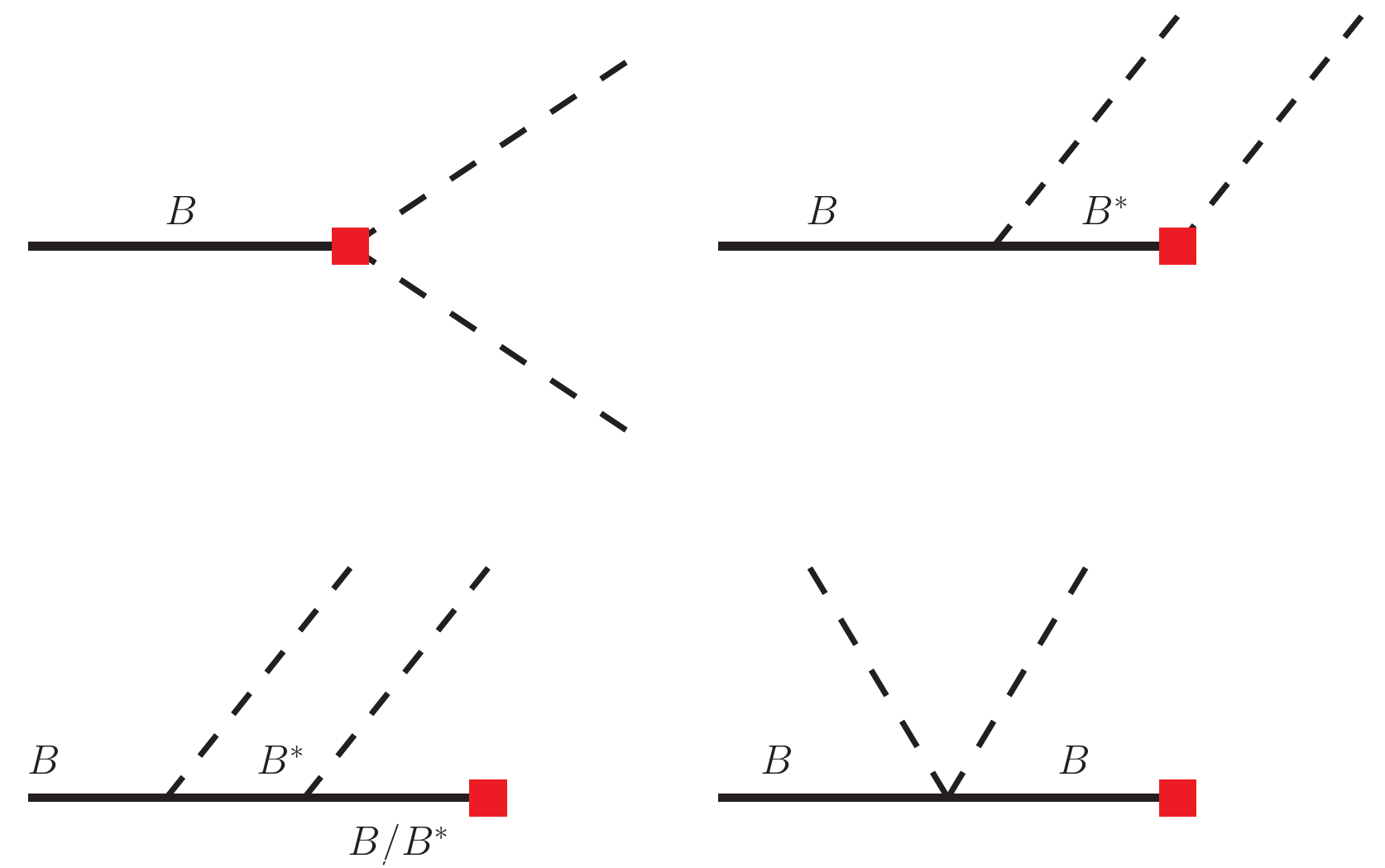} 
\caption{At the low recoil where all pseudo-Goldstone bosons are having small momentum, the  LO Feynman  diagrams for the $B\to \pi\pi \ell\bar\nu$ at hadron level are shown. Solid lines and dashed lines represent the heavy mesons and pseudo-Goldstone bosons, respectively. Shaded square denotes an insertion of the weak current.   }
\label{fig:hmchipt}
\end{center}
\end{figure} 
%%%%%%%%%%%%%%%%%%%%%%
%%%%%%%%%%%%%%%%%%%%%%

In Ref.~\cite{Kang:2013jaa}   the form factors for the $B\to\pi\pi$ system are explored  in dispersion theory and heavy meson $\chi$PT at   low recoil, following the technique employed in Ref.~\cite{Lee:1992ih}.  At  hadron level,   Feynman diagrams have been shown in Fig.~\ref{fig:hmchipt}. Solid lines and dashed lines represent   heavy mesons and pseudo-Goldstone bosons, respectively. Shaded square denotes an insertion of the weak current. This  analysis has taken into account the $\pi-\pi$ rescattering effects, as well as the effect of the $\rho$ meson.

%%%%%%%%%%%%%%%%%%%%%%
\section{$\gamma$ }
\label{sec:gamma}
%%%%%%%%%%%%%%%%%%%%%%

$\beta$  is extracted from the  golden mode $B\to J/\psi K_S$,  which is dominated by the $b\to c\bar c s$ transition.  Penguin contaminations in this mode are found to be ${\cal O}(10^{-3})$ (see Ref.~\cite{Li:2006vq} for a recent discussion). 
In the case of $\alpha$, penguins pollutions in $B\to (\pi, \rho, a_{1})\pi$  and $B\to \rho\rho$ may  be sizeable~\cite{Beringer:1900zz}.  The inclusion of   isospin related processes, however some of which have   small branching ratios,  may refine the analysis.

\begin{figure}[ht]
\begin{center}
\includegraphics[width=0.8\columnwidth]{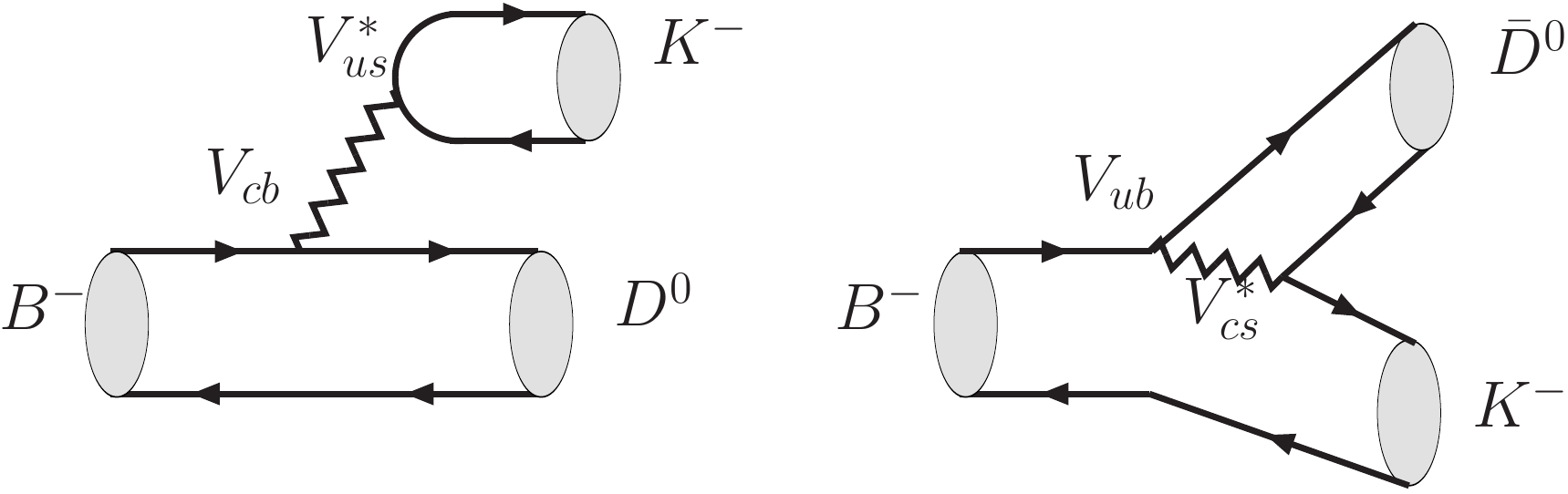} 
\caption{ Feynman diagrams for $B\to DK$ that can be used  to extract the $\gamma$ angle. }
\label{fig:FeynGraphGamma}
\end{center}
\end{figure}

The angle $\gamma\equiv arg(- V_{ud}V_{ub}^*/(V_{cd}V_{cb}^*))$ is the relative weak phase of   decays  induced by the $b\to c\bar us$ and $b\to u\bar cs$ transitions.  
It can be extracted from    tree-dominated modes $B\to DK$~\cite{Gronau:1990ra,Gronau:1991dp,Atwood:1996ci,Atwood:2000ck,Giri:2003ty} whose Feynman diagrams are depicted in Fig.~\ref{fig:FeynGraphGamma}.    The GLW  method~\cite{Gronau:1990ra,Gronau:1991dp}  uses the fact that the six decay amplitudes of $B^\pm \to (D^0, \bar D^0, D_{CP}^0)K^\pm$ form two triangles in the complex plane,  graphically representing 
\begin{eqnarray}
 \sqrt 2 A(B^+\to D_\pm^0 K^+) = A(B^+ \to D^0K^+)  \pm A(B^+\to \bar D^0K^+), \nonumber\\
 %\end{eqnarray}
 %\begin{eqnarray}
 \sqrt 2 A(B^-\to D_\pm^0 K^-) = A(B^- \to D^0K^-)  \pm A(B^-\to \bar D^0K^-),\label{eq:identity}
\end{eqnarray} 
where the convention $CP|D^0\rangle =|\bar D^0\rangle$ has been used and $D^0_+(D^0_-)$ is  the CP even (odd) eigenstate.   Measurements of the six decay rates  will  fully determine the sides and apexes of the two triangles, 
%$A(B^- \to D^0K^-) [A(B^- \to \bar D^0K^-)] $ is described by 
%the $b\to c \bar us(b\to u\bar cs)$ transition, which contains a weak phase $0(\gamma)$.
in particular the relative phase between $A(B^- \to \bar D^0K^-)$ and $A(B^+ \to D^0K^+)$ is $2\gamma$.

%%%%%%%%%%%%%%%%%%%%%%
%%%%%%%%%%%%%%%%%%%%%%
%%%%%%%%%%%%%%%%%%%%%%
%%%%%%%%%%%%%%%%%%%%%%

\subsection{CP violation effects and errors in $B\to DK$} 

Since the identities in Eq.~\eqref{eq:identity} holds irrespective of the strong phase in the decay, this method is free of hadronic uncertainties and  is  believed  theoretically clean. Thus the measurement of $\gamma$ provides a benchmark of extraction of the CKM parameters.

However the GLW method is based on the neglect of the direct CP asymmetry in $D^{0}$ and $\bar D^{0}$ decays. For instance the $K^+K^-$ and $\pi^+\pi^-$ final states can  project out the same   $D^0_+$. 
One of the most exciting measurements by LHCb collaboration~\cite{Aaij:2011in}, confirmed by CDF~\cite{Collaboration:2012qw} and Belle~\cite{Ko:2012px} collaborations,  was   CP violation in charm sector. 
These three collaborations have found nonzero  difference of  CP asymmetries (CPAs) which are much larger than  SM expectation.   
The direct CP violation of $D^0$ decays was extracted as~\cite{Amhis:2012bh}
\begin{eqnarray}
 \Delta A^{\rm dir}_{CP} = (-0.678 \pm 0.147 )\%. \label{delta_ACP}
 \end{eqnarray} 
However this large value is not confirmed in later analysis by   LHCb collaboration~\cite{Aaij:2013bra}
and the new average  is~\cite{Amhis:2012bh} 
 \begin{eqnarray}
 \Delta A^{\rm dir}_{CP} = (-0.329\pm0.121)\%.  \label{new_delta_ACP}
\end{eqnarray}
  
Though the new result in Eq.~\eqref{new_delta_ACP} has a smaller central  value,  the CPA in $D$ decays may play an important role in measuring the $\gamma$~\cite{Wang:2012ie,Martone:2012nj,Bhattacharya:2013vc,Bondar:2013jxa} ( see also Ref.~\cite{Meca:1998ee,Silva:1999bd}). Physical observables  are given as
\begin{eqnarray}
 R_{+}^K &=&2\frac{{\cal B}(B^-\to D^0_{+} K^-)+{\cal B}(B^+\to D^0_{+} K^+)  }{{\cal B}(B^-\to D^0K^-) +{\cal B}(B^+\to \bar D^0 K^+) }\nonumber\\
&=&1+(r_{B}^K)^2  + \frac{2r_B^K \cos\delta_B[ (1+(r_D^f)^2)\cos\gamma+ 2 r_D^f   \cos\delta_D^f]}{1+ (r_D^{f})^2 +2r_D^f \cos\gamma \cos \delta_{D}^f}, \nonumber\\
&\equiv & 1+(r_{B}^K )^2+ 2r_{B}^K  \cos\delta_{B}^K \cos\gamma_{eff},\label{eq:RplusK_corrections}
\end{eqnarray}
\begin{eqnarray}
 A_{+}^K &=&\frac{{\cal B}(B^-\to D^0_{+} K^-)-{\cal B}(B^+\to D^0_{+} K^+)  }{{\cal B}(B^-\to D^0_{+} K^-) +{\cal B}(B^+\to D^0_{+} K^+) }\nonumber\\
&=& \frac{1}{ R_{+}^K }  \bigg[(1-(r_B^K)^2)  A_{CP}^{dir} (D^0\to f)  +  \frac{ 2r_B^K  (1+(r_D^f)^2) \sin\delta_{B}^K \sin\gamma }{ 1+(r_D^f)^2 +2r_D^f \cos\delta_D^f \cos\gamma}\bigg] \nonumber\\
&\equiv &  2r_B^K \sin\delta_{B}^K \sin\gamma_{eff} /R_{+}^K, \label{eq:AplusK_corrections}
\end{eqnarray}
where the last lines in the above equations correspond to the case with no CPA.   $r_D^f$ is the ratio of   penguin and tree amplitudes in $D\to f$ decays, with $f=K^+K^-,\pi^+\pi^-$.  $\gamma$ and $\delta_D^f$ is the weak phase difference and strong phase difference respectively.   These two  equations explicitly show the CPA effects on   experimental observables.

%%%%%%%%%%%%%%%%%%%%%%%%%%%%%%%%%%%%%%%%%%%%%%%%%%%%
\begin{figure}\begin{center}
\includegraphics[width=0.5\columnwidth]{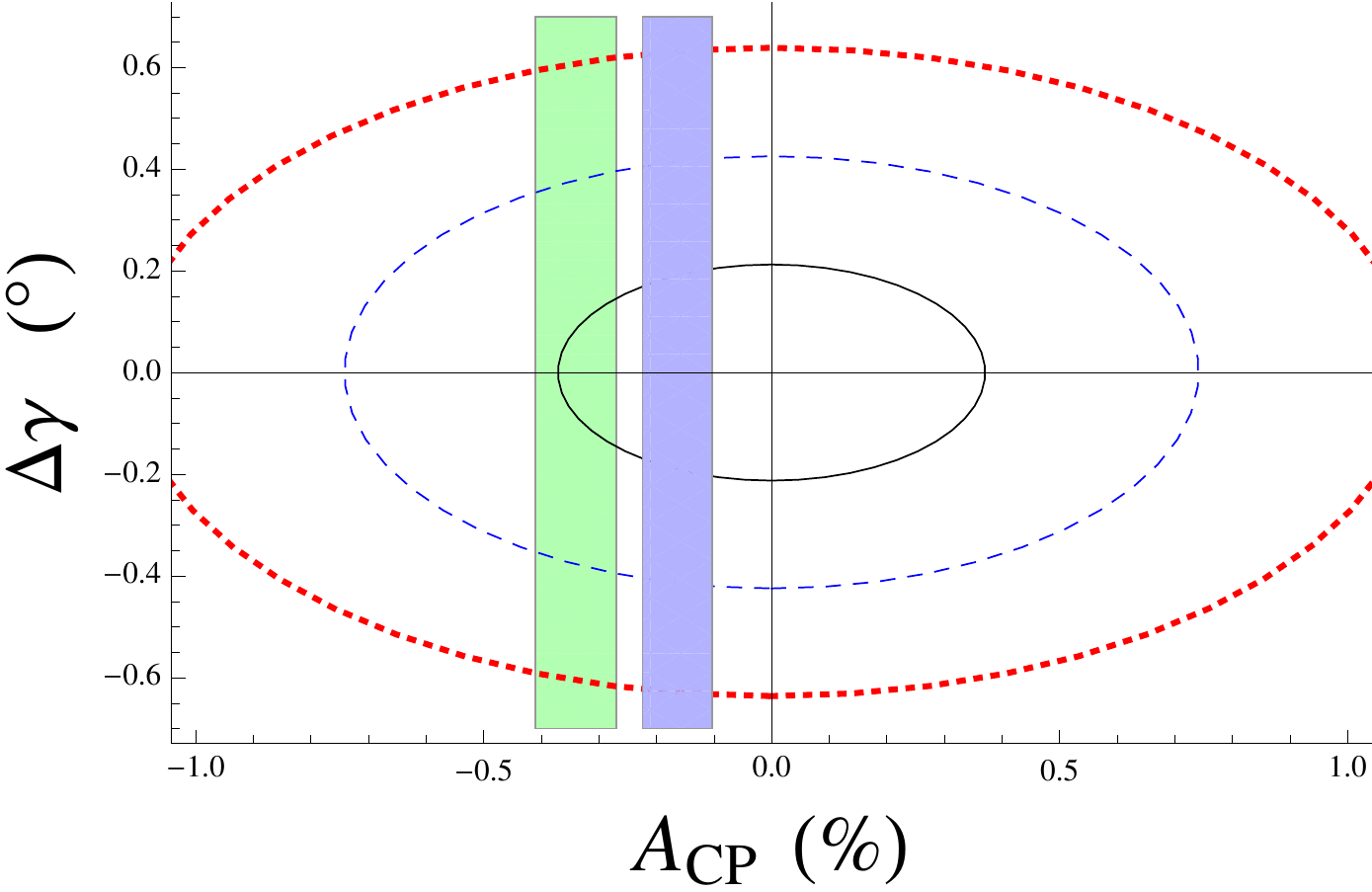} 
\caption{Effect of CP violation  on the extraction of   $\gamma$ via the $R_{+}^K $. The sold (black), dashed(blue), and dotted (red) lines correspond to $r_{D}^f$=0.002, 0.004 and  0.006 respectively.  The shadowed region  corresponds to  $A_{CP}^{K^+K^-}= (-0.34\pm 0.07)\%$ (left) and $A_{CP}^{K^+K^-}= (-0.16\pm 0.06)\%$ (right).  } \label{fig:gamma_R_CP}
\end{center}
\end{figure}
%%%%%%%%%%%%%%%%%%%%%%%%%%%%%%%%%%%%%%%%%%%%%%%%%%%%%

The $ A_{+}^K $ in Eq.~\eqref{eq:AplusK_corrections} receives new contributions proportional to the direct CPA in $D$ decays. 
Neglecting terms suppressed by ${\cal O}(r_D^f)$, the dominant  correction to  $\sin\gamma$  is proportional to $A_{CP}^{dir} (D^0\to f)/(2r_B^K \sin\delta_{B})$.  
The value for $\gamma$ obtained from $K^+K^-$ final state and the one from  $\pi^+\pi^-$ final states can  differ   by  $3^\circ$. Fortunately  such  effects can   be incorporated   once the data on the direct CPA is available.

The effects on $ R_{+}^K $ are shown in Fig.~\ref{fig:gamma_R_CP}. The sold (black), dashed(blue), and dotted (red) lines correspond to $r_{D}^f$=0.002, 0.004 and  0.006 respectively.  The shadowed region is the CPA for $D^0\to K^+K^-$ from the experimental data: $A_{CP}^{K^+K^-}= (-0.34\pm 0.07)\%$ (left) and $A_{CP}^{K^+K^-}= (-0.16\pm 0.06)\%$ (right), where   the U-spin symmetry has been assumed  for the CP asymmetry $A_{CP}^{K^+K^-}= -A_{CP}^{\pi^+\pi^-}= \Delta A_{CP}/2$.

Recent theoretical works~\cite{Rama:2013voa,Brod:2013sga} have investigated other errors to the $\gamma$ angle and found that these uncertainties are tiny.

\subsection{Experimental  results} 

\begin{figure}
\begin{center}
\includegraphics[width=0.45\columnwidth]{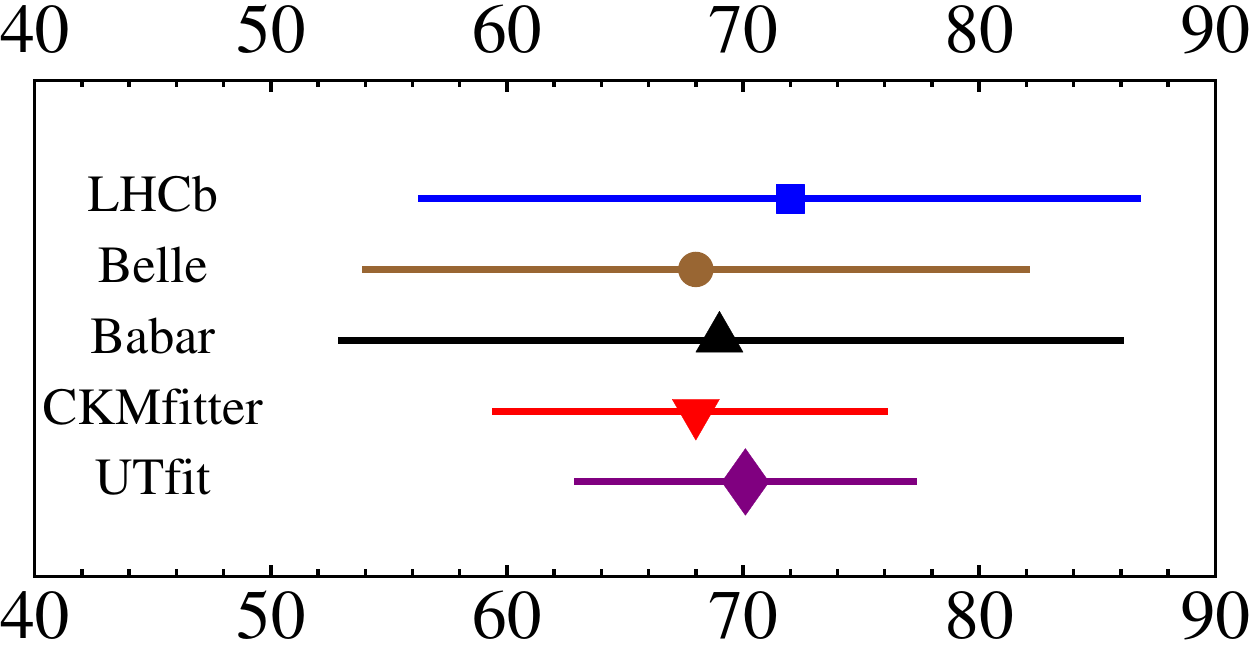}
\caption{Recent experimental data on the $\gamma$ angle vs indirect determination via global fits} \label{fig:gammaData}
\end{center}
\end{figure}

Latest experimental measurements by Belle~\cite{Trabelsi:2013uj} and BaBar\cite{Lees:2013nha} collaborations are given as 
\begin{eqnarray}
\gamma&=&(68^{+15}_{-14})^{\circ}, \;\;\;{\rm Belle}\nonumber\\
\gamma&=&(69^{+17}_{-16})^{\circ}. \;\;\;{\rm BaBar}
\end{eqnarray}
Recently a combination of three LHCb measurements of the CKM angle $\gamma$ has been presented~\cite{Aaij:2013zfa} 
\begin{eqnarray}
\gamma=(72.0^{+14.7}_{-15.6})^{\circ}. \;\;\;{\rm LHCb}
\end{eqnarray}
These results, all of which are consistent with each other,  are displayed in Fig.~\ref{fig:gammaData}, in which we have also shown the global fitting results from   CKMfitter\cite{Charles:2004jd} and UTfit~\cite{Ciuchini:2000de}.

%%%%%%%%%%%%%%%%%%%%%%%%%%%%%%%%%%%%%%%%%%%%%%%%%%%%
%%%%%%%%%%%%%%%%%%%%%%%%%%%%%%%%%%%%%%%%%%%%%%%%%%%%%

\subsection{B decays into a scalar/tesnor state} 
 
In the GLW method, the shape of the two triangles formed by   decay amplitudes is governed by  two quantities  
\begin{eqnarray}
 r_{B}^K \equiv\left|{A(B^-\to \bar D^0 {K^-})}/{A(B^-\to D^0 K^{-})}\right|,\nonumber\\
 \delta_{B}^{K} \equiv arg\left[{e^{i\gamma} A(B^-\to \bar D^0 K^{-})}/{A(B^-\to D^0 K^{-})}\right],\nonumber
\end{eqnarray}
with the world averages   from the CKMfitter~\cite{Charles:2004jd}
\begin{eqnarray}
 r_B^K= 0.0956^{+0.0062}_{-0.0064},\;\;
 \delta_B^K=( 114.8^{+9.0}_{-9.7})^\circ. 
\end{eqnarray}
The smallness of $r_B^K$ implies the mild sensitivity to $\gamma$  and thereby the two triangles formed by decay amplitudes are very squashed.
$ R_{CP\pm}^K $ and $ A_{CP\pm}^K$ have a mild sensitivity to the angle $\gamma$, inducing large experimental uncertainties.

%%%%%%%%%%%%%%%%%%%%%%%%%%%%%%%%%%%%%%%%%%%%%%%%%%%%
\begin{figure}[h]
\begin{center}
\includegraphics[scale=0.35]{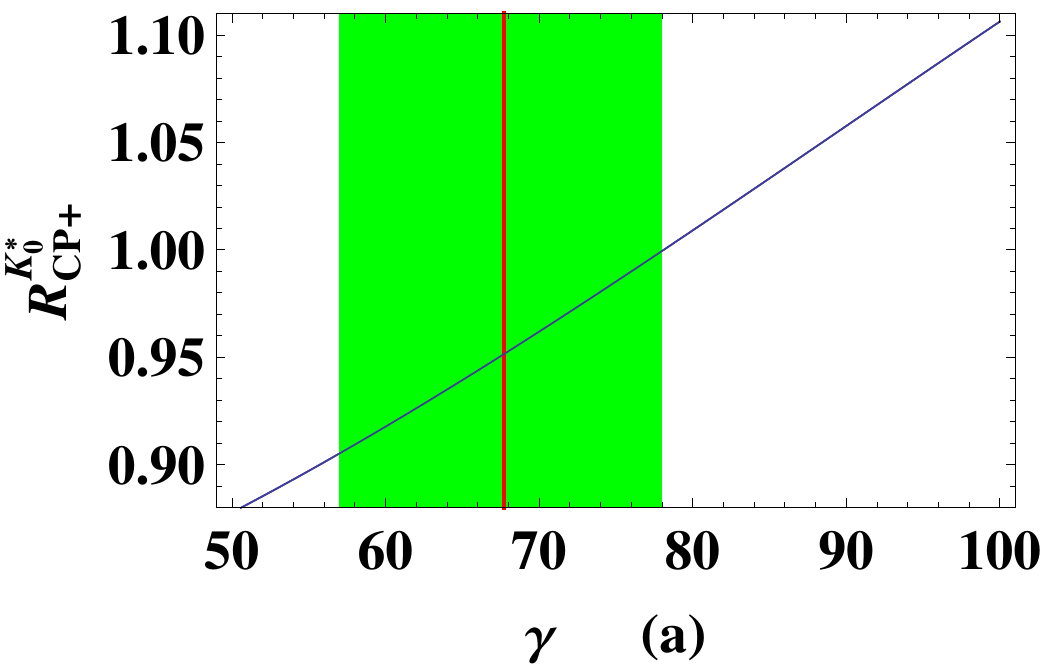}
\includegraphics[scale=0.38]{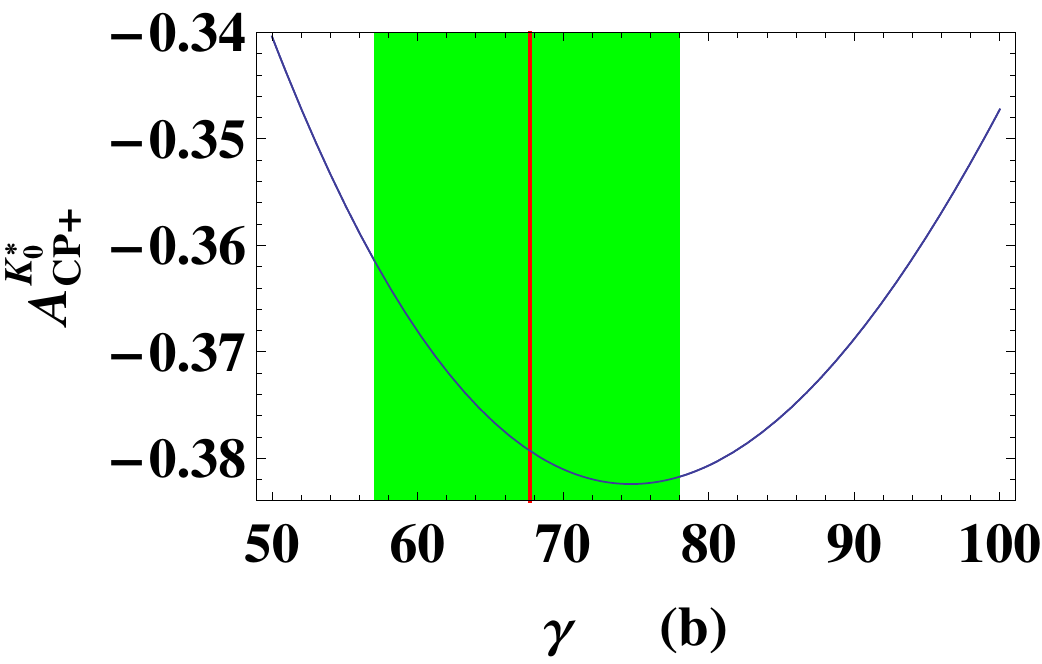}
\includegraphics[scale=0.35]{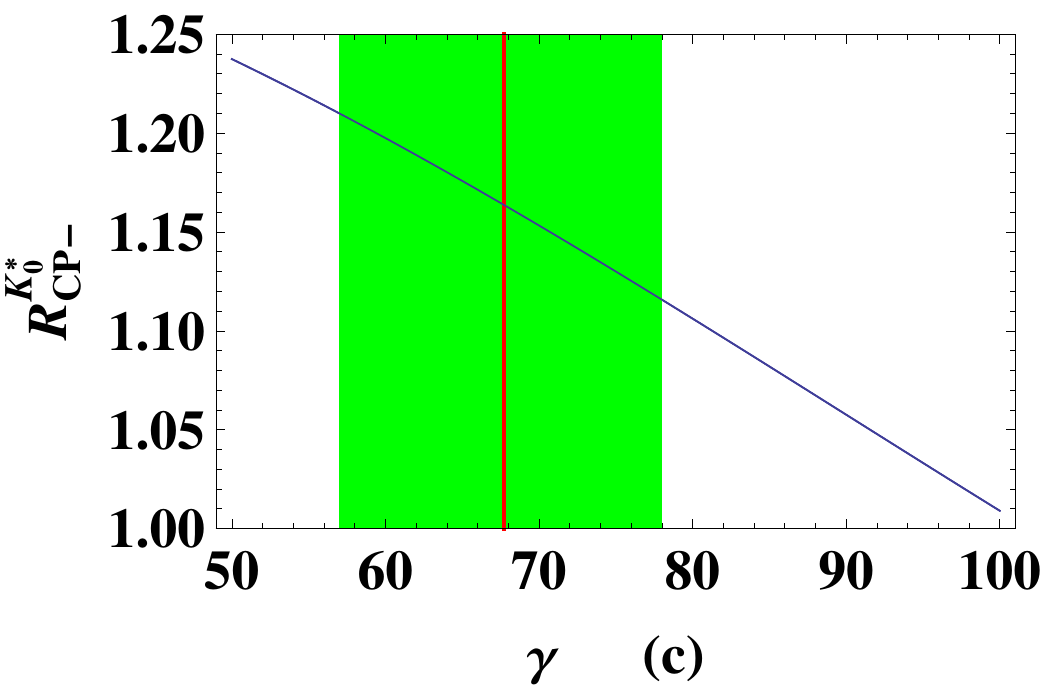}
\includegraphics[scale=0.38]{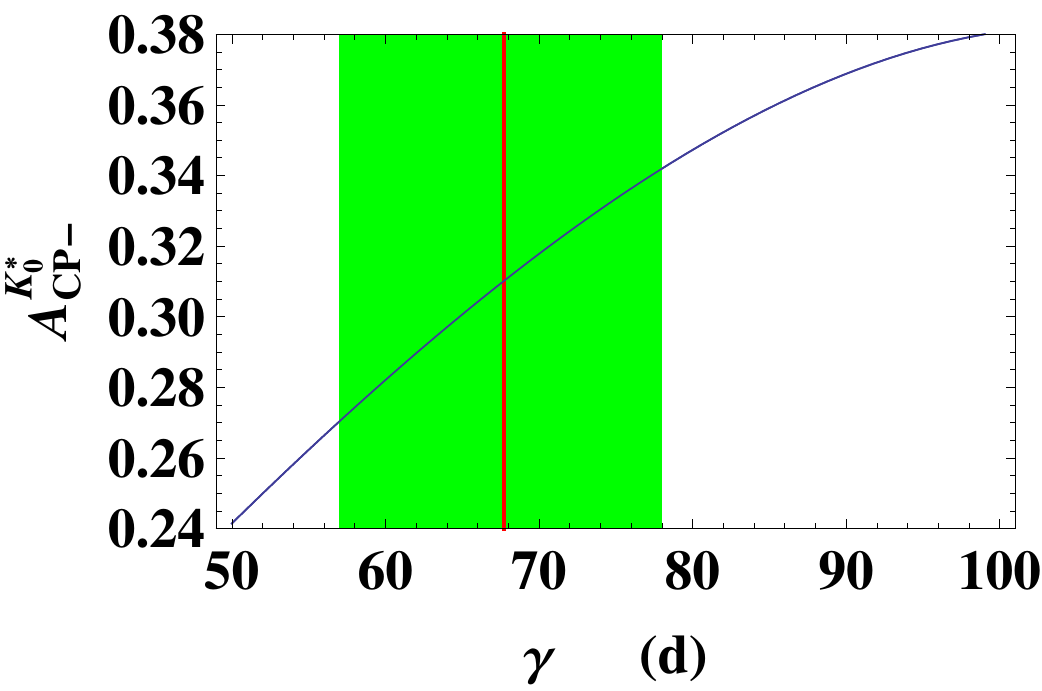}
\includegraphics[scale=0.35]{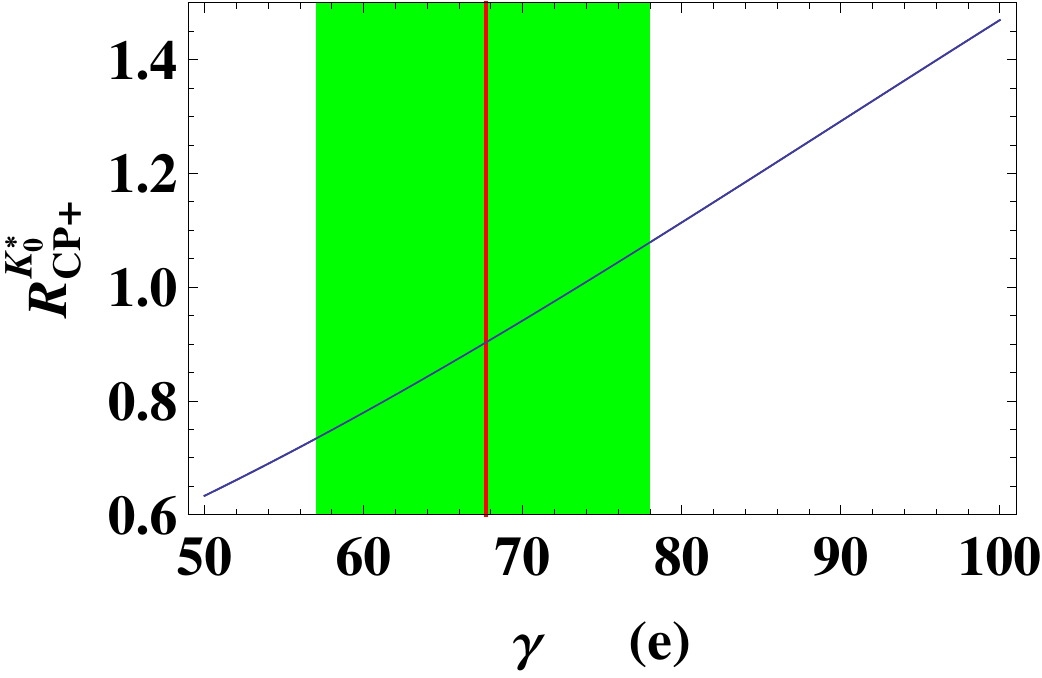}
\includegraphics[scale=0.38]{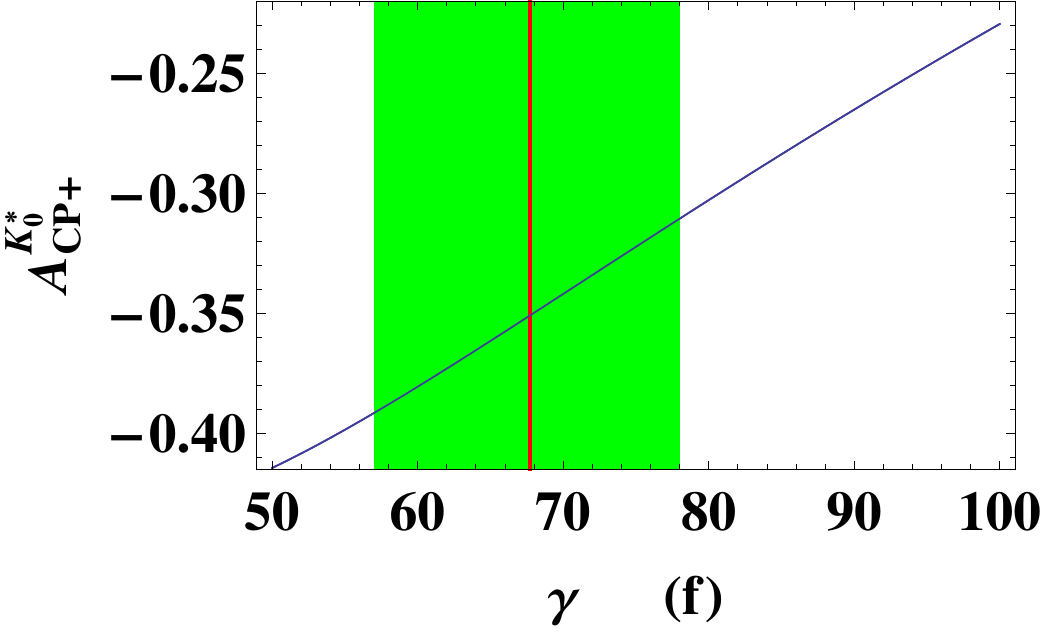}
\includegraphics[scale=0.35]{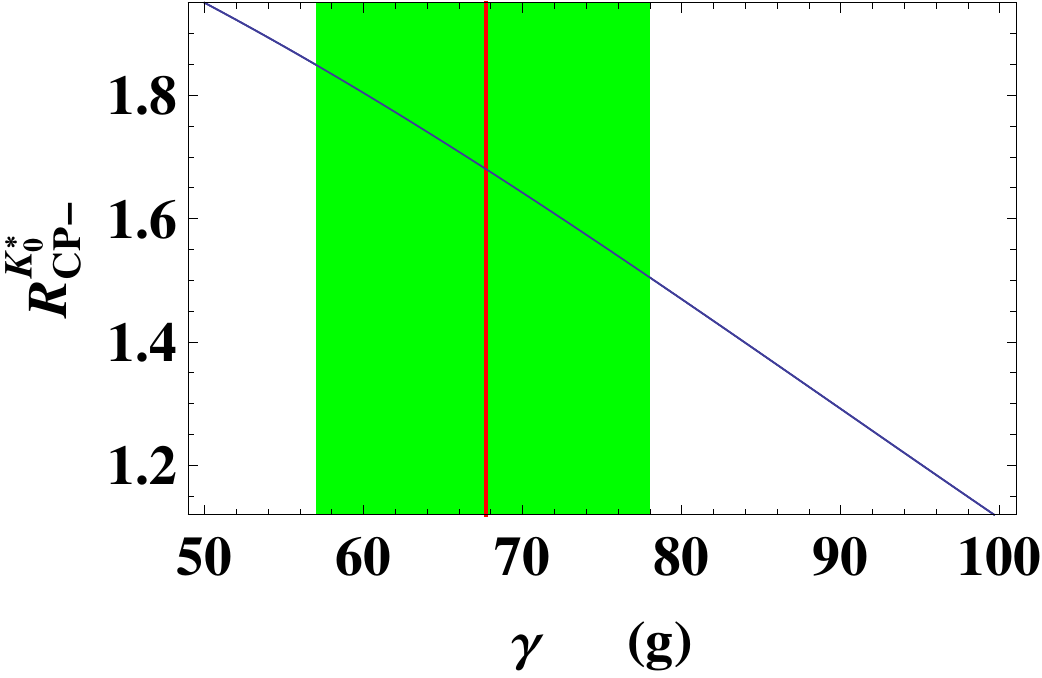}
\includegraphics[scale=0.38]{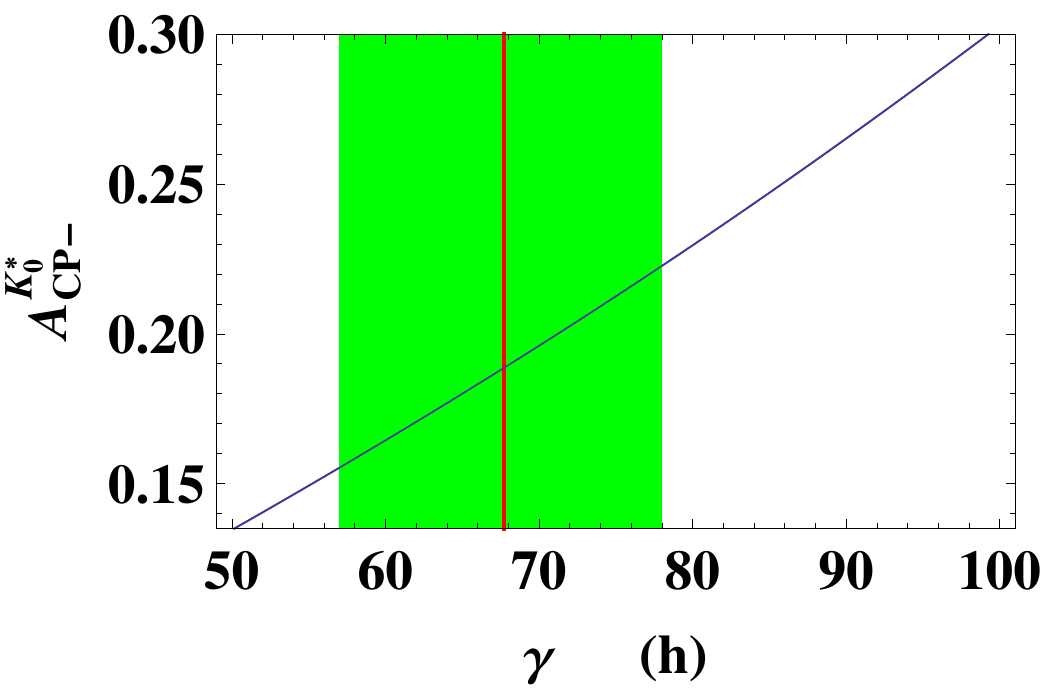}
\includegraphics[scale=0.35]{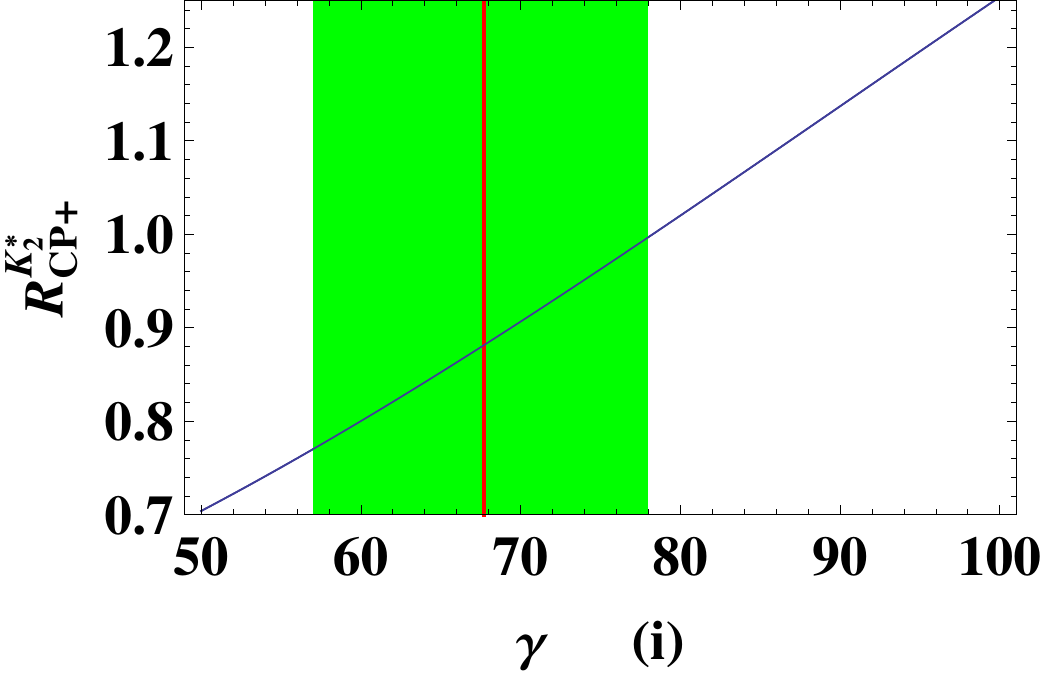}
\includegraphics[scale=0.38]{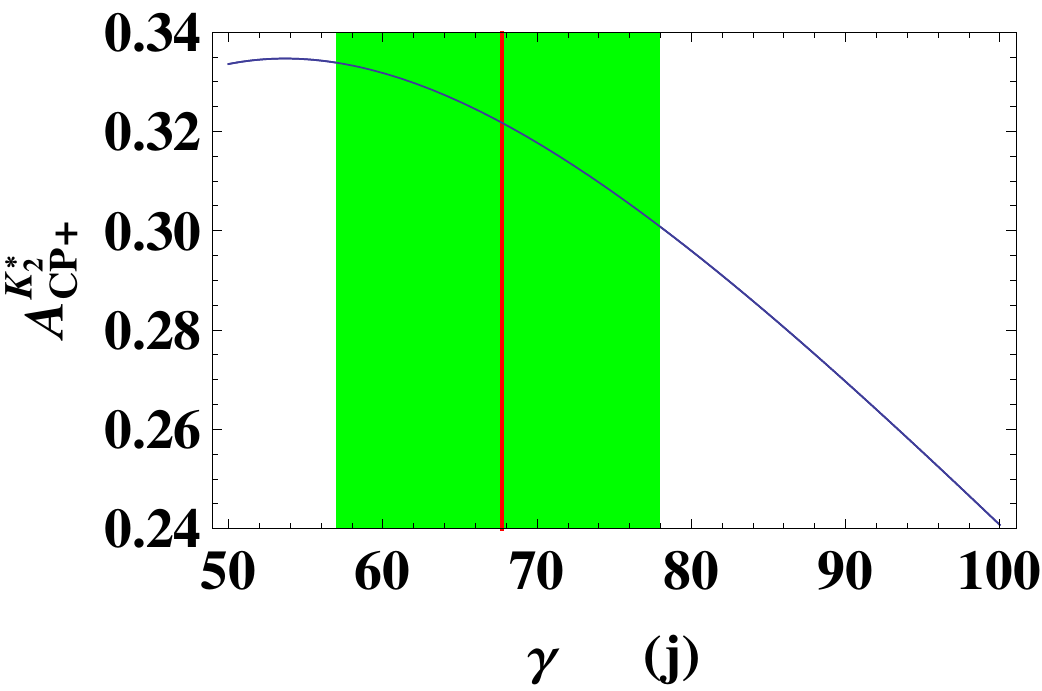}
\includegraphics[scale=0.35]{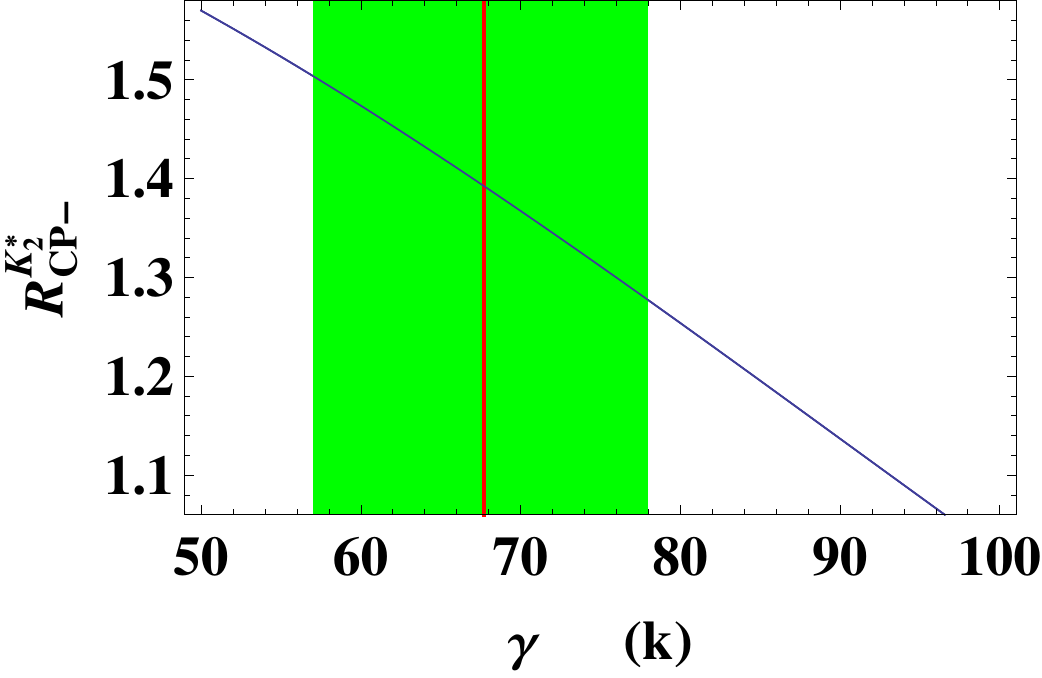}
\includegraphics[scale=0.38]{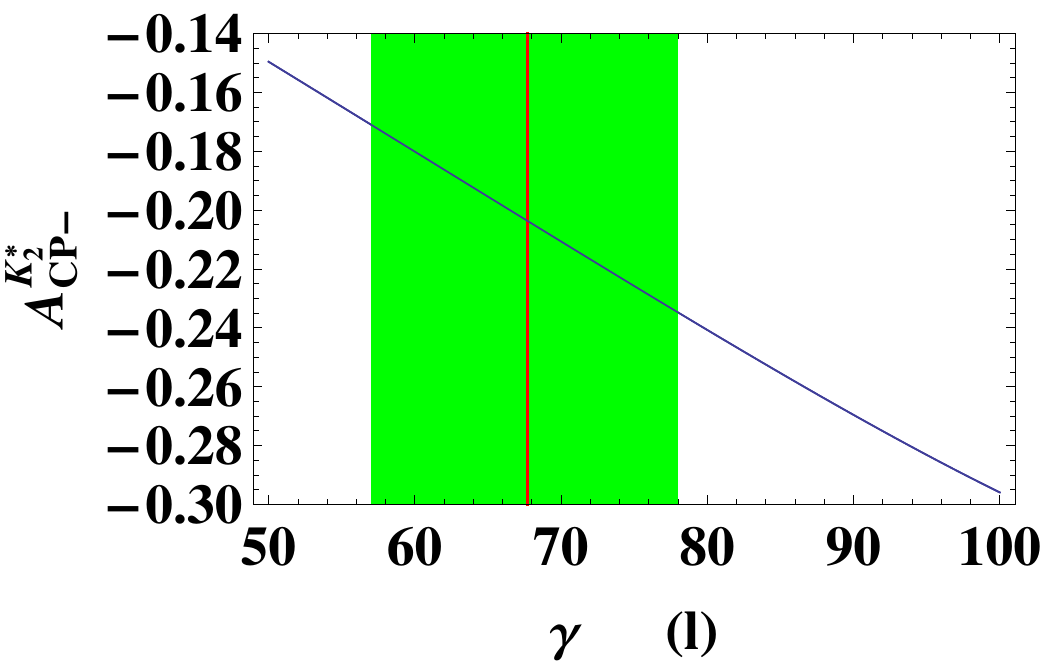}
\caption{The dependence of $R_{CP}$ and $A_{CP}$ on $\gamma$. Diagrams (a)-(d) show $R^{K_0^*}_{CP}$ and $A^{K_0^*}_{CP}$ in $S1$, (e)-(h) in $S2$, and diagrams (i)-(l) show $R^{K_2^*}_{CP}$ and $A^{K_2^*}_{CP}$. The shadowed (green) region denotes the current bounds on $\gamma=(68_{-11}^{+10})^\circ$ from a combined analysis of $B^\pm\to DK^\pm$ \cite{Charles:2004jd}, and the vertical (red) line represents the central value. }
\label{fig:dependenceInBtoDK02}
\end{center}
\end{figure}
%%%%%%%%%%%%%%%%%%%%%%%%%%%%%%%%%%%%%%%%%%%%%%%%%%%%%

It has been shown in Ref.~\cite{Wang:2011zw,Wang:2012jba} (see  also \cite{Diehl:2001xe,Diehl:2001ey}) that  the low sensitivity problem can be highly improved  in $B^\pm\to  D {K^{*\pm}_{0,2} }$ due to $r_{K_{0,2}^*}\sim 1$, where $K_{0,2}^{*}$ is a scalar/tensor strange meson. This meson can also be replaced by the $K\pi$ state with the same quantum numbers.   Though the color-allowed diagram has a large Wilson coefficient $a_1\sim 1$,  the emitted ${K_{0,2}^*}$ meson is produced from a local vector or  axial-vector current (at the lowest order in $\alpha_s$), whose matrix element between the QCD vacuum and  the $K^*_{0}$($K^*_{2}$) state is small (identically zero).  Due to this suppression, the color-allowed amplitude can  be comparable to the color-suppressed one. 
This has been validated  by  explicit pQCD calculations~\cite{Kim:2013ria,Zou:2012sx,Zou:2012xk,Zou:2014aba}, using   scalar and tensor meson LCDA~\cite{Li:2008tk,Cheng:2010hn,Wang:2010ni,Han:2013zg,Lu:2006fr,Cheng:2013fba,Cheng:2005nb}. 
Physical observables are  defined by: 
\begin{eqnarray}
 R_{CP\pm}^{K_J} &=&2\frac{{\cal B}(B^-\to D_{CP\pm} K_J^-)+{\cal B}(B^+\to D_{CP\pm} K_J^+)  }{{\cal B}(B^-\to D^0K_J^-) +{\cal B}(B^+\to \bar D^0 K_J^+) } \nonumber\\
 &=& 1+(r_{B}^{K_J})^2\pm 2r_{B}^{K_J} \cos\delta_{B}^{K_J} \cos\gamma,\nonumber\\
 A_{CP\pm}^{K_J} &=&\frac{{\cal B}(B^-\to D_{CP\pm} K_J^-)-{\cal B}(B^+\to D_{CP\pm} K_J^+)  }{{\cal B}(B^-\to D_{CP\pm} K_J^-) +{\cal B}(B^+\to D_{CP\pm} K_J^+) } \nonumber\\
 &=&\pm 2r_B^{K_J} \sin\delta_{B}^{K_J} \sin\gamma /R_{CP\pm}^{K_J}.  \nonumber 
 \label{eq:experiments}
\end{eqnarray}

The dependence of $R_{CP}$ and $A_{CP}$ on $\gamma$ is shown in Fig.~\ref{fig:dependenceInBtoDK02}. Diagrams (a)-(d) show $R^{K_0^*}_{CP}$ and $A^{K_0^*}_{CP}$ in first scenario for the structure of  scalar mesons, (e)-(h) in second scenario, and diagrams (i)-(l) show $R^{K_2^*}_{CP}$ and $A^{K_2^*}_{CP}$. The shadowed (green) region denotes the current bounds on $\gamma=(68_{-11}^{+10})^\circ$ from a combined analysis of $B^\pm\to DK^\pm$ \cite{Charles:2004jd}, and the vertical (red) line represents the central value.

Properties of useful $B$ decay channels into a scalar/tensor meson towards the extraction of the CKM angle $\gamma$ are  summarised in Tab.~\ref{Tab:gammaAnglesFuture}.

 %%%%%%%%%%%%%%%%%%%%%
 \begin{table}[http]
\tbl{Properties of useful $B$ decay channels into a scalar/tensor meson towards the extraction of the CKM angle $\gamma$. All these  modes are expected to have larger $r_f$, ratios of decay amplitudes, compared to the corresponding channels in which the scalar/tensor meson is replaced by a pseudo-scalar meson.  Branching fractions and ratios of decay amplitudes of $B\to DT$ are taken from the perturbative QCD calculation~\cite{Zou:2012sx} while the rest  entries  when available are obtained in the factorization approximation~\cite{Wang:2011zw}.      }
%\begin{center}
{\begin{tabular}{|c|c|c|c|}\hline
 Channel & CKM angle to access & BRs for suppressed and allowed modes  & $r_f$ \\ \hline
 $B^\pm \to D^\pm K_{0}^*$ & $\gamma$  & [$4\times 10^{-6}$,  $4\times 10^{-5}$] & 0.3 \\ 
 $B^\pm \to D^\pm K_{2}^*$ & $\gamma$  & [$3\times 10^{-6}$,  $3\times 10^{-5}$] & 0.3  \\\hline
 $B\to D^\pm a_{0}^{\mp} $    & $\gamma+2\beta$ &   &   \\
% $B\to D^\pm b_{1}^{\mp} $    & $\gamma+2\beta$ & [1] &  0.1 \\
 $B\to D^\pm a_{2}^{\mp} $    & $\gamma+2\beta$ & $[2\times 10^{-6}, 4\times 10^{-4}]$ &  0.1 \\\hline
 $B_s\to D_s^\pm K_{0}^{*\mp} $    & $\gamma+2\beta_s$ &   &    \\
 $B_s\to D_s^\pm K_{2}^{*\mp} $    & $\gamma+2\beta_s$ &$ [2\times 10^{-5},2\times 10^{-5}]$ &  1 \\\hline
 $B_s\to D f_0(980) $    & $\gamma+2\beta_s$ &$[1\times 10^{-6}, 3\times 10^{-6}]$ &  0.5 \\
 $B_s\to D f_2'(1525)$    & $\gamma+2\beta_s$ & $[3\times 10^{-6}, 1.4\times 10^{-5}]$ &  0.5\\\hline
 \end{tabular}}
%\end{center}
\label{Tab:gammaAnglesFuture}
\end{table}%
 %%%%%%%%%%%%%%%%%%%%%

\subsection{Three-body $B$ decays}

It has been proposed in Ref.~\cite{Bhattacharya:2013cla} (see also Ref.~\cite{Bediaga:2006jk,Bediaga:2008zz}) that the $B\to K\bar KK$ and $B\to K\pi\pi$ can be used to extract the CKM angle $\gamma$. These channels are related by   the SU(3) symmetry and therefore the extracted result will rely on the symmetry breaking effects. 

 %%%%%%%%%%%%%%%%%%%%%%
%%%%%%%%%%%%%%%%%%%%%%
\section{$\beta_s$}
\label{sec:betas}
%%%%%%%%%%%%%%%%%%%%%%
%%%%%%%%%%%%%%%%%%%%%%

In the SM, the non-vanishing phase in ($bs$) triangle is related to the $B_s-\bar B_s$ mixing phase.
 States of $B_s^0$ or $\bar B_s^0$ at $t=0$ can evolve in time and thus be mixed with  each other.
These states at $t$ will  be denoted as $B_s(t)$ and $\bar B_s(t)$.  Since both the $B_s^0$ and $\bar B_s^0$ can decay into the same final state for instance  $J/\psi \phi$, there is an indirect CP asymmetry between the rates of $B_s(t)\to  J/\psi \phi$ and $\bar B_s(t) \to  J/\psi \phi$,   quantified by
\begin{eqnarray}
  {\rm Im}\left[\frac{q}{ p} \frac{ \bar A_f}{A_f}\right].   \label{eq:indirectCPA}
\end{eqnarray}
Here the $A_f$ and $\bar A_f$ are the   $B_s$ and $\bar B_s$ decay amplitudes  which  are dominated  by the $b\to c\bar cs$ transition.  Since the  CKM factors $V_{cb}V_{cs}^*$  are real in the standard parametrization of CKM,   the indirect CPA defined in Eq.~\eqref{eq:indirectCPA} measures the  phase of  $q/p$: $\phi_s = - {\rm arg} (q/p)$. The $\phi_s$ is tiny in SM,  and in particular $ \phi_s=-2\beta_s = -2 {\rm arg} [- V_{ts}V_{tb}^*/ (V_{cs}V_{cb}^*)]$~\cite{Charles:2011va}:
\begin{eqnarray}
\phi_s= (-0.036 \pm 0.002) \quad { \rm rad}.
\end{eqnarray}
Any observation of a significant non-zero  value   would be a  NP signal.

The  $\phi_s$ extraction   has    benefited a lot from the  $B_s/\bar B_s\to J/\psi \phi$.   Thanks to the large amount of data sample collected by Tevatron and LHC experiments,  the result for $\phi_s$  is getting more and more precise~\cite{LHCb:2011aa,Aaltonen:2012ie,Abazov:2011ry,Aad:2012kba,Bediaga:2012py}. Recently  based on the $1.0 {\rm fb}^{-1}$ data   collected at 7 TeV in 2011, the LHCb collaboration gives~\cite{Aaij:2013oba}
\begin{eqnarray}
\phi_s^{J/\psi \phi}= (0.07\pm 0.09\pm 0.01 ) \quad { \rm rad},
\end{eqnarray}
which is in agreement with the SM when  errors are taken into account.
In addition  new alternative  channels are proposed and  the $B_s\to J/\psi f_0(980) $ is  powerful in reducing the error~\cite{Stone:2008ak,Stone:2009hd}.
Since the $f_0(980)$ is a  $0^{++}$ scalar meson,  the final state $J/\psi f_0$ is a CP eigenstate and no angular decomposition is requested.   Agreement  on  branching fractions  is found between   theoretical calculation~\cite{Colangelo:2010bg,Colangelo:2010wg,Leitner:2010fq,Fleischer:2011au} and experimental measurements~\cite{Aaij:2011fx,Li:2011pg,Aaltonen:2011nk}. The $\phi_s$ is extracted  by the LHCb collaboration~\cite{Aaij:2013oba}
\begin{eqnarray}
\phi_s^{J/\psi f_0} =(-0.14^{+0.17}_{-0.16} \pm 0.01) \quad { \rm rad}.
\end{eqnarray}

The   decay distributions  can be derived   using  helicity amplitudes and for a detailed discussion we   refer the reader to Refs.~\cite{Xie:2009fs,Zhang:2012zk,Dighe:1998vk,Fleischer:1999zi,Liu:2013nea}.   In the presence of   S-wave $K^+K^-$ the angular distribution for $B_s\to J/\psi (\to l^+l^-)\phi(\to K^+K^-)$ at the time $t$ of the state that was a pure $B_s$ at $t=0$ is derived  as
\begin{eqnarray}
 \frac{d^4 \Gamma( t) }{dm_{K\bar K }^2 d\cos\theta_K d\cos\theta_l d\phi}&=&  \sum_{k=1}^{10} h_k(t) f_k(\theta_K, \theta_l, \phi),  \label{eq:time-dependent-angular}
 \end{eqnarray}
where the time-dependent functions $h_k(t)$ are given as
\begin{eqnarray}
 h_k(t)= \frac{3}{ 4\pi}
   e^{-\Gamma t}\left\{ a_k \cosh\frac{\Delta \Gamma t}{2} + b_k \sinh\frac{\Delta \Gamma t}{2} + c_k \cos(\Delta mt)  +  d_k \sin(\Delta mt) \right\}.\label{eq:time-dependent-angular2}
\end{eqnarray}
Here $\Delta m= m_H-m_L$, $\Delta \Gamma = \Gamma_L-\Gamma_H$, and $\Gamma = (\Gamma_L+\Gamma_H)/2$.
For the state that was a $\bar B_s$ at $t=0$, the signs in front of $c_k$ and $d_k$ should be reversed. Explicit  results for these coefficients are given for instance in Ref.~\cite{Liu:2013nea}.

On the  theoretical side, although decays of the $B_s/B_s^0$ meson into $J/\psi (\phi/f_0)$ are mainly governed by the $b\to c\bar cs$ transition at the quark level,  there are   penguin  contributions   with non-vanishing different weak phases. Thus the indirect CPA can be moved  away from the $\phi_s$.  Intuitively penguin contribution is expected to be small in the SM, but a complete and reliable estimate of its effects by some QCD-inspired approach is requested. Such estimate will become  mandatory soon.  As a reference,  after the upgrade of  LHC  the error can be diminished to $\Delta \phi_s \sim 0.008$~\cite{Bediaga:2012py}.

Ref.~\cite{Liu:2013nea} has attempted  to  estimate the penguin contributions in the $B\to J/\psi V$ decays and explore  the impact to the CPA measurement, following the calculation in Ref.~\cite{Li:2006vq,Chen:2005ht,Liu:2010zh,Liu:2012ib}.   Instead of using the flavor SU(3) symmetry to relate the effects in $B_s\to J/\psi \phi$ and the counterpart of $B$ decay modes~\cite{Bhattacharya:2012ph,Faller:2008gt},  the pQCD approach  has been used  to directly compute both tree amplitudes and penguin amplitudes. Apart from  LO contributions,   NLO order corrections in $\alpha_s$ have been included, which are sizeable especially to penguin contributions.

The mixing phase is given as
\begin{eqnarray}
 \phi_s^{\rm eff} =-  {\rm arg} \left[\frac{q}{p} \frac{\bar {\cal A}_{f}^\alpha}{{\cal A}_{f}^\alpha} \right] =\phi_s +\Delta \phi_s,
\end{eqnarray}
 where $\alpha$ denotes three polarization configurations $L$, $\parallel$, and $\perp$, and
 %$j(=L, \parallel, \perp)$ denotes three polarization configurations, and
 ${\cal A}_f^\alpha (\bar {\cal A}_f^\alpha)$ stands
 for the decay amplitude of $B_s \to J/\psi \phi (\bar B_s \to J/\psi \phi)$,
 which  can be decomposed into~\cite{Faller:2008gt}
  \begin{eqnarray}
  {\cal A}_f^\alpha (B_s \to J/\psi \phi) & = &  V_{cb}^* V_{cs} (T_c^\alpha + P_c^\alpha+P_t^\alpha) + V_{ub}^* V_{us} (P_u^\alpha + P_t^\alpha).
  \end{eqnarray}
Here, the unitarity relation $V_{tb}^* V_{ts} = - V_{cb}^*V_{cs} - V_{ub}^*V_{us}$ for the CKM matrix elements has been used.
The  tree amplitude $T_c^\alpha$ is dominant to  ${\cal B}(B_s \to J/\psi \phi)$, while  $P_c^\alpha$, $P_u^\alpha$, and $P_t^\alpha$ are   penguin pollutions.
The u-quark and c-quark penguin were not included in Ref.~\cite{Liu:2013nea}.  Then the charge conjugation amplitude for $B_s \to J/\psi \phi$ decay is
  \begin{eqnarray}
  \bar {\cal A}_f^\alpha(\bar B_s \to J/\psi \phi) &=&   V_{cb} V_{cs}^* (T_c^\alpha + P_c^\alpha+P_t^\alpha) + V_{ub} V_{us}^* (P_u^\alpha + P_t^\alpha) \;.
  \end{eqnarray}
For simplicity, one can introduce the ratio
\begin{eqnarray}
 a_{f}e^{i\delta_{f}+i\gamma} \equiv \frac{V_{ub}^* V_{us} (P_u^\alpha + P_t^\alpha)}{ V_{cb}^* V_{cs} (T_c^\alpha + P_c^\alpha+P_t^\alpha)},
\end{eqnarray}
which leads to  
\begin{eqnarray}
\frac{\bar {\cal A}_{f}^\alpha}{{\cal A}_{f}^\alpha}=  \frac{1+ a_{f}e^{i\delta_{f}-i\gamma}}{1+a_{f}e^{i\delta_{f}+i\gamma}} \simeq 1-2ia_{f}\cos\delta_{f} \sin\gamma,
\end{eqnarray}
and the phase shift:
\begin{eqnarray}
 \Delta\phi_{s} \simeq \arcsin(2a_{f}\cos\delta_{f}\sin\gamma).
 \end{eqnarray}

With the inclusion of various parametric errors,  the quantity $\Delta\phi_s$
from the  pQCD calculation is predicted  as follows~\cite{Liu:2013nea} 
\begin{eqnarray}
 \Delta\phi_s(L) &\approx&
 0.96^{+0.04}_{-0.03 }(\omega_B) ^{+0.02}_{-0.00}(f_M)^{+0.01}_{-0.01}(a_i)^{+0.03}_{-0.02}(m_c)\;\;
 [0.96^{+0.05}_{-0.04}]\times 10^{-3}  \;;  \nonumber\\
 \Delta\phi_s(\parallel) &\approx&
 0.84^{+0.02 }_{-0.02 }(\omega_B)^{+0.00}_{-0.00}(f_M)^{+0.01}_{-0.01}(a_i) ^{+0.00}_{-0.01}(m_c)\;\;
 [0.84^{+0.02}_{-0.02}]\times 10^{-3} \;;  \nonumber\\
 \Delta\phi_s(\perp) &\approx &
 0.80^{+0.01 }_{-0.01 }(\omega_B) ^{+0.00}_{-0.00}(f_M)^{+0.01}_{-0.01}(a_i)^{+0.00}_{-0.02}(m_c)\;\;
 [0.80^{+0.01}_{-0.02}]\times 10^{-3} \;.
  \end{eqnarray}
%the errors are from the variation of the shape parameter $\omega_{B}$ in the distribution amplitude
%of $B_s$ meson, the combined decay constants $f_M$ of $J/\psi$ and $\phi$, the Gegenbauer moments $a_i$ from the light-cone distribution amplitude
%of $\phi$ on both longitudinal and transverse polarizations, the charm quark mass $m_c$, and the CKM matrix elements ($\bar\rho, \bar \eta$), respectively, and
The values as given in the parentheses   have been added in quadrature.

Thanks to the large amount of data sample, the LHCb experiment is able to perform an analysis of the angular distribution of $B_{s}\to J/\psi\phi$.   Predictions for the P-wave   coefficients  (in units of $10^{-3}$)   are as follows:
 \begin{equation}{
\begin{array}{|c|c|c|c|c|c|c}
 f_{k} &\Delta a_{k} & \Delta b_{k} & \Delta c_{k} & \Delta d_{k}
 \\ \hline
 c_{K}^{2}s_{l}^{2} & 0.28 & 0.32& -0.3 & 1.0
 \\ \hline
 \frac{s_{K}^{2}(1-c_{\phi}^{2}c_{l}^{2)}}{2} & -0.83 & 1.52& 0.84& -1.0
 \\ \hline
 \frac{s_{K}^{2}(1-s_{\phi}^{2}c_{l}^{2)}}{2}  & -1.1 & -1.8& 1.2 & 1.1
 \\ \hline
 s_{K}^{2}s_{l}^{2} s_{\phi}c_{\phi} & -0.1 & 1.1& 6.4 &1.0
 \\ \hline
 \sqrt{2} s_{K} c_{K} s_{l}c_{l}c_{\phi}& -1.0 & 1.0& 0.3 & 0.03
 \\ \hline
 \sqrt{2} s_{K} c_{K} s_{l}c_{l}s_{\phi}& 1.1 &-0.02& -44& -1.4
\end{array}.
}\nonumber
\end{equation}
Most of the results for the other coefficients are of order $10^{-3}$.
The other four angular  coefficients  are the S-wave and the interference terms. The study of them requests the calculation of $B_{s}\to J/\psi (K^{+}K^{-})_{S}$, presumably dominated by the $f_{0}(980)$.

Though these results should be taken with caution as only part of  the known NLO contributions  is included, 
the deviation is found to be of ${\cal O}(10^{-3})$, and it may provide
an important SM reference for verifying the existing NP from
the $B_s \to J/\psi \phi$ data. 

%This finding  can be   examined  in the ongoing LHCb experiment and under-designed Super B factory and

%%%%%%%%%%%%%%%%%%%%%%%
%%%%%%%%%%%%%%%%%%%%%%%
\section{Summary and Outlook} 
\label{sec:conclusions}
%%%%%%%%%%%%%%%%%%%%%%%
%%%%%%%%%%%%%%%%%%%%%%%

Up to this date,  the CKM mechanism continues to give a consistent explanation of most available data on the flavor observables and CP violation with an incredible  accuracy.  This great success, together with the observation of   Higgs boson consistent with the SM~\cite{Aad:2012tfa,Chatrchyan:2012ufa},  implies  that the NP effects should be tiny and  renders the precision predictions for the involved quantities particularly important.

Despite the success, there are still unsatisfactory in this mechanism and thus perhaps some room for  NP.    In this review, 
we have focused on the heavy  flavour sector and  have discussed three quantities of the CKM matrix  in order to give an idea of the
current status of the field.  These quantities are quite uncertain at this stage. 
\begin{itemize}
\item The $|V_{ub}|$ can be extracted from inclusive and exclusive semileptonic $B$ decays. Deviations are found in the two determinations  but the significance, at about $3\sigma$ level, is still low. 
\item The angle $\gamma$ has the largest error, about $10^{\circ}$, compared to the other two CKM angles $\alpha,\beta$. This is the main source of errors in the unitarity triangle and currently  does not allow  us for  a direct test of CKM unitarity.
\item The $B_{s}-\bar B_{s}$ mixing phase is predicted  tiny in the SM. The current experimental  results are extracted from $B_{s}\to J/\psi\phi$ and $B_{s}\to J/\psi f_{0}(980)$.
\end{itemize}
Improving the knowledge on the CKM  renders the precise theoretical  predictions and experimental  measurements  important. The reduction of  experimental uncertainties  seems to have a promising prospect in   near future, due to the  large amount of  data sample (to be) collected at LHC~\cite{Bediaga:2012py} and the forthcoming Super KEKB factory~\cite{Aushev:2010bq}. 
So we are heading towards exciting  times in   CKM and 
flavour physics.

\section*{Acknowledgments}
The author is very grateful  to    Yu-Ming Wang and  Yue-Hong Xie  for  useful discussions, collaborations and carefully  reading the manuscript.  He also thanks Hai-Yang Cheng, Feng-Kun Guo, Christoph Hanhart, Bastian Kubis,  Gang Li, Hsiang-Nan Li,  Run-Hui Li, Ying Li, Xin Liu,  Cai-Dian L\"u,    Ulf-G. Mei{\ss}ner, Qin Qin, Yue-Long Shen, Wen-Fei Wang,  Fu-Sheng Yu,  Zhi-Qing Zhang, and Zhi-Tian Zou  for useful discussions. This work is supported in part by the DFG and the NSFC through funds provided to
the Sino-German CRC 110 ``Symmetries and the Emergence of Structure in QCD''.

%\end{appendix}

\bibliographystyle{ws-mpla}

\begin{thebibliography}{100}
 

%\cite{Cabibbo:1963yz}
\bibitem{Cabibbo:1963yz} 
  N.~Cabibbo,
  %\emph{Unitary Symmetry and Leptonic Decays},
  Phys.\ Rev.\ Lett.\  {\bf 10}, 531 (1963).

%\cite{Kobayashi:1973fv}
\bibitem{Kobayashi:1973fv} 
  M.~Kobayashi and T.~Maskawa,
  %\emph{CP Violation in the Renormalizable Theory of Weak Interaction},
  Prog.\ Theor.\ Phys.\  {\bf 49}, 652 (1973).
  
  
%\cite{Wolfenstein:1983yz}
\bibitem{Wolfenstein:1983yz} 
  L.~Wolfenstein,
   %\emph{Parametrization of the Kobayashi-Maskawa Matrix},
  Phys.\ Rev.\ Lett.\  {\bf 51}, 1945 (1983).

%\cite{Ahn:2011fg}
\bibitem{Ahn:2011fg} 
  Y.~H.~Ahn, H.~-Y.~Cheng and S.~Oh,
   %\emph{Wolfenstein Parametrization at Higher Order: Seeming Discrepancies and Their Resolution},
  Phys.\ Lett.\ B {\bf 703}, 571 (2011)
  [arXiv:1106.0935 [hep-ph]].
 
  
  %\cite{Ricciardi:2013cda}
\bibitem{Ricciardi:2013cda} 
  G.~Ricciardi,
   %\emph{Determination of the CKM matrix elements $|V_{(xb)}|$},
  Mod.\ Phys.\ Lett.\ A {\bf 28}, 1330016 (2013)
  [arXiv:1305.2844 [hep-ph]].
 
 %\cite{Ricciardi:2014iga}
\bibitem{Ricciardi:2014iga} 
  G.~Ricciardi,
   %\emph{Progress on semi-leptonic $B_{(s)}$ decays},
  arXiv:1403.7750 [hep-ph].

%\cite{Xiao:2014ana}
\bibitem{Xiao:2014ana} 
  Z.~-J.~Xiao, Y.~-Y.~Fan, W.~-F.~Wang and S.~Cheng,
  %\emph{The semileptonic decays of $B/B_s$ meson in the perturbative QCD approach: A short review},
  arXiv:1401.0571 [hep-ph].


%\cite{Lenz:2014nka}
\bibitem{Lenz:2014nka} 
  A.~Lenz,
  %``Selected Topics in Heavy Flavour Physics,''
  arXiv:1404.6197 [hep-ph].


%\cite{Beringer:1900zz}
\bibitem{Beringer:1900zz} 
  J.~Beringer {\it et al.}  [Particle Data Group Collaboration],
   %\emph{Review of Particle Physics (RPP)},
  Phys.\ Rev.\ D {\bf 86}, 010001 (2012). 
  


%\cite{Charles:2004jd}
\bibitem{Charles:2004jd} 
  J.~Charles {\it et al.}  [CKMfitter Group Collaboration],
   %\emph{CP violation and the CKM matrix: Assessing the impact of the asymmetric $B$ factories},
  Eur.\ Phys.\ J.\ C {\bf 41}, 1 (2005)
  [hep-ph/0406184],updated results and plots available at: http://ckmfitter.in2p3.fr. 

%\cite{Ciuchini:2000de}
\bibitem{Ciuchini:2000de} 
  M.~Ciuchini{\it et al.}  [UTfit Group Collaboration],
   %\emph{2000 CKM triangle analysis: A Critical review with updated experimental inputs and theoretical parameters},
  JHEP {\bf 0107}, 013 (2001)
  [hep-ph/0012308],
  updated results and plots available at: http://www.utfit.org/UTfit/WebHome.  
  
  
  

%\cite{Eigen:2013cv}
\bibitem{Eigen:2013cv} 
  G.~Eigen, G.~Dubois-Felsmann, D.~G.~Hitlin and F.~C.~Porter,
   %\emph{Global CKM Fits with the Scan Method},
  Phys.\ Rev.\ D {\bf 89}, 033004 (2014)
  [arXiv:1301.5867 [hep-ex]]. 
  
  
%\cite{Amhis:2012bh}
\bibitem{Amhis:2012bh} 
  Y.~Amhis {\it et al.}  [Heavy Flavor Averaging Group Collaboration],
   %\emph{Averages of B-Hadron, C-Hadron, and tau-lepton properties as of early 2012},
  arXiv:1207.1158 [hep-ex].


%\cite{Bediaga:2012py}
\bibitem{Bediaga:2012py} 
  R. Aaij {\it et al.}  [LHCb Collaboration],
   %%\emph{Implications of LHCb measurements and future prospects},
  Eur.\ Phys.\ J.\ C {\bf 73}, 2373 (2013)
  [arXiv:1208.3355 [hep-ex]]. 


%\cite{Bona:2007qt}
\bibitem{Bona:2007qt} 
  M.~Bona {\it et al.}  [SuperB Collaboration],
   %\emph{SuperB: A High-Luminosity Asymmetric e+ e- Super Flavor Factory. Conceptual Design Report},
  Pisa, Italy: INFN (2007) 453 p. www.pi.infn.it/SuperB/?q=CDR
  [arXiv:0709.0451 [hep-ex]]. 


%\cite{Aushev:2010bq}
\bibitem{Aushev:2010bq} 
  T.~Aushev, W.~Bartel, A.~Bondar, J.~Brodzicka, T.~E.~Browder, P.~Chang, Y.~Chao and K.~F.~Chen {\it et al.},
     %\emph{Physics at Super B Factory},
  arXiv:1002.5012 [hep-ex]. 


%\cite{Charles:2011va}
\bibitem{Charles:2011va} 
  J.~Charles, O.~Deschamps, S.~Descotes-Genon, R.~Itoh, H.~Lacker, A.~Menzel, S.~Monteil and V.~Niess {\it et al.},
   %\emph{Predictions of selected flavour observables within the Standard Model},
  Phys.\ Rev.\ D {\bf 84}, 033005 (2011)
  [arXiv:1106.4041 [hep-ph]]. 


%\cite{Aaltonen:2007he}
\bibitem{Aaltonen:2007he} 
  T.~Aaltonen {\it et al.}  [CDF Collaboration],
     %\emph{First Flavor-Tagged Determination of Bounds on Mixing-Induced CP Violation in $B^0_{s} \to J/\psi \phi$ Decays},
  Phys.\ Rev.\ Lett.\  {\bf 100}, 161802 (2008)
  [arXiv:0712.2397 [hep-ex]]. 

%\cite{Abazov:2008af}
\bibitem{Abazov:2008af} 
  V.~M.~Abazov {\it et al.}  [D0 Collaboration],
     %\emph{Measurement of $B^0_{s}$ mixing parameters from the flavor-tagged decay $B^0_{s} \to J/\psi \phi$},
  Phys.\ Rev.\ Lett.\  {\bf 101}, 241801 (2008)
  [arXiv:0802.2255 [hep-ex]]. 



%\cite{Kim:2010wr}
\bibitem{Kim:2010wr} 
  C.~S.~Kim and Y.~Li,
     %\emph{Comments on $\left|\frac{V_{ub}}{V_{cb}}\right|$ and $|V_{ub}|$ from Non-leptonic $B$ Decays within the Perturbative QCD Approach},
  Eur.\ Phys.\ J.\ C {\bf 71}, 1531 (2011)
  [arXiv:1007.2291 [hep-ph]]. 
  
  
  %\cite{Crivellin:2014zpa}
\bibitem{Crivellin:2014zpa} 
  A.~Crivellin and S.~Pokorski,
  %``Can the differences in the determinations of $V_{ub}$ and $V_{cb}$ be explained by New Physics?,''
  arXiv:1407.1320 [hep-ph].
  %%CITATION = ARXIV:1407.1320;%%
  %1 citations counted in INSPIRE as of 05 Aug 2014
     
     %\cite{Brucherseifer:2013cu}
\bibitem{Brucherseifer:2013cu} 
  M.~Brucherseifer, F.~Caola and K.~Melnikov,
     %\emph{On the $O(\alpha_s^2)$ corrections to $b \to X_u e \bar \nu$ inclusive decays},
  Phys.\ Lett.\ B {\bf 721}, 107 (2013)
  [arXiv:1302.0444 [hep-ph]]. 
     
     %\cite{Lange:2005yw}
\bibitem{Lange:2005yw} 
  B.~O.~Lange, M.~Neubert and G.~Paz,
     %\emph{Theory of charmless inclusive B decays and the extraction of $V_{ub}$},
  Phys.\ Rev.\ D {\bf 72}, 073006 (2005)
  [hep-ph/0504071]. 
     
%\cite{Bosch:2004th}
\bibitem{Bosch:2004th} 
  S.~W.~Bosch, B.~O.~Lange, M.~Neubert and G.~Paz,
     %\emph{Factorization and shape function effects in inclusive B meson decays},
  Nucl.\ Phys.\ B {\bf 699}, 335 (2004)
  [hep-ph/0402094]. 

%\cite{Gambino:2007rp}
\bibitem{Gambino:2007rp} 
  P.~Gambino, P.~Giordano, G.~Ossola and N.~Uraltsev,
     %\emph{Inclusive semileptonic B decays and the determination of $|V_{ub}|$},
  JHEP {\bf 0710}, 058 (2007)
  [arXiv:0707.2493 [hep-ph]]. 

%\cite{Andersen:2005mj}
\bibitem{Andersen:2005mj} 
  J.~R.~Andersen and E.~Gardi,
     %\emph{Inclusive spectra in charmless semileptonic B decays by dressed gluon exponentiation},
  JHEP {\bf 0601}, 097 (2006)
  [hep-ph/0509360]. 
%\cite{Aglietti:2004fz}
\bibitem{Aglietti:2004fz} 
  U.~Aglietti and G.~Ricciardi,
     %\emph{A Model for next-to-leading order threshold resummed form-factors},
  Phys.\ Rev.\ D {\bf 70}, 114008 (2004)
  [hep-ph/0407225]. 

%\cite{Aglietti:2006yb}
\bibitem{Aglietti:2006yb} 
  U.~Aglietti, G.~Ferrera and G.~Ricciardi,
     %\emph{Semi-Inclusive B Decays and a Model for Soft-Gluon Effects},
  Nucl.\ Phys.\ B {\bf 768}, 85 (2007)
  [hep-ph/0608047]. 
%\cite{Aglietti:2007ik}
\bibitem{Aglietti:2007ik} 
  U.~Aglietti, F.~Di Lodovico, G.~Ferrera and G.~Ricciardi,
     %\emph{Inclusive measure of $|V_{ub}|$ with the analytic coupling model},
  Eur.\ Phys.\ J.\ C {\bf 59}, 831 (2009)
  [arXiv:0711.0860 [hep-ph]]. 



%\cite{Urquijo:2009tp}
\bibitem{Urquijo:2009tp} 
  P.~Urquijo {\it et al.}  [Belle Collaboration],
     %\emph{Measurement Of $|V_{ub}|$ From Inclusive Charmless Semileptonic B Decays},
  Phys.\ Rev.\ Lett.\  {\bf 104}, 021801 (2010)
  [arXiv:0907.0379 [hep-ex]]. 


%\cite{Lees:2011fv}
\bibitem{Lees:2011fv} 
  J.~P.~Lees {\it et al.}  [BaBar Collaboration],
     %\emph{Study of $\bar{B}\to X_u \ell \bar{\nu}$ decays in $B\bar{B}$ events tagged by a fully reconstructed B-meson decay and determination of $\|V_{ub}\|$},
  Phys.\ Rev.\ D {\bf 86}, 032004 (2012)
  [arXiv:1112.0702 [hep-ex]]. 
  
  
%\cite{Lees:2012vv}
\bibitem{Lees:2012vv} 
  J.~P.~Lees {\it et al.}  [BaBar Collaboration],
     %\emph{Branching fraction and form-factor shape measurements of exclusive charmless semileptonic B decays, and determination of $|V_{ub}|$},
  Phys.\ Rev.\ D {\bf 86}, 092004 (2012)
  [arXiv:1208.1253 [hep-ex]].  

%\cite{Sibidanov:2013rkk}
\bibitem{Sibidanov:2013rkk} 
  A.~Sibidanov {\it et al.}  [Belle Collaboration],
     %\emph{Study of Exclusive $B \to X_u \ell \nu$ Decays and Extraction of $\|V_{ub}\|$ using Full Reconstruction Tagging at the Belle Experiment},
  Phys.\ Rev.\ D {\bf 88}, no. 3, 032005 (2013)
  [arXiv:1306.2781 [hep-ex]]. 

%\cite{Bharucha:2012wy}
\bibitem{Bharucha:2012wy} 
  A.~Bharucha,
     %\emph{Two-loop Corrections to the B to pi Form Factor from QCD Sum Rules on the Light-Cone and $|V_{ub}$},
  JHEP {\bf 1205}, 092 (2012)
  [arXiv:1203.1359 [hep-ph]]. 
  
  %\cite{Li:2012gr}
\bibitem{Li:2012gr} 
  Z.~-H.~Li, N.~Zhu, X.~-J.~Fan and T.~Huang,
     %\emph{Form Factors $f^{B\to \pi}_+(0)$ and $f^{D\to \pi}_+(0)$ in $QCD$ and Determination of $|V_{ub}|$ and $|V_{cd}|$},
  JHEP {\bf 1205}, 160 (2012)
  [arXiv:1206.0091 [hep-ph]]. 
  
  
%\cite{Khodjamirian:2011ub}
\bibitem{Khodjamirian:2011ub} 
  A.~Khodjamirian, T.~.Mannel, N.~Offen and Y.~-M.~Wang,
     %\emph{$B \to \pi \ell \nu_l$ Width and $|V_{ub}|$ from QCD Light-Cone Sum Rules},
  Phys.\ Rev.\ D {\bf 83}, 094031 (2011)
  [arXiv:1103.2655 [hep-ph]]. 
  
  
%\cite{Keum:2000ph}
\bibitem{Keum:2000ph} 
  Y.~-Y.~Keum, H.~-n.~Li and A.~I.~Sanda,
  %``Fat penguins and imaginary penguins in perturbative QCD,''
  Phys.\ Lett.\ B {\bf 504}, 6 (2001)
  [hep-ph/0004004]. 


%\cite{Keum:2000wi}
\bibitem{Keum:2000wi} 
  Y.~Y.~Keum, H.~-N.~Li and A.~I.~Sanda,
  %``Penguin enhancement and $B \to K \pi$ decays in perturbative QCD,''
  Phys.\ Rev.\ D {\bf 63}, 054008 (2001)
  [hep-ph/0004173].
  %%CITATION = HEP-PH/0004173;%%
  %628 citations counted in INSPIRE as of 14 Feb 2014


%\cite{Lu:2000em}
\bibitem{Lu:2000em} 
  C.~-D.~Lu, K.~Ukai and M.~-Z.~Yang,
  %``Branching ratio and CP violation of B ---> pi pi decays in perturbative QCD approach,''
  Phys.\ Rev.\ D {\bf 63}, 074009 (2001)
  [hep-ph/0004213].
  %%CITATION = HEP-PH/0004213;%%
  %338 citations counted in INSPIRE as of 14 Feb 2014


%\cite{Lu:2000hj}
\bibitem{Lu:2000hj} 
  C.~-D.~Lu and M.~-Z.~Yang,
  %``B ---> pi rho, pi omega decays in perturbative QCD approach,''
  Eur.\ Phys.\ J.\ C {\bf 23}, 275 (2002)
  [hep-ph/0011238].
  %%CITATION = HEP-PH/0011238;%%
  %123 citations counted in INSPIRE as of 14 Feb 2014


%\cite{Kurimoto:2001zj}
\bibitem{Kurimoto:2001zj} 
  T.~Kurimoto, H.~-n.~Li and A.~I.~Sanda,
  %``Leading power contributions to B ---> pi, rho transition form-factors,''
  Phys.\ Rev.\ D {\bf 65}, 014007 (2002)
  [hep-ph/0105003].
  %%CITATION = HEP-PH/0105003;%%
  %163 citations counted in INSPIRE as of 14 Feb 2014


  %\cite{Li:2010nn}
\bibitem{Li:2010nn} 
  H.~-n.~Li, Y.~-L.~Shen, Y.~-M.~Wang and H.~Zou,
  %``Next-to-leading-order correction to pion form factor in $k_T$ factorization,''
  Phys.\ Rev.\ D {\bf 83}, 054029 (2011)
  [arXiv:1012.4098 [hep-ph]].
  %%CITATION = ARXIV:1012.4098;%%
  %18 citations counted in INSPIRE as of 24 Apr 2014
  
%\cite{Li:2012nk}
\bibitem{Li:2012nk} 
  H.~-n.~Li, Y.~-L.~Shen and Y.~-M.~Wang,
     %\emph{Next-to-leading-order corrections to $B \to \pi$ form factors in $k_T$ factorization},
  Phys.\ Rev.\ D {\bf 85}, 074004 (2012)
  [arXiv:1201.5066 [hep-ph]].
  %%CITATION = ARXIV:1201.5066;%%
  %23 citations counted in INSPIRE as of 22 Apr 2014
  
  %\cite{Hu:2012cp}
\bibitem{Hu:2012cp} 
  H.~-C.~Hu and H.~-n.~Li,
  %``Next-to-leading-order time-like pion form factors in $k_T$ factorization,''
  Phys.\ Lett.\ B {\bf 718}, 1351 (2013)
  [arXiv:1204.6708 [hep-ph]].
  %%CITATION = ARXIV:1204.6708;%%
  %2 citations counted in INSPIRE as of 24 Apr 2014
  
  %\cite{Li:2012md}
\bibitem{Li:2012md} 
  H.~-N.~Li, Y.~-L.~Shen and Y.~-M.~Wang,
  %``Resummation of rapidity logarithms in $B$ meson wave functions,''
  JHEP {\bf 1302}, 008 (2013)
  [arXiv:1210.2978 [hep-ph]].
  %%CITATION = ARXIV:1210.2978;%%
  %7 citations counted in INSPIRE as of 05 May 2014
  
  %\cite{Li:2013xna}
\bibitem{Li:2013xna} 
  H.~-N.~Li, Y.~-L.~Shen and Y.~-M.~Wang,
  %``Joint resummation for pion wave function and pion transition form factor,''
  JHEP {\bf 1401}, 004 (2014)
  [arXiv:1310.3672 [hep-ph]].
  %%CITATION = ARXIV:1310.3672;%%
  %5 citations counted in INSPIRE as of 05 May 2014
  
%\cite{Wang:2012ab}
\bibitem{Wang:2012ab} 
  W.~-F.~Wang and Z.~-J.~Xiao,
     %\emph{The semileptonic decays $B/B_s \to (\pi, K)(\ell^+\ell^-,\ell\nu,\nu\bar{\nu})$ in the perturbative QCD approach beyond the leading-order},
  Phys.\ Rev.\ D {\bf 86}, 114025 (2012)
  [arXiv:1207.0265 [hep-ph]].
  %%CITATION = ARXIV:1207.0265;%%
  %23 citations counted in INSPIRE as of 22 Apr 2014

%\cite{Cheng:2014fwa}
\bibitem{Cheng:2014fwa} 
  S.~Cheng, Y.~-Y.~Fan, X.~Yu, C.~-D.~L\"u and Z.~-J.~Xiao,
     %\emph{The NLO twist-3 contributions to $B \to \pi$ form factors in $k_{T}$ factorization},
  arXiv:1402.5501 [hep-ph].
  %%CITATION = ARXIV:1402.5501;%%
  %1 citations counted in INSPIRE as of 22 Apr 2014



%\cite{Bailey:2008wp}
\bibitem{Bailey:2008wp} 
  J.~A.~Bailey, C.~Bernard, C.~E.~DeTar, M.~Di Pierro, A.~X.~El-Khadra, R.~T.~Evans, E.~D.~Freeland and E.~Gamiz {\it et al.},
     %\emph{The $B \to \pi \ell \nu$ semileptonic form factor from three-flavor lattice QCD: A Model-independent determination of $|V_{ub}|$},
  Phys.\ Rev.\ D {\bf 79}, 054507 (2009)
  [arXiv:0811.3640 [hep-lat]].
  %%CITATION = ARXIV:0811.3640;%%
  %111 citations counted in INSPIRE as of 14 Feb 2014


%\cite{Dalgic:2006dt}
\bibitem{Dalgic:2006dt} 
  E.~Dalgic, A.~Gray, M.~Wingate, C.~T.~H.~Davies, G.~P.~Lepage and J.~Shigemitsu,
     %\emph{B meson semileptonic form-factors from unquenched lattice QCD},
  Phys.\ Rev.\ D {\bf 73}, 074502 (2006)
  [Erratum-ibid.\ D {\bf 75}, 119906 (2007)]
  [hep-lat/0601021].
  %%CITATION = HEP-LAT/0601021;%%
  %176 citations counted in INSPIRE as of 14 Feb 2014


%\cite{AlHaydari:2009zr}
\bibitem{AlHaydari:2009zr} 
  A.~Al-Haydari {\it et al.}  [QCDSF Collaboration],
     %\emph{Semileptonic form factors D ---> pi, K and B ---> pi, K from a fine lattice},
  Eur.\ Phys.\ J.\ A {\bf 43}, 107 (2010)
  [arXiv:0903.1664 [hep-lat]].
  %%CITATION = ARXIV:0903.1664;%%
  %31 citations counted in INSPIRE as of 14 Feb 2014



%\cite{Du:2013kea}
\bibitem{Du:2013kea} 
  D.~Du, J.~A.~Bailey, A.~Bazavov, C.~Bernard, A.~X.~El-Khadra, S.~Gottlieb, R.~D.~Jain and A.~S.~Kronfeld {\it et al.},
     %\emph{$B\to\pi\ell\nu$ and $B\to\pi\ell^+\ell^-$ semileptonic form factors from unquenched lattice QCD},
  PoS LATTICE {\bf 2013}, 383 (2013)
  [arXiv:1311.6552 [hep-lat]].
  %%CITATION = ARXIV:1311.6552;%%
  %5 citations counted in INSPIRE as of 22 Apr 2014

%\cite{Kawanai:2013qxa}
\bibitem{Kawanai:2013qxa} 
  T.~Kawanai, R.~S.~Van de Water and O.~Witzel,
     %\emph{The form factors for $B \to \pi \ell \nu$ semileptonic decay from 2+1 flavors of domain-wall fermions},
  arXiv:1311.1143 [hep-lat].
  %%CITATION = ARXIV:1311.1143;%%
  %5 citations counted in INSPIRE as of 22 Apr 2014
  
  
  
%\cite{Meissner:2013pba}
\bibitem{Meissner:2013pba} 
  U.~-G.~Mei{\ss}ner and W.~Wang,
  %``${\bf B_s\to K^{(*)} \ell\bar \nu}$, Angular Analysis, S-wave Contributions and ${\bf |V_{ub}|}$,''
  JHEP {\bf 1401}, 107 (2014)
  [arXiv:1311.5420 [hep-ph]].
  %%CITATION = ARXIV:1311.5420;%%
  %2 citations counted in INSPIRE as of 14 Feb 2014

  
%\cite{Bauer:2000ew}
\bibitem{Bauer:2000ew} 
  C.~W.~Bauer, S.~Fleming and M.~E.~Luke,
  %``Summing Sudakov logarithms in B ---> X(s gamma) in effective field theory,''
  Phys.\ Rev.\ D {\bf 63}, 014006 (2000)
  [hep-ph/0005275].
  %%CITATION = HEP-PH/0005275;%%
  %475 citations counted in INSPIRE as of 14 Feb 2014




%\cite{Bauer:2000yr}
\bibitem{Bauer:2000yr} 
  C.~W.~Bauer, S.~Fleming, D.~Pirjol and I.~W.~Stewart,
  %``An Effective field theory for collinear and soft gluons: Heavy to light decays,''
  Phys.\ Rev.\ D {\bf 63}, 114020 (2001)
  [hep-ph/0011336].
  %%CITATION = HEP-PH/0011336;%%
  %751 citations counted in INSPIRE as of 14 Feb 2014


%\cite{Bauer:2001yt}
\bibitem{Bauer:2001yt} 
  C.~W.~Bauer, D.~Pirjol and I.~W.~Stewart,
  %``Soft collinear factorization in effective field theory,''
  Phys.\ Rev.\ D {\bf 65}, 054022 (2002)
  [hep-ph/0109045].
  %%CITATION = HEP-PH/0109045;%%
  %625 citations counted in INSPIRE as of 14 Feb 2014


%\cite{Beneke:2003pa}
\bibitem{Beneke:2003pa} 
  M.~Beneke and T.~Feldmann,
  %``Factorization of heavy to light form-factors in soft collinear effective theory,''
  Nucl.\ Phys.\ B {\bf 685}, 249 (2004)
  [hep-ph/0311335].
  %%CITATION = HEP-PH/0311335;%%
  %132 citations counted in INSPIRE as of 14 Feb 2014
  
  


%\cite{Becher:2005bg}
\bibitem{Becher:2005bg} 
  T.~Becher and R.~J.~Hill,
     %\emph{Comment on form-factor shape and extraction of |V(ub)| from B ---> pi l nu},
  Phys.\ Lett.\ B {\bf 633}, 61 (2006)
  [hep-ph/0509090].
  %%CITATION = HEP-PH/0509090;%%
  %117 citations counted in INSPIRE as of 14 Feb 2014


%\cite{Arnesen:2005ez}
\bibitem{Arnesen:2005ez} 
  M.~C.~Arnesen, B.~Grinstein, I.~Z.~Rothstein and I.~W.~Stewart,
     %\emph{A Precision model independent determination of |V(ub)| from B ---> pi e nu},
  Phys.\ Rev.\ Lett.\  {\bf 95}, 071802 (2005)
  [hep-ph/0504209].
  %%CITATION = HEP-PH/0504209;%%
  %110 citations counted in INSPIRE as of 14 Feb 2014


%\cite{Bourrely:2008za}
\bibitem{Bourrely:2008za} 
  C.~Bourrely, I.~Caprini and L.~Lellouch,
     %\emph{Model-independent description of B ---> pi l nu decays and a determination of |V(ub)|},
  Phys.\ Rev.\ D {\bf 79}, 013008 (2009)
  [Erratum-ibid.\ D {\bf 82}, 099902 (2010)]
  [arXiv:0807.2722 [hep-ph]].
  %%CITATION = ARXIV:0807.2722;%%
  %76 citations counted in INSPIRE as of 14 Feb 2014


%\cite{Becirevic:1999kt}
\bibitem{Becirevic:1999kt} 
  D.~Becirevic and A.~B.~Kaidalov,
     %\emph{Comment on the heavy ---> light form-factors},
  Phys.\ Lett.\ B {\bf 478}, 417 (2000)
  [hep-ph/9904490].
  %%CITATION = HEP-PH/9904490;%%
  %239 citations counted in INSPIRE as of 14 Feb 2014



%\cite{Shifman:1978bx}
\bibitem{Shifman:1978bx} 
  M.~A.~Shifman, A.~I.~Vainshtein and V.~I.~Zakharov,
     %\emph{QCD and Resonance Physics. Sum Rules},
  Nucl.\ Phys.\ B {\bf 147}, 385 (1979).
  %%CITATION = NUPHA,B147,385;%%
  %4171 citations counted in INSPIRE as of 14 Feb 2014


%\cite{Shifman:1978by}
\bibitem{Shifman:1978by} 
  M.~A.~Shifman, A.~I.~Vainshtein and V.~I.~Zakharov,
     %\emph{QCD and Resonance Physics: Applications},
  Nucl.\ Phys.\ B {\bf 147}, 448 (1979).
  %%CITATION = NUPHA,B147,448;%%
  %2384 citations counted in INSPIRE as of 14 Feb 2014


%\cite{Craigie:1982ng}
\bibitem{Craigie:1982ng} 
  N.~S.~Craigie and J.~Stern,
     %\emph{What Can We Learn From Sum Rules for Vertex Functions in {QCD}},
  Nucl.\ Phys.\ B {\bf 216}, 209 (1983).
  %%CITATION = NUPHA,B216,209;%%
  %37 citations counted in INSPIRE as of 14 Feb 2014


%\cite{Braun:1988qv}
\bibitem{Braun:1988qv} 
  V.~M.~Braun and I.~E.~Filyanov,
     %\emph{QCD Sum Rules in Exclusive Kinematics and Pion Wave Function},
  Z.\ Phys.\ C {\bf 44}, 157 (1989)
  [Sov.\ J.\ Nucl.\ Phys.\  {\bf 50}, 511 (1989)]
  [Yad.\ Fiz.\  {\bf 50}, 818 (1989)].
  %%CITATION = ZEPYA,C44,157;%%
  %336 citations counted in INSPIRE as of 14 Feb 2014


%\cite{Chernyak:1990ag}
\bibitem{Chernyak:1990ag} 
  V.~L.~Chernyak and I.~R.~Zhitnitsky,
     %\emph{B meson exclusive decays into baryons},
  Nucl.\ Phys.\ B {\bf 345}, 137 (1990).
  %%CITATION = NUPHA,B345,137;%%
  %340 citations counted in INSPIRE as of 14 Feb 2014


%\cite{Belyaev:1994zk}
\bibitem{Belyaev:1994zk} 
  V.~M.~Belyaev, V.~M.~Braun, A.~Khodjamirian and R.~Ruckl,
     %\emph{D* D pi and B* B pi couplings in QCD},
  Phys.\ Rev.\ D {\bf 51}, 6177 (1995)
  [hep-ph/9410280].
  %%CITATION = HEP-PH/9410280;%%
  %387 citations counted in INSPIRE as of 14 Feb 2014


%\cite{Colangelo:2000dp}
\bibitem{Colangelo:2000dp} 
  P.~Colangelo and A.~Khodjamirian,
     %\emph{QCD sum rules, a modern perspective},
  In *Shifman, M. (ed.): At the frontier of particle physics, vol. 3* 1495-1576
  [hep-ph/0010175].
  %%CITATION = HEP-PH/0010175;%%
  %360 citations counted in INSPIRE as of 14 Feb 2014




%\cite{Lees:2013gja}
\bibitem{Lees:2013gja} 
  J.~P.~Lees {\it et al.}  [BaBar Collaboration],
  %\emph{Measurement of the B+ --> omega l+ nu branching fraction with semileptonically tagged B mesons},
  Phys.\ Rev.\ D {\bf 88}, 072006 (2013)
  [arXiv:1308.2589 [hep-ex]].
  %%CITATION = ARXIV:1308.2589;%%
  %1 citations counted in INSPIRE as of 23 Apr 2014
  

 
%\cite{Ball:2004rg}
\bibitem{Ball:2004rg} 
  P.~Ball and R.~Zwicky,
  %``B(D,S) ---> rho, omega, K*, phi decay form-factors from light-cone sum rules revisited,''
  Phys.\ Rev.\ D {\bf 71}, 014029 (2005)
  [hep-ph/0412079].
  %%CITATION = HEP-PH/0412079;%%
  %378 citations counted in INSPIRE as of 14 Feb 2014
 
 %\cite{Del Debbio:1997kr}
\bibitem{DelDebbio:1997kr} 
  L.~Del Debbio {\it et al.}  [UKQCD Collaboration],
  %``Lattice constrained parametrizations of form-factors for semileptonic and rare radiative B decays,''
  Phys.\ Lett.\ B {\bf 416}, 392 (1998)
  [hep-lat/9708008].
  %%CITATION = HEP-LAT/9708008;%%
  %168 citations counted in INSPIRE as of 24 Apr 2014  
  
  %\cite{Chen:2009qk}
\bibitem{Chen:2009qk} 
  C.~-H.~Chen, Y.~-L.~Shen and W.~Wang,
  %``|V(ub)| and B ---> eta(') Form Factors in Covariant Light Front Approach,''
  Phys.\ Lett.\ B {\bf 686}, 118 (2010)
  [arXiv:0911.2875 [hep-ph]].
  %%CITATION = ARXIV:0911.2875;%%
  %7 citations counted in INSPIRE as of 22 Apr 2014


%\cite{Wang:2013ix}
\bibitem{Wang:2013ix} 
  W.~-F.~Wang, Y.~-Y.~Fan, M.~Liu and Z.~-J.~Xiao,
  %``Semileptonic decays $B/B_s \to (\eta,\etar, G)(\ell^+\ell^-,\ell\bar{\nu},\nu\bar{\nu})$ in the perturbative QCD approach beyond the leading order,''
  Phys.\ Rev.\ D {\bf 87}, 097501 (2013)
  [arXiv:1301.0197].
  %%CITATION = ARXIV:1301.0197;%%
  %4 citations counted in INSPIRE as of 14 Feb 2014




%\cite{Horgan:2013hoa}
\bibitem{Horgan:2013hoa} 
  R.~R.~Horgan, Z.~Liu, S.~Meinel and M.~Wingate,
  %``Lattice QCD calculation of form factors describing the rare decays $B \to K^* \ell^+ \ell^-$ and $B_s \to \phi \ell^+ \ell^-$,''
  arXiv:1310.3722 [hep-lat].
  %%CITATION = ARXIV:1310.3722;%%
  %7 citations counted in INSPIRE as of 14 Feb 2014

 
%\cite{Faustov:2013ima}
\bibitem{Faustov:2013ima} 
  R.~N.~Faustov and V.~O.~Galkin,
  %``Charmless weak $B_s$ decays in the relativistic quark model,''
  Phys.\ Rev.\ D {\bf 87}, no. 9, 094028 (2013)
  [arXiv:1304.3255].
  %%CITATION = ARXIV:1304.3255;%%
  %3 citations counted in INSPIRE as of 14 Feb 2014


%\cite{Cheng:2003sm}
\bibitem{Cheng:2003sm} 
  H.~-Y.~Cheng, C.~-K.~Chua and C.~-W.~Hwang,
  %``Covariant light front approach for s wave and p wave mesons: Its application to decay constants and form-factors,''
  Phys.\ Rev.\ D {\bf 69}, 074025 (2004)
  [hep-ph/0310359].
  %%CITATION = HEP-PH/0310359;%%
  %194 citations counted in INSPIRE as of 14 Feb 2014


%\cite{Lu:2007sg}
\bibitem{Lu:2007sg} 
  C.~-D.~Lu, W.~Wang and Z.~-T.~Wei,
  %``Heavy-to-light form factors on the light cone,''
  Phys.\ Rev.\ D {\bf 76}, 014013 (2007)
  [hep-ph/0701265 [HEP-PH]].
  %%CITATION = HEP-PH/0701265;%%
  %39 citations counted in INSPIRE as of 14 Feb 2014


%\cite{Verma:2011yw}
\bibitem{Verma:2011yw} 
  R.~C.~Verma,
  %``Decay constants and form factors of s-wave and p-wave mesons in the covariant light-front quark model,''
  J.\ Phys.\ G {\bf 39}, 025005 (2012)
  [arXiv:1103.2973 [hep-ph]].
  %%CITATION = ARXIV:1103.2973;%%
  %8 citations counted in INSPIRE as of 14 Feb 2014

 
%\cite{Hussain:1992rb}
\bibitem{Hussain:1992rb} 
  F.~Hussain, D.~-S.~Liu, M.~Kramer, J.~G.~Korner and S.~Tawfiq,
  %``General analysis of weak decay form-factors in heavy to heavy and heavy to light baryon transitions,''
  Nucl.\ Phys.\ B {\bf 370}, 259 (1992).
  %%CITATION = NUPHA,B370,259;%%
  %47 citations counted in INSPIRE as of 14 Feb 2014


%\cite{Mannel:1990vg}
\bibitem{Mannel:1990vg} 
  T.~Mannel, W.~Roberts and Z.~Ryzak,
  %``Baryons in the heavy quark effective theory,''
  Nucl.\ Phys.\ B {\bf 355}, 38 (1991).
  %%CITATION = NUPHA,B355,38;%%
  %227 citations counted in INSPIRE as of 14 Feb 2014


%\cite{Hussain:1990uu}
\bibitem{Hussain:1990uu} 
  F.~Hussain, J.~G.~Korner, M.~Kramer and G.~Thompson,
  %``On heavy baryon decay form-factors,''
  Z.\ Phys.\ C {\bf 51}, 321 (1991).
  %%CITATION = ZEPYA,C51,321;%%
  %115 citations counted in INSPIRE as of 14 Feb 2014


%\cite{Feldmann:2011xf}
\bibitem{Feldmann:2011xf} 
  T.~Feldmann and M.~W.~Y.~Yip,
  %``Form Factors for Lambda_b -> Lambda Transitions in SCET,''
  Phys.\ Rev.\ D {\bf 85}, 014035 (2012)
  [Erratum-ibid.\ D {\bf 86}, 079901 (2012)]
  [arXiv:1111.1844 [hep-ph]].
  %%CITATION = ARXIV:1111.1844;%%
  %15 citations counted in INSPIRE as of 14 Feb 2014


%\cite{Mannel:2011xg}
\bibitem{Mannel:2011xg} 
  T.~Mannel and Y.~-M.~Wang,
  %``Heavy-to-light baryonic form factors at large recoil,''
  JHEP {\bf 1112}, 067 (2011)
  [arXiv:1111.1849 [hep-ph]].
  %%CITATION = ARXIV:1111.1849;%%
  %13 citations counted in INSPIRE as of 14 Feb 2014


%\cite{Wang:2011uv}
\bibitem{Wang:2011uv} 
  W.~Wang,
  %``Factorization of Heavy-to-Light Baryonic Transitions in SCET,''
  Phys.\ Lett.\ B {\bf 708}, 119 (2012)
  [arXiv:1112.0237 [hep-ph]].
  %%CITATION = ARXIV:1112.0237;%%
  %5 citations counted in INSPIRE as of 14 Feb 2014

%\cite{Wei:2009np}
\bibitem{Wei:2009np} 
  Z.~-T.~Wei, H.~-W.~Ke and X.~-Q.~Li,
  %``Evaluating decay Rates and Asymmetries of Lambda(b) into Light Baryons in LFQM,''
  Phys.\ Rev.\ D {\bf 80}, 094016 (2009)
  [arXiv:0909.0100 [hep-ph]].
  %%CITATION = ARXIV:0909.0100;%%

%\cite{Huang:1998rq}
\bibitem{Huang:1998rq} 
  C.~-S.~Huang, C.~-F.~Qiao and H.~-G.~Yan,
  %``Decay Lambda(b) ---> p lepton anti-neutrino in QCD sum rules,''
  Phys.\ Lett.\ B {\bf 437}, 403 (1998)
  [hep-ph/9805452].
  %%CITATION = HEP-PH/9805452;%%
  %17 citations counted in INSPIRE as of 14 Feb 2014

 
%\cite{Marques de Carvalho:1999ia}
\bibitem{Carvalho:1999ia} 
  R.~S.~Marques de Carvalho, F.~S.~Navarra, M.~Nielsen, E.~Ferreira and H.~G.~Dosch,
  %``Form-factors and decay rates for heavy Lambda semileptonic decays from QCD sum rules,''
  Phys.\ Rev.\ D {\bf 60}, 034009 (1999)
  [hep-ph/9903326].
  %%CITATION = HEP-PH/9903326;%%
  %34 citations counted in INSPIRE as of 24 Apr 2014
 
%\cite{Huang:2004vf}
\bibitem{Huang:2004vf} 
  M.~-q.~Huang and D.~-W.~Wang,
  %``Light cone QCD sum rules for the semileptonic decay Lambda(b) -> p l anti-nu,''
  Phys.\ Rev.\ D {\bf 69}, 094003 (2004)
  [hep-ph/0401094].
  %%CITATION = HEP-PH/0401094;%%
  %34 citations counted in INSPIRE as of 14 Feb 2014


%\cite{Wang:2009hra}
\bibitem{Wang:2009hra} 
  Y.~-M.~Wang, Y.~-L.~Shen and C.~-D.~Lu,
  %``Lambda(b) ---> p, Lambda transition form factors from QCD light-cone sum rules,''
  Phys.\ Rev.\ D {\bf 80}, 074012 (2009)
  [arXiv:0907.4008 [hep-ph]].
  %%CITATION = ARXIV:0907.4008;%%
  %16 citations counted in INSPIRE as of 14 Feb 2014


%\cite{Azizi:2009wn}
\bibitem{Azizi:2009wn} 
  K.~Azizi, M.~Bayar, Y.~Sarac and H.~Sundu,
  %``Semileptonic Lambda(b,c) to Nucleon Transitions in Full QCD at Light Cone,''
  Phys.\ Rev.\ D {\bf 80}, 096007 (2009)
  [arXiv:0908.1758 [hep-ph]].
  %%CITATION = ARXIV:0908.1758;%%
  %8 citations counted in INSPIRE as of 14 Feb 2014

%\cite{Khodjamirian:2011jp}
\bibitem{Khodjamirian:2011jp} 
  A.~Khodjamirian, C.~.Klein, T.~.Mannel and Y.~-M.~Wang,
  %``Form Factors and Strong Couplings of Heavy Baryons from QCD Light-Cone Sum Rules,''
  JHEP {\bf 1109}, 106 (2011)
  [arXiv:1108.2971 [hep-ph]].
  %%CITATION = ARXIV:1108.2971;%%
  %24 citations counted in INSPIRE as of 24 Apr 2014

%\cite{Detmold:2013nia}
\bibitem{Detmold:2013nia} 
  W.~Detmold, C.~-J.~D.~Lin, S.~Meinel and M.~Wingate,
  %``$\Lambda_b \to p l^- \bar{\nu}$ form factors from lattice QCD with static b quarks,''
  Phys.\ Rev.\ D {\bf 88}, 014512 (2013)
  [arXiv:1306.0446 [hep-lat]].
  %%CITATION = ARXIV:1306.0446;%%
  %5 citations counted in INSPIRE as of 14 Feb 2014


%\cite{Soffer:2014kxa}
\bibitem{Soffer:2014kxa} 
  A.~Soffer,
  %``B-Meson Decays into Final States with a $\tau$ Lepton,''
  Mod.\ Phys.\ Lett.\ A {\bf 29}, no. 7, 1430007 (2014)
  [arXiv:1401.7947 [hep-ex]].
  %%CITATION = ARXIV:1401.7947;%%






%\cite{Ikado:2006un}
\bibitem{Ikado:2006un} 
  K.~Ikado {\it et al.}  [Belle Collaboration],
     %\emph{Evidence of the Purely Leptonic Decay B- ---> tau- anti-nu(tau)},
  Phys.\ Rev.\ Lett.\  {\bf 97}, 251802 (2006)
  [hep-ex/0604018].
  %%CITATION = HEP-EX/0604018;%%
  %297 citations counted in INSPIRE as of 14 Feb 2014


%\cite{Aubert:2009wt}
\bibitem{Aubert:2009wt} 
  B.~Aubert {\it et al.}  [BaBar Collaboration],
     %\emph{A Search for $B^+ \to \ell^+ \nu_{\ell}$ Recoiling Against $B^{-}\to D^{0} \ell^{-}\bar{\nu} X$},
  Phys.\ Rev.\ D {\bf 81}, 051101 (2010)
  [arXiv:0912.2453 [hep-ex]].
  %%CITATION = ARXIV:0912.2453;%%
  %92 citations counted in INSPIRE as of 22 Apr 2014

%\cite{Lees:2012ju}
\bibitem{Lees:2012ju} 
  J.~P.~Lees {\it et al.}  [BaBar Collaboration],
     %\emph{Evidence of B+→τ+ν decays with hadronic B tags},
  Phys.\ Rev.\ D {\bf 88}, no. 3, 031102 (2013)
  [arXiv:1207.0698 [hep-ex]].
  %%CITATION = ARXIV:1207.0698;%%
  %58 citations counted in INSPIRE as of 22 Apr 2014


%\cite{Hara:2010dk}
\bibitem{Hara:2010dk} 
  K.~Hara {\it et al.}  [Belle Collaboration],
     %\emph{Evidence for $B^- -> \tau^- \bar{\nu}$ with a Semileptonic Tagging Method},
  Phys.\ Rev.\ D {\bf 82}, 071101 (2010)
  [arXiv:1006.4201 [hep-ex]].
  %%CITATION = ARXIV:1006.4201;%%
  %84 citations counted in INSPIRE as of 22 Apr 2014

%\cite{Adachi:2012mm}
\bibitem{Adachi:2012mm} 
  I.~Adachi {\it et al.}  [Belle Collaboration],
  %``Measurement of $B^- \to \tau^- \bar{\nu}_\tau$ with a Hadronic Tagging Method Using the Full Data Sample of Belle,''
  Phys.\ Rev.\ Lett.\  {\bf 110}, 131801 (2013)
  [arXiv:1208.4678 [hep-ex]].
  %%CITATION = ARXIV:1208.4678;%%
  %80 citations counted in INSPIRE as of 14 Feb 2014



  
  
  %\cite{Bazavov:2011aa}
\bibitem{Bazavov:2011aa} 
  A.~Bazavov {\it et al.}  [Fermilab Lattice and MILC Collaborations],
  %``B- and D-meson decay constants from three-flavor lattice QCD,''
  Phys.\ Rev.\ D {\bf 85}, 114506 (2012)
  [arXiv:1112.3051 [hep-lat]].
  %%CITATION = ARXIV:1112.3051;%%
  %115 citations counted in INSPIRE as of 22 Apr 2014


 %\cite{Na:2012kp}
\bibitem{Na:2012kp} 
  H.~Na, C.~J.~Monahan, C.~T.~H.~Davies, R.~Horgan, G.~P.~Lepage and J.~Shigemitsu,
  %``The B and B_s Meson Decay Constants from Lattice QCD,''
  Phys.\ Rev.\ D {\bf 86}, 034506 (2012)
  [arXiv:1202.4914 [hep-lat]].
  %%CITATION = ARXIV:1202.4914;%%
  %84 citations counted in INSPIRE as of 22 Apr 2014 
  
  
  %\cite{Carrasco:2013zta}
\bibitem{Carrasco:2013zta} 
  N.~Carrasco {\it et al.}  [ETM Collaboration],
  %``B-physics from $N_f$ = 2 tmQCD: the Standard Model and beyond,''
  JHEP {\bf 1403}, 016 (2014)
  [arXiv:1308.1851 [hep-lat]].
  %%CITATION = ARXIV:1308.1851;%%
  %21 citations counted in INSPIRE as of 22 Apr 2014
 

  %\cite{Bernardoni:2014fva}
\bibitem{Bernardoni:2014fva} 
  F.~Bernardoni, B.~Blossier, J.~Bulava, M.~Della Morte, P.~Fritzsch, N.~Garron, A.~Gerardin and J.~Heitger {\it et al.},
  %``Decay constants of B-mesons from non-perturbative HQET with two light dynamical quarks,''
  arXiv:1404.3590 [hep-lat].
  %%CITATION = ARXIV:1404.3590;%%
  %1 citations counted in INSPIRE as of 22 Apr 2014
  

  %\cite{Christ:2014uea}
\bibitem{Christ:2014uea} 
  N.~H.~Christ, J.~M.~Flynn, T.~Izubuchi, T.~Kawanai, C.~Lehner, A.~Soni, R.~S.~Van de Water and O.~Witzel,
  %``B-meson decay constants from 2+1-flavor lattice QCD with domain-wall light quarks and relativistic heavy quarks,''
  arXiv:1404.4670 [hep-lat].
  %%CITATION = ARXIV:1404.4670;%%




%\cite{delAmoSanchez:2010zd}
%\bibitem{delAmoSanchez:2010zd} 
%  P.~del Amo Sanchez {\it et al.}  [BaBar Collaboration],
  %``Measurement of the $B^0 \to \pi^\ell \ell^+ \nu$ and $B^+ \to \eta^{(')} \ell^+ \nu$ Branching Fractions, the $B^0 \to \pi^- \ell^+ \nu$ and $B^+ \to \eta \ell^+ \nu$ Form-Factor Shapes, and Determination of $|V_{ub}|$,''
%  Phys.\ Rev.\ D {\bf 83}, 052011 (2011)
 % [arXiv:1010.0987 [hep-ex]].
  %%CITATION = ARXIV:1010.0987;%%
  %37 citations counted in INSPIRE as of 14 Feb 2014






%\cite{Lunghi:2010gv}
%\bibitem{Lunghi:2010gv} 
%  E.~Lunghi and A.~Soni,
  %``Possible evidence for the breakdown of the CKM-paradigm of CP-violation,''
%  Phys.\ Lett.\ B {\bf 697}, 323 (2011)
 % [arXiv:1010.6069 [hep-ph]].
  %%CITATION = ARXIV:1010.6069;%%
  %91 citations counted in INSPIRE as of 14 Feb 2014




 



%\cite{Kruger:1999xa}
\bibitem{Kruger:1999xa} 
  F.~Kruger, L.~M.~Sehgal, N.~Sinha and R.~Sinha,
  %``Angular distribution and CP asymmetries in the decays anti-B ---> K- pi+ e- e+ and anti-B ---> pi- pi+ e- e+,''
  Phys.\ Rev.\ D {\bf 61}, 114028 (2000)
  [Erratum-ibid.\ D {\bf 63}, 019901 (2001)]
  [hep-ph/9907386].
  %%CITATION = HEP-PH/9907386;%%
  %130 citations counted in INSPIRE as of 14 Feb 2014


%\cite{Lu:2011jm}
\bibitem{Lu:2011jm} 
  C.~-D.~Lu and W.~Wang,
  %``Analysis of $B\to K^*_J (\to K \pi) \mu^+\mu^-$ in the higher kaon resonance region,''
  Phys.\ Rev.\ D {\bf 85}, 034014 (2012)
  [arXiv:1111.1513 [hep-ph]].
  %%CITATION = ARXIV:1111.1513;%%
  %15 citations counted in INSPIRE as of 14 Feb 2014


%\cite{Doring:2013wka}
\bibitem{Doring:2013wka} 
  M.~Doring, U.~-G.~Mei{\ss}ner and W.~Wang,
  %``Chiral Dynamics and S-wave Contributions in Semileptonic B decays,''
  JHEP {\bf 1310}, 011 (2013)
  [arXiv:1307.0947 [hep-ph]].
  %%CITATION = ARXIV:1307.0947;%%
  %7 citations counted in INSPIRE as of 14 Feb 2014


%\cite{Li:2010ra}
\bibitem{Li:2010ra} 
  R.~-H.~Li, C.~-D.~Lu and W.~Wang,
  %``Branching ratios, forward-backward asymmetries and angular distributions of $B\to K_2^*l^+l^-$ in the standard model and new physics scenarios,''
  Phys.\ Rev.\ D {\bf 83}, 034034 (2011)
  [arXiv:1012.2129 [hep-ph]].
  %%CITATION = ARXIV:1012.2129;%%
  %14 citations counted in INSPIRE as of 14 Feb 2014


%\cite{Becirevic:2012dp}
\bibitem{Becirevic:2012dp} 
  D.~Becirevic and A.~Tayduganov,
  %``Impact of $B\to K^\ast_0 \ell^+\ell^-$ on the New Physics search in $B\to K^\ast \ell^+\ell^-$ decay,''
  Nucl.\ Phys.\ B {\bf 868}, 368 (2013)
  [arXiv:1207.4004 [hep-ph]].
  %%CITATION = ARXIV:1207.4004;%%
  %29 citations counted in INSPIRE as of 14 Feb 2014


%\cite{Matias:2012qz}
\bibitem{Matias:2012qz} 
  J.~Matias,
  %``On the S-wave pollution of B-> K* l+l- observables,''
  Phys.\ Rev.\ D {\bf 86}, 094024 (2012)
  [arXiv:1209.1525 [hep-ph]].
  %%CITATION = ARXIV:1209.1525;%%
  %25 citations counted in INSPIRE as of 14 Feb 2014


%\cite{Blake:2012mb}
\bibitem{Blake:2012mb} 
  T.~Blake, U.~Egede and A.~Shires,
  %``The effect of S-wave interference on the $B^0 \to K^{\ast 0}\ell^+\ell^-$ angular observables,''
  JHEP {\bf 1303}, 027 (2013)
  [arXiv:1210.5279 [hep-ph]].
  %%CITATION = ARXIV:1210.5279;%%
  %19 citations counted in INSPIRE as of 14 Feb 2014


%\cite{Bobeth:2012vn}
\bibitem{Bobeth:2012vn} 
  C.~Bobeth, G.~Hiller and D.~van Dyk,
  %``General Analysis of $\bar{B} \to \bar{K}^{(*)}\ell^+ \ell^-$ Decays at Low Recoil,''
  Phys.\ Rev.\ D {\bf 87}, 034016 (2013)
  [arXiv:1212.2321 [hep-ph]].
  %%CITATION = ARXIV:1212.2321;%%
  %43 citations counted in INSPIRE as of 14 Feb 2014


%\cite{Descotes-Genon:2013vna}
\bibitem{Descotes-Genon:2013vna} 
  S.~Descotes-Genon, T.~Hurth, J.~Matias and J.~Virto,
  %``Optimizing the basis of ${B} \to {K}^{*}\ell^+ \ell^-$ observables in the full kinematic range,''
  JHEP {\bf 1305}, 137 (2013)
  [arXiv:1303.5794 [hep-ph]].
  %%CITATION = ARXIV:1303.5794;%%
  %24 citations counted in INSPIRE as of 14 Feb 2014


%\cite{Descotes-Genon:2013wba}
\bibitem{Descotes-Genon:2013wba} 
  S.~Descotes-Genon, J.~Matias and J.~Virto,
  %``Understanding the $B \to K^*\mu^+\mu^-$ Anomaly,''
  Phys.\ Rev.\ D {\bf 88}, 074002 (2013)
  [arXiv:1307.5683 [hep-ph]].
  %%CITATION = ARXIV:1307.5683;%%
  %26 citations counted in INSPIRE as of 14 Feb 2014


%\cite{Kopp:1990yx}
\bibitem{Kopp:1990yx} 
  G.~Kopp, G.~Kramer, G.~A.~Schuler and W.~F.~Palmer,
  %``Angular correlations for semileptonic D meson decays,''
  Z.\ Phys.\ C {\bf 48}, 327 (1990).
  %%CITATION = ZEPYA,C48,327;%%
  %17 citations counted in INSPIRE as of 14 Feb 2014


%\cite{Lee:1992ih}
\bibitem{Lee:1992ih} 
  C.~L.~Y.~Lee, M.~Lu and M.~B.~Wise,
  %``B(l4) and D(l4) decay,''
  Phys.\ Rev.\ D {\bf 46}, 5040 (1992).
  %%CITATION = PHRVA,D46,5040;%%
  %55 citations counted in INSPIRE as of 14 Feb 2014


%\cite{Ananthanarayan:2005us}
\bibitem{Ananthanarayan:2005us} 
  B.~Ananthanarayan and K.~Shivaraj,
  %``Comment on evidence for new interference phenomena in the decay D+ ---> K- pi+ mu+ nu,''
  Phys.\ Lett.\ B {\bf 628}, 223 (2005)
  [hep-ph/0508116].
  %%CITATION = HEP-PH/0508116;%%
  %2 citations counted in INSPIRE as of 14 Feb 2014


%\cite{Faller:2013dwa}
\bibitem{Faller:2013dwa} 
  S.~Faller, T.~Feldmann, A.~Khodjamirian, T.~Mannel and D.~van Dyk,
  %``Disentangling the Decay Observables in $B^- \to \pi^+\pi^-\ell^-\bar\nu_\ell$,''
  Phys.\ Rev.\ D {\bf 89}, 014015 (2014)
  [arXiv:1310.6660 [hep-ph]].
  %%CITATION = ARXIV:1310.6660;%%
  %3 citations counted in INSPIRE as of 14 Feb 2014




%\cite{Buettiker:2003pp}
\bibitem{Buettiker:2003pp} 
  P.~Buettiker, S.~Descotes-Genon and B.~Moussallam,
  %``A new analysis of pi K scattering from Roy and Steiner type equations,''
  Eur.\ Phys.\ J.\ C {\bf 33}, 409 (2004)
  [hep-ph/0310283].
  %%CITATION = HEP-PH/0310283;%%
  %137 citations counted in INSPIRE as of 14 Feb 2014


%\cite{DescotesGenon:2006uk}
\bibitem{DescotesGenon:2006uk} 
  S.~Descotes-Genon and B.~Moussallam,
  %``The K*0 (800) scalar resonance from Roy-Steiner representations of pi K scattering,''
  Eur.\ Phys.\ J.\ C {\bf 48}, 553 (2006)
  [hep-ph/0607133].
  %%CITATION = HEP-PH/0607133;%%
  %117 citations counted in INSPIRE as of 14 Feb 2014


  %\cite{Meissner:2013hya}
\bibitem{Meissner:2013hya} 
  U.~-G.~Mei{\ss}ner and W.~Wang,
  %``Generalized Heavy-to-Light Form Factors in Light-Cone Sum Rules,''
  Phys.\ Lett.\ B {\bf 730}, 336 (2014)
  [arXiv:1312.3087 [hep-ph]].
  %%CITATION = ARXIV:1312.3087;%%
  %3 citations counted in INSPIRE as of 23 Apr 2014

%\cite{Watson:1952ji}
\bibitem{Watson:1952ji} 
  K.~M.~Watson,
  %``The Effect of final state interactions on reaction cross-sections,''
  Phys.\ Rev.\  {\bf 88}, 1163 (1952).
  %%CITATION = PHRVA,88,1163;%%
  %469 citations counted in INSPIRE as of 24 Apr 2014



%\cite{Bar:2012ce}
\bibitem{Bar:2012ce} 
  O.~Bar and M.~Golterman,
  %``Excited-state contribution to the axial-vector and pseudo-scalar correlators with two extra pions,''
  Phys.\ Rev.\ D {\bf 87}, 014505 (2013)
  [arXiv:1209.2258 [hep-lat]].
  %%CITATION = ARXIV:1209.2258;%%
  %2 citations counted in INSPIRE as of 14 Feb 2014


%\cite{Gasser:1990bv}
\bibitem{Gasser:1990bv} 
  J.~Gasser and U.~G.~Mei{\ss}ner,
  %``Chiral expansion of pion form-factors beyond one loop,''
  Nucl.\ Phys.\ B {\bf 357}, 90 (1991).
  %%CITATION = NUPHA,B357,90;%%
  %166 citations counted in INSPIRE as of 14 Feb 2014


%\cite{Oller:2000ug}
\bibitem{Oller:2000ug} 
  J.~A.~Oller, E.~Oset and J.~E.~Palomar,
  %``Pion and kaon vector form-factors,''
  Phys.\ Rev.\ D {\bf 63}, 114009 (2001)
  [hep-ph/0011096].
  %%CITATION = HEP-PH/0011096;%%
  %64 citations counted in INSPIRE as of 14 Feb 2014


%\cite{Gardner:2001gc}
\bibitem{Gardner:2001gc} 
  S.~Gardner and U.~-G.~Mei{\ss}ner,
  %``Rescattering and chiral dynamics in B ---> rho pi decay,''
  Phys.\ Rev.\ D {\bf 65}, 094004 (2002)
  [hep-ph/0112281].
  %%CITATION = HEP-PH/0112281;%%
  %77 citations counted in INSPIRE as of 14 Feb 2014


%\cite{Meissner:2000bc}
\bibitem{Meissner:2000bc} 
  U.~-G.~Mei{\ss}ner and J.~A.~Oller,
  %``J / psi ---> phi pi pi (K anti-K) decays, chiral dynamics and OZI violation,''
  Nucl.\ Phys.\ A {\bf 679}, 671 (2001)
  [hep-ph/0005253].
  %%CITATION = HEP-PH/0005253;%%
  %86 citations counted in INSPIRE as of 14 Feb 2014


%\cite{Frink:2002ht}
\bibitem{Frink:2002ht} 
  M.~Frink, B.~Kubis and U.~-G.~Mei{\ss}ner,
  %``Analysis of the pion kaon sigma term and related topics,''
  Eur.\ Phys.\ J.\ C {\bf 25}, 259 (2002)
  [hep-ph/0203193].
  %%CITATION = HEP-PH/0203193;%%
  %21 citations counted in INSPIRE as of 14 Feb 2014


%\cite{Bijnens:2003uy}
\bibitem{Bijnens:2003uy} 
  J.~Bijnens and P.~Talavera,
  %``K(l3) decays in chiral perturbation theory,''
  Nucl.\ Phys.\ B {\bf 669}, 341 (2003)
  [hep-ph/0303103].
  %%CITATION = HEP-PH/0303103;%%
  %172 citations counted in INSPIRE as of 14 Feb 2014


%\cite{Lahde:2006wr}
\bibitem{Lahde:2006wr} 
  T.~A.~Lahde and U.~-G.~Mei{\ss}ner,
  %``Improved Analysis of J/psi Decays into a Vector Meson and Two Pseudoscalars,''
  Phys.\ Rev.\ D {\bf 74}, 034021 (2006)
  [hep-ph/0606133].
  %%CITATION = HEP-PH/0606133;%%
  %32 citations counted in INSPIRE as of 14 Feb 2014


%\cite{Guo:2012yt}
\bibitem{Guo:2012yt} 
  Z.~-H.~Guo, J.~A.~Oller and J.~Ruiz de Elvira,
  %``Chiral dynamics in form factors, spectral-function sum rules, meson-meson scattering and semi-local duality,''
  Phys.\ Rev.\ D {\bf 86}, 054006 (2012)
  [arXiv:1206.4163 [hep-ph]].
  %%CITATION = ARXIV:1206.4163;%%
  %14 citations counted in INSPIRE as of 14 Feb 2014


%\cite{Donoghue:1990xh}
\bibitem{Donoghue:1990xh} 
  J.~F.~Donoghue, J.~Gasser and H.~Leutwyler,
  %``The Decay of a Light Higgs Boson,''
  Nucl.\ Phys.\ B {\bf 343}, 341 (1990).
  %%CITATION = NUPHA,B343,341;%%
  %138 citations counted in INSPIRE as of 14 Feb 2014


%\cite{Jamin:2000wn}
\bibitem{Jamin:2000wn} 
  M.~Jamin, J.~A.~Oller and A.~Pich,
  %``S wave K pi scattering in chiral perturbation theory with resonances,''
  Nucl.\ Phys.\ B {\bf 587}, 331 (2000)
  [hep-ph/0006045].
  %%CITATION = HEP-PH/0006045;%%
  %166 citations counted in INSPIRE as of 14 Feb 2014


%\cite{Jamin:2001zq}
\bibitem{Jamin:2001zq} 
  M.~Jamin, J.~A.~Oller and A.~Pich,
  %``Strangeness changing scalar form-factors,''
  Nucl.\ Phys.\ B {\bf 622}, 279 (2002)
  [hep-ph/0110193].
  %%CITATION = HEP-PH/0110193;%%
  %118 citations counted in INSPIRE as of 14 Feb 2014


%\cite{Jamin:2006tj}
\bibitem{Jamin:2006tj} 
  M.~Jamin, J.~A.~Oller and A.~Pich,
  %``Scalar K pi form factor and light quark masses,''
  Phys.\ Rev.\ D {\bf 74}, 074009 (2006)
  [hep-ph/0605095].
  %%CITATION = HEP-PH/0605095;%%
  %111 citations counted in INSPIRE as of 14 Feb 2014


%\cite{Bernard:2007tk}
\bibitem{Bernard:2007tk} 
  V.~Bernard and E.~Passemar,
  %``Matching chiral perturbation theory and the dispersive representation of the scalar K pi form-factor,''
  Phys.\ Lett.\ B {\bf 661}, 95 (2008)
  [arXiv:0711.3450 [hep-ph]].
  %%CITATION = ARXIV:0711.3450;%%
  %23 citations counted in INSPIRE as of 14 Feb 2014


%\cite{Bernard:2009ds}
\bibitem{Bernard:2009ds} 
  V.~Bernard and E.~Passemar,
  %``Chiral Extrapolation of the Strangeness Changing K pi Form Factor,''
  JHEP {\bf 1004}, 001 (2010)
  [arXiv:0912.3792 [hep-ph]].
  %%CITATION = ARXIV:0912.3792;%%
  %19 citations counted in INSPIRE as of 14 Feb 2014
  
  
    
\bibitem{twoHadronFormFactor}
F.~K.~Guo, B.~Kubis, U.~-G.~Mei{\ss}ner, and W.~Wang, in preparation.   


%\cite{Chen:2002th}
\bibitem{Chen:2002th} 
  C.-H.~Chen and H.-N.~Li,
  %``Three body nonleptonic B decays in perturbative QCD,''
  Phys.\ Lett.\ B {\bf 561}, 258 (2003)
  [hep-ph/0209043].
  %%CITATION = HEP-PH/0209043;%%
  %15 citations counted in INSPIRE as of 14 Nov 2013


%\cite{ElBennich:2009da}
\bibitem{ElBennich:2009da} 
 B.~El-Bennich, A.~Furman, R.~Kaminski, L.~Lesniak, B.~Loiseau and B.~Moussallam,
  %``CP violation and kaon-pion interactions in B ---> K pi+ pi- decays,''
  Phys.\ Rev.\ D {\bf 79}, 094005 (2009)
  [Erratum-ibid.\ D {\bf 83}, 039903 (2011)]
  [arXiv:0902.3645 [hep-ph]].
  %%CITATION = ARXIV:0902.3645;%%
  %26 citations counted in INSPIRE as of 14 Nov 2013

%\cite{Zhang:2013oqa}
\bibitem{Zhang:2013oqa} 
  Z.~-H.~Zhang, X.~-H.~Guo and Y.~-D.~Yang,
  %``CP violation in $B^{\pm} \rightarrow \pi^{\pm}\pi^{+}\pi^{-}$ in the region with low invariant mass of one $\pi^{+}\pi^{-}$ pair,''
  Phys.\ Rev.\ D {\bf 87}, no. 7, 076007 (2013)
  [arXiv:1303.3676 [hep-ph]].
  %%CITATION = ARXIV:1303.3676;%%
  %10 citations counted in INSPIRE as of 24 Apr 2014
%\cite{Bediaga:2013ela}
\bibitem{Bediaga:2013ela} 
  I.~Bediaga, T.~Frederico and O.~Lourenzo,
  %``CP violation and CPT invariance in B+- decays with final state interactions,''
  arXiv:1307.8164 [hep-ph].
  %%CITATION = ARXIV:1307.8164;%%
  %2 citations counted in INSPIRE as of 14 Nov 2013

 

%\cite{Cheng:2013dua}
\bibitem{Cheng:2013dua} 
  H.~-Y.~Cheng and C.~-K.~Chua,
  %``Branching Fractions and Direct CP Violation in Charmless Three-body Decays of B Mesons,''
  Phys.\ Rev.\ D {\bf 88}, 114014 (2013)
  [arXiv:1308.5139 [hep-ph]].
  %%CITATION = ARXIV:1308.5139;%%
  %3 citations counted in INSPIRE as of 09 Dec 2013
%\cite{Xu:2013rua}
\bibitem{Xu:2013rua} 
  D.~Xu, G.~-N.~Li and X.~-G.~He,
  %``U-spin analysis of CP violation in $B^-$ decays into three charged light pseudoscalar mesons,''
  Phys.\ Lett.\ B {\bf 728}, 579 (2014)
  [arXiv:1311.3714 [hep-ph]].
  %%CITATION = ARXIV:1311.3714;%%
  %5 citations counted in INSPIRE as of 24 Apr 2014
  
  
%\cite{Xu:2013dta}
\bibitem{Xu:2013dta} 
  D.~Xu, G.~-N.~Li and X.~-G.~He,
  %``Large SU(3) breaking effects and CP violation in $B^+ $ decays into three charged octet pseudoscalar mesons,''
  Int.\ J.\ Mod.\ Phys.\ A {\bf 29}, 1450011 (2014)
  [arXiv:1307.7186 [hep-ph]].
  %%CITATION = ARXIV:1307.7186;%%
  %9 citations counted in INSPIRE as of 24 Apr 2014
  
  %\cite{Cheng:2014uga}
\bibitem{Cheng:2014uga} 
  H.~-Y.~Cheng and C.~-K.~Chua,
  %``Charmless Three-body Decays of B_s Mesons,''
  Phys.\ Rev.\ D {\bf 89}, 074025 (2014)
  [arXiv:1401.5514 [hep-ph]].
  %%CITATION = ARXIV:1401.5514;%%
  %3 citations counted in INSPIRE as of 24 Apr 2014


%\cite{Li:2014fla}
\bibitem{Li:2014fla} 
  Y.~Li,
  %``Branching Fractions and Direct $CP$ Asymmetries of $\overline B_s ^0 \to K^0 h^+h^{\prime -}(h^{(\prime)}=K,\pi)$ Decays,''
  arXiv:1401.5948 [hep-ph].
  %%CITATION = ARXIV:1401.5948;%%
  %2 citations counted in INSPIRE as of 24 Apr 2014
  
  %\cite{Wang:2014ira}
\bibitem{Wang:2014ira} 
  W.~-F.~Wang, H.~-C.~Hu, H.~-n.~Li and C.~-D.~L\"u
  %``Direct CP asymmetries of three-body $B$ decays in perturbative QCD,''
  Phys.\ Rev.\ D {\bf 89}, 074031 (2014)
  [arXiv:1402.5280 [hep-ph]].
  %%CITATION = ARXIV:1402.5280;%%
  %1 citations counted in INSPIRE as of 24 Apr 2014

%\cite{Li:2014oca}
\bibitem{Li:2014oca} 
  Y.~Li,
  %``Comprehensive Study of $\overline B^0\to \stackrel{(-)}{K^0} K^\mp\pi^\pm$ Decays in the Factorization Approach,''
  arXiv:1402.6052 [hep-ph].
  %%CITATION = ARXIV:1402.6052;%%
  
  %\cite{Bhattacharya:2013cvn}
\bibitem{Bhattacharya:2013cvn} 
  B.~Bhattacharya, M.~Gronau and J.~L.~Rosner,
  %``CP asymmetries in three-body $B^\pm$ decays to charged pions and kaons,''
  Phys.\ Lett.\ B {\bf 726}, 337 (2013)
  [arXiv:1306.2625 [hep-ph]].
  %%CITATION = ARXIV:1306.2625;%%
  %14 citations counted in INSPIRE as of 24 Apr 2014


%\cite{Charles:1998dr}
\bibitem{Charles:1998dr} 
  J.~Charles, A.~Le Yaouanc, L.~Oliver, O.~Pene and J.~C.~Raynal,
  %``Heavy to light form-factors in the heavy mass to large energy limit of QCD,''
  Phys.\ Rev.\ D {\bf 60}, 014001 (1999)
  [hep-ph/9812358].
  %%CITATION = HEP-PH/9812358;%%
  %293 citations counted in INSPIRE as of 14 Feb 2014

%\cite{Beneke:2000wa}
\bibitem{Beneke:2000wa} 
  M.~Beneke and T.~Feldmann,
  %``Symmetry breaking corrections to heavy to light B meson form-factors at large recoil,''
  Nucl.\ Phys.\ B {\bf 592}, 3 (2001)
  [hep-ph/0008255].
  %%CITATION = HEP-PH/0008255;%%
  %378 citations counted in INSPIRE as of 14 Feb 2014


%\cite{Bauer:2002aj}
\bibitem{Bauer:2002aj} 
  C.~W.~Bauer, D.~Pirjol and I.~W.~Stewart,
  %``Factorization and endpoint singularities in heavy to light decays,''
  Phys.\ Rev.\ D {\bf 67}, 071502 (2003)
  [hep-ph/0211069].
  %%CITATION = HEP-PH/0211069;%%
  %155 citations counted in INSPIRE as of 14 Feb 2014




%\cite{Beneke:2004rc}
\bibitem{Beneke:2004rc} 
  M.~Beneke, Y.~Kiyo and D.~s.~Yang,
  %``Loop corrections to subleading heavy quark currents in SCET,''
  Nucl.\ Phys.\ B {\bf 692}, 232 (2004)
  [hep-ph/0402241].
  %%CITATION = HEP-PH/0402241;%%
  %33 citations counted in INSPIRE as of 14 Feb 2014


%\cite{Beneke:2005gs}
\bibitem{Beneke:2005gs} 
  M.~Beneke and D.~Yang,
  %``Heavy-to-light B meson form-factors at large recoil energy: Spectator-scattering corrections,''
  Nucl.\ Phys.\ B {\bf 736}, 34 (2006)
  [hep-ph/0508250].
  %%CITATION = HEP-PH/0508250;%%
  %59 citations counted in INSPIRE as of 14 Feb 2014


 

%\cite{Diehl:1998dk}
\bibitem{Diehl:1998dk} 
  M.~Diehl, T.~Gousset, B.~Pire and O.~Teryaev,
  %``Probing partonic structure in gamma* gamma ---> pi pi near threshold,''
  Phys.\ Rev.\ Lett.\  {\bf 81}, 1782 (1998)
  [hep-ph/9805380].
  %%CITATION = HEP-PH/9805380;%%
  %172 citations counted in INSPIRE as of 14 Feb 2014


%\cite{Polyakov:1998ze}
\bibitem{Polyakov:1998ze} 
  M.~V.~Polyakov,
  %``Hard exclusive electroproduction of two pions and their resonances,''
  Nucl.\ Phys.\ B {\bf 555}, 231 (1999)
  [hep-ph/9809483].
  %%CITATION = HEP-PH/9809483;%%
  %142 citations counted in INSPIRE as of 14 Feb 2014
  
  
%\cite{Kivel:1999sd}
\bibitem{Kivel:1999sd} 
  N.~Kivel, L.~Mankiewicz and M.~V.~Polyakov,
  %``NLO corrections and contribution of a tensor gluon operator to the process gamma* gamma ---> pi pi,''
  Phys.\ Lett.\ B {\bf 467}, 263 (1999)
  [hep-ph/9908334].
  %%CITATION = HEP-PH/9908334;%%
  %31 citations counted in INSPIRE as of 14 Feb 2014


%\cite{Diehl:2003ny}
\bibitem{Diehl:2003ny} 
  M.~Diehl,
  %``Generalized parton distributions,''
  Phys.\ Rept.\  {\bf 388}, 41 (2003)
  [hep-ph/0307382].
  %%CITATION = HEP-PH/0307382;%%
  %614 citations counted in INSPIRE as of 14 Feb 2014
  
   
  %\cite{Charng:2006zj}
\bibitem{Charng:2006zj} 
  Y.~-Y.~Charng, T.~Kurimoto and H.~-n.~Li,
  %``Gluonic contribution to B ---> eta-(prime) form factors,''
  Phys.\ Rev.\ D {\bf 74}, 074024 (2006)
  [Erratum-ibid.\ D {\bf 78}, 059901 (2008)]
  [hep-ph/0609165].
  %%CITATION = HEP-PH/0609165;%%
  %48 citations counted in INSPIRE as of 22 Apr 2014

%\cite{Wang:2009rc}
\bibitem{Wang:2009rc} 
  W.~Wang, Y.~-L.~Shen and C.~-D.~Lu,
  %\emph{$B$-to-Glueball form factor and Glueball production in $B$ decays},
  J.\ Phys.\ G {\bf 37}, 085006 (2010)
  [arXiv:0908.2216 [hep-ph]].
  %%CITATION = ARXIV:0908.2216;%%
  %5 citations counted in INSPIRE as of 22 Apr 2014
  
  
  
%\cite{Braun:2003rp}
%\bibitem{Braun:2003rp} 
%  V.~M.~Braun, G.~P.~Korchemsky and D.~Mueller,
  %``The Uses of conformal symmetry in QCD,''
%  Prog.\ Part.\ Nucl.\ Phys.\  {\bf 51}, 311 (2003)
%  [hep-ph/0306057].
  %%CITATION = HEP-PH/0306057;%%
  %173 citations counted in INSPIRE as of 14 Feb 2014



%\cite{Kang:2013jaa}
\bibitem{Kang:2013jaa} 
  X.~-W.~Kang, B.~Kubis, C.~Hanhart and U.~-G.~Mei{\ss}ner,
  %``B_l4 decays and the extraction of |V_ub|,''
  arXiv:1312.1193 [hep-ph].
  %%CITATION = ARXIV:1312.1193;%%
  %1 citations counted in INSPIRE as of 14 Feb 2014


  %\cite{Li:2006vq}
\bibitem{Li:2006vq} 
  H.~-n.~Li and S.~Mishima,
  %``Penguin pollution in the B0 ---> J/psi K(S) decay,''
  JHEP {\bf 0703}, 009 (2007)
  [hep-ph/0610120].
  %%CITATION = HEP-PH/0610120;%%
  %57 citations counted in INSPIRE as of 27 Feb 2014


%\cite{Gronau:1990ra}
\bibitem{Gronau:1990ra} 
  M.~Gronau and D.~London,
  %``How to determine all the angles of the unitarity triangle from B(d)0 ---> D K(s) and B(s)0 ---> D0,''
  Phys.\ Lett.\ B {\bf 253}, 483 (1991).
  %%CITATION = PHLTA,B253,483;%%
  %601 citations counted in INSPIRE as of 25 Apr 2014
%\cite{Gronau:1991dp}
\bibitem{Gronau:1991dp} 
  M.~Gronau and D.~Wyler,
  %``On determining a weak phase from CP asymmetries in charged B decays,''
  Phys.\ Lett.\ B {\bf 265}, 172 (1991).
  %%CITATION = PHLTA,B265,172;%%
  %782 citations counted in INSPIRE as of 14 Feb 2014


%\cite{Atwood:1996ci}
\bibitem{Atwood:1996ci} 
  D.~Atwood, I.~Dunietz and A.~Soni,
  %``Enhanced CP violation with B ---> K D0 (anti-D0) modes and extraction of the CKM angle gamma,''
  Phys.\ Rev.\ Lett.\  {\bf 78}, 3257 (1997)
  [hep-ph/9612433].
  %%CITATION = HEP-PH/9612433;%%
  %632 citations counted in INSPIRE as of 14 Feb 2014


%\cite{Atwood:2000ck}
\bibitem{Atwood:2000ck} 
  D.~Atwood, I.~Dunietz and A.~Soni,
  %``Improved methods for observing CP violation in B+- ---> K D and measuring the CKM phase gamma,''
  Phys.\ Rev.\ D {\bf 63}, 036005 (2001)
  [hep-ph/0008090].
  %%CITATION = HEP-PH/0008090;%%
  %327 citations counted in INSPIRE as of 25 Apr 2014

%\cite{Giri:2003ty}
\bibitem{Giri:2003ty} 
  A.~Giri, Y.~Grossman, A.~Soffer and J.~Zupan,
  %``Determining gamma using B+- ---> DK+- with multibody D decays,''
  Phys.\ Rev.\ D {\bf 68}, 054018 (2003)
  [hep-ph/0303187].
  %%CITATION = HEP-PH/0303187;%%
  %426 citations counted in INSPIRE as of 14 Feb 2014



 \bibitem{Aaij:2011in} 
  R.~Aaij {\it et al.}  [LHCb Collaboration],
  %``Evidence for CP violation in time-integrated D0 -> h-h+ decay rates,''
  Phys.\ Rev.\ Lett.\  {\bf 108}, 111602 (2012)
  [arXiv:1112.0938 [hep-ex]].
  %%CITATION = ARXIV:1112.0938;%%
 
%\cite{Collaboration:2012qw}
\bibitem{Collaboration:2012qw} 
  T.~Aaltonen {\it et al.}  [CDF Collaboration],
  %``Measurement of the difference of CP--violating asymmetries in $D^0 \to K^+K^-$ and $D^0 \to \pi^+\pi^-$ decays at CDF,''
  Phys.\ Rev.\ Lett.\  {\bf 109}, 111801 (2012)
  [arXiv:1207.2158 [hep-ex]].
  %%CITATION = ARXIV:1207.2158;%%
%CDF note 10784 at http://www-cdf.fnal.gov/physics/new/bottom/120216.blessed-CPVcharm10fb/cdf10784.pdf. 

%\cite{Ko:2012px}
\bibitem{Ko:2012px} 
  B.~R.~Ko [Belle Collaboration],
  %``Direct CP violation in charm at Belle,''
  arXiv:1212.1975 [hep-ex].
  %%CITATION = ARXIV:1212.1975;%%


%\cite{Aaij:2013bra}
\bibitem{Aaij:2013bra} 
  RAaij {\it et al.}  [LHCb Collaboration],
  %``Search for direct $CP$ violation in $D^0 \rightarrow h^- h^+$ modes using semileptonic $B$ decays,''
  Phys.\ Lett.\ B {\bf 723}, 33 (2013)
  [arXiv:1303.2614 [hep-ex]].
  %%CITATION = ARXIV:1303.2614;%%
  %42 citations counted in INSPIRE as of 24 Apr 2014
    
 %\cite{Wang:2012ie}
\bibitem{Wang:2012ie} 
  W.~Wang,
  %``CP violation effects on the measurement of $\gamma$ from $B\to DK$,''
  Phys.\ Rev.\ Lett.\  {\bf 110}, no. 6, 061802 (2013)
  [arXiv:1211.4539 [hep-ph]].
  %%CITATION = ARXIV:1211.4539;%%
  %7 citations counted in INSPIRE as of 23 Apr 2014
 
 
 %\cite{Martone:2012nj}
\bibitem{Martone:2012nj} 
  M.~Martone and J.~Zupan,
  %``$B^\pm \to D K^\pm$ with direct CP violation in charm,''
  Phys.\ Rev.\ D {\bf 87}, no. 3, 034005 (2013)
  [arXiv:1212.0165 [hep-ph]].
  %%CITATION = ARXIV:1212.0165;%%
  %6 citations counted in INSPIRE as of 23 Apr 2014
  
  
  %\cite{Bhattacharya:2013vc}
\bibitem{Bhattacharya:2013vc} 
  B.~Bhattacharya, D.~London, M.~Gronau and J.~L.~Rosner,
  %``Shift in weak phase $\gamma$ due to CP asymmetries in $D$ decays to two pseudoscalar mesons,''
  Phys.\ Rev.\ D {\bf 87}, 074002 (2013)
  [arXiv:1301.5631 [hep-ph]].
  %%CITATION = ARXIV:1301.5631;%%
  %7 citations counted in INSPIRE as of 23 Apr 2014
 
 %\cite{Bondar:2013jxa}
\bibitem{Bondar:2013jxa} 
  A.~Bondar, A.~Dolgov, A.~Poluektov and V.~Vorobiev,
  %``Effect of direct CP violation in charm on $\gamma$ extraction from $B \to DK^{\pm}, D \to K^0_S \pi^+ \pi^-$ Dalitz plot analysis,''
  Eur.\ Phys.\ J.\ C {\bf 73}, 2476 (2013)
  [arXiv:1303.6305 [hep-ph]].
  %%CITATION = ARXIV:1303.6305;%%
  %3 citations counted in INSPIRE as of 23 Apr 2014 
  
  %\cite{Meca:1998ee}
\bibitem{Meca:1998ee} 
  C.~C.~Meca and J.~P.~Silva,
  %``Detecting new physics contributions to the D0 - anti-D0 mixing through their effects on B decays,''
  Phys.\ Rev.\ Lett.\  {\bf 81}, 1377 (1998)
  [hep-ph/9807320].
  %%CITATION = HEP-PH/9807320;%%
  %30 citations counted in INSPIRE as of 05 Aug 2014
  
  %\cite{Silva:1999bd}
\bibitem{Silva:1999bd} 
  J.~P.~Silva and A.~Soffer,
  %``Impact of D0 - anti-D0 mixing on the experimental determination of gamma,''
  Phys.\ Rev.\ D {\bf 61}, 112001 (2000)
  [hep-ph/9912242].
  %%CITATION = HEP-PH/9912242;%%
  %64 citations counted in INSPIRE as of 05 Aug 2014
  
 %\cite{Rama:2013voa}
\bibitem{Rama:2013voa} 
  M.~Rama,
  %``Effect of D-Dbar mixing in the extraction of gamma with B- -> D0 K- and B- -> D0 pi- decays,''
  Phys.\ Rev.\ D {\bf 89}, 014021 (2014)
  [arXiv:1307.4384 [hep-ex]].
  %%CITATION = ARXIV:1307.4384;%%
  %5 citations counted in INSPIRE as of 23 Apr 2014 
  
  
  %\cite{Brod:2013sga}
\bibitem{Brod:2013sga} 
  J.~Brod and J.~Zupan,
  %``The ultimate theoretical error on $\gamma$ from $B \to DK$ decays,''
  JHEP {\bf 1401}, 051 (2014)
  [arXiv:1308.5663 [hep-ph]].
  %%CITATION = ARXIV:1308.5663;%%
  %3 citations counted in INSPIRE as of 23 Apr 2014
  
  %\cite{Trabelsi:2013uj}
\bibitem{Trabelsi:2013uj} 
  K.~Trabelsi [Belle Collaboration],
  %``Study of direct CP in charmed B decays and measurement of the CKM angle gamma at Belle,''
  arXiv:1301.2033 [hep-ex].
  %%CITATION = ARXIV:1301.2033;%%
  %19 citations counted in INSPIRE as of 25 Apr 2014
  
  %\cite{Lees:2013nha}
\bibitem{Lees:2013nha} 
  J.~P.~Lees {\it et al.}  [BaBar Collaboration],
  %``Observation of direct CP violation in the measurement of the Cabibbo-Kobayashi-Maskawa angle gamma with $B^\pm\to D^{(*)}K^{(*)\pm}$ decays,''
  Phys.\ Rev.\ D {\bf 87}, no. 5, 052015 (2013)
  [arXiv:1301.1029 [hep-ex]].
  %%CITATION = ARXIV:1301.1029;%%
  %21 citations counted in INSPIRE as of 25 Apr 2014
  
  %\cite{Aaij:2013zfa}
\bibitem{Aaij:2013zfa} 
  RAaij {\it et al.}  [LHCb Collaboration],
  %``Measurement of the CKM angle $\gamma$ from a combination of $B^{\pm} \to Dh^{\pm}$ analyses,''
  Phys.\ Lett.\ B {\bf 726}, 151 (2013)
  [arXiv:1305.2050 [hep-ex]].
  %%CITATION = ARXIV:1305.2050;%%
  %19 citations counted in INSPIRE as of 23 Apr 2014
  
  
%\cite{Wang:2011zw}
\bibitem{Wang:2011zw} 
  W.~Wang,
  %``Determining CP violation angle $\gamma$ with B decays into a scalar/tensor meson,''
  Phys.\ Rev.\ D {\bf 85}, 051301 (2012)
  [arXiv:1110.5194 [hep-ph]].
  %%CITATION = ARXIV:1110.5194;%%
  %6 citations counted in INSPIRE as of 14 Feb 2014
  
  
%\cite{Wang:2012jba}
\bibitem{Wang:2012jba} 
  W.~Wang,
  %``B decays into a scalar/tensor meson in pursuit of determining the CKM angle $\gamma$,''
  AIP Conf.\ Proc.\  {\bf 1492}, 117 (2012)
  [arXiv:1209.1244 [hep-ph]].
  %%CITATION = ARXIV:1209.1244;%%
  %3 citations counted in INSPIRE as of 24 Apr 2014  
  
  %\cite{Diehl:2001xe,Diehl:2001ey}
\bibitem{Diehl:2001xe} 
  M.~Diehl and G.~Hiller,
  %``New ways to explore factorization in b decays,''
  JHEP {\bf 0106}, 067 (2001)
  [hep-ph/0105194].
  %%CITATION = HEP-PH/0105194;%%
  %79 citations counted in INSPIRE as of 24 Apr 2014
%\cite{Diehl:2001ey}
\bibitem{Diehl:2001ey} 
  M.~Diehl and G.~Hiller,
  %``Yet another way to measure gamma,''
  Phys.\ Lett.\ B {\bf 517}, 125 (2001)
  [hep-ph/0105213].
  %%CITATION = HEP-PH/0105213;%%
  %37 citations counted in INSPIRE as of 24 Apr 2014

 


%\cite{Kim:2013ria}
\bibitem{Kim:2013ria} 
  C.~S.~Kim, R.~-H.~Li and W.~Wang,
  %``$B\to DK^*_{0,2}$ Decays: PQCD analysis to determine CP violation phase angle $\gamma$,''
  Phys.\ Rev.\ D {\bf 88}, 034003 (2013)
  [arXiv:1305.5320 [hep-ph]].
  %%CITATION = ARXIV:1305.5320;%%
  %1 citations counted in INSPIRE as of 14 Feb 2014



%\cite{Zou:2012sx}
\bibitem{Zou:2012sx} 
  Z.~-T.~Zou, X.~Yu and C.~-D.~Lu,
  %``The $B(B_{s})\rightarrow D_{(s)}(\bar{D}_{(s)}) T$ and $D_{(s)}^{*}(\bar{D}_{(s)}^{*})T$ Decays in Perturbative QCD Approach,''
  Phys.\ Rev.\ D {\bf 86}, 094001 (2012)
  [arXiv:1205.2971 [hep-ph]].
  %%CITATION = ARXIV:1205.2971;%%
  %9 citations counted in INSPIRE as of 24 Apr 2014
  
 %\cite{Zou:2014aba}
\bibitem{Zou:2014aba} 
  Z.~-T.~Zou and C.~-D.~Lu,
  %``The study of the nonleptonic two body B decays involving a light tensor meson in final states,''
  arXiv:1401.1298 [hep-ph].
  %%CITATION = ARXIV:1401.1298;%% 
  
 %\cite{Zou:2012xk}
\bibitem{Zou:2012xk} 
  Z.~-T.~Zou, X.~Yu and C.~-D.~Lu,
  %``Charmed B($B_{s}$) decays involving a light tensor meson in PQCD approach,''
  arXiv:1209.3369 [hep-ph].
  %%CITATION = ARXIV:1209.3369;%%
  %1 citations counted in INSPIRE as of 28 Apr 2014 

%\cite{Cheng:2010hn}
\bibitem{Cheng:2010hn} 
  H.~-Y.~Cheng, Y.~Koike and K.~-C.~Yang,
  %``Two-parton Light-cone Distribution Amplitudes of Tensor Mesons,''
  Phys.\ Rev.\ D {\bf 82}, 054019 (2010)
  [arXiv:1007.3541 [hep-ph]].
  %%CITATION = ARXIV:1007.3541;%%
  %17 citations counted in INSPIRE as of 28 Apr 2014

%\cite{Wang:2010ni}
\bibitem{Wang:2010ni} 
  W.~Wang,
  %``B to tensor meson form factors in the perturbative QCD approach,''
  Phys.\ Rev.\ D {\bf 83}, 014008 (2011)
  [arXiv:1008.5326 [hep-ph]].
  %%CITATION = ARXIV:1008.5326;%%
  %30 citations counted in INSPIRE as of 14 Feb 2014
  
  
%\cite{Li:2008tk}
\bibitem{Li:2008tk} 
  R.~-H.~Li, C.~-D.~Lu, W.~Wang and X.~-X.~Wang,
  %``B ---> S Transition Form Factors in the PQCD approach,''
  Phys.\ Rev.\ D {\bf 79}, 014013 (2009)
  [arXiv:0811.2648 [hep-ph]].
  %%CITATION = ARXIV:0811.2648;%%
  %40 citations counted in INSPIRE as of 24 Apr 2014




%\cite{Cheng:2005nb}
\bibitem{Cheng:2005nb}
  H.~-Y.~Cheng, C.~-K.~Chua, K.~-C.~Yang,
  %``Charmless hadronic B decays involving scalar mesons: Implications to the nature of light scalar mesons,''
  Phys.\ Rev.\  {\bf D73}, 014017 (2006)
  [hep-ph/0508104].
  
  %\cite{Cheng:2013fba,Cheng:2005nb}
\bibitem{Cheng:2013fba}
  H.~-Y.~Cheng, C.~-K.~Chua, K.~-C.~Yang and Z.~-Q.~Zhang,
  %``Revisiting charmless hadronic B decays to scalar mesons,''
  arXiv:1303.4403 [hep-ph].
  %%CITATION = ARXIV:1303.4403;%%

%\cite{Lu:2006fr,Cheng:2013fba,Cheng:2005nb}
\bibitem{Lu:2006fr}
  C.~-D.~Lu, Y.~-M.~Wang and H.~Zou,
  %``Twist-3 distribution amplitudes of scalar mesons from QCD sum rules,''
  Phys.\ Rev.\ D {\bf 75}, 056001 (2007)
  [hep-ph/0612210].
  %%CITATION = HEP-PH/0612210;%%

%\cite{Han:2013zg}
\bibitem{Han:2013zg}
  H.~-Y.~Han, X.~-G.~Wu, H.~-B.~Fu, Q.~-L.~Zhang and T.~Zhong,
  %``Twist-3 Distribution Amplitudes of Scalar Mesons within the QCD Sum Rules and Its Application to the $B \to S$ Transition Form Factors,''
  arXiv:1301.3978 [hep-ph].
  %%CITATION = ARXIV:1301.3978;%%
  %2 citations counted in INSPIRE as of 01 May 2013

 


%\cite{Bhattacharya:2013cla}
\bibitem{Bhattacharya:2013cla} 
  B.~Bhattacharya, M.~Imbeault and D.~London,
  %``Extraction of the CP-violating phase $\gamma$ using $B \to K \pi \pi$ and $B \to K K {\bar K}$ decays,''
  Phys.\ Lett.\ B {\bf 728}, 206 (2014)
  [arXiv:1303.0846 [hep-ph]].
  %%CITATION = ARXIV:1303.0846;%%
  %6 citations counted in INSPIRE as of 24 Apr 2014


 
%\cite{Bediaga:2006jk}
\bibitem{Bediaga:2006jk} 
  I.~Bediaga, G.~Guerrer and J.~M.~de Miranda,
  %``Extracting the quark mixing phase $\gamma$ from $B^\pm \to K^\pm \pi^+ \pi^-$, $B^0 \to K_S \pi^+ \pi^-$, and $\bar{B}^0 \to K_S \pi^+ \pi^-$,''
  Phys.\ Rev.\ D {\bf 76}, 073011 (2007)
  [hep-ph/0608268].
  %%CITATION = HEP-PH/0608268;%%
  %11 citations counted in INSPIRE as of 02 May 2014
   
 %\cite{Bediaga:2008zz}
\bibitem{Bediaga:2008zz} 
  I.~Bediaga, D.~R.~Boito, G.~Guerrer, F.~S.~Navarra and M.~Nielsen,
  %``Final state hadronic interactions and non-resonant B+- ---> K+- pi+ pi- decays,''
  Phys.\ Lett.\ B {\bf 665}, 30 (2008)
  [arXiv:0709.0075 [hep-ph]].
  %%CITATION = ARXIV:0709.0075;%%
  %5 citations counted in INSPIRE as of 02 May 2014


 

%\cite{LHCb:2011aa}
\bibitem{LHCb:2011aa} 
  R.~Aaij {\it et al.}  [LHCb Collaboration],
  %``Measurement of the CP-violating phase $\phi_s$ in the decay $B^0_s\to J/\psi \phi$,''
  Phys.\ Rev.\ Lett.\  {\bf 108}, 101803 (2012)
  [arXiv:1112.3183 [hep-ex]].
  %%CITATION = ARXIV:1112.3183;%%
  %120 citations counted in INSPIRE as of 14 Feb 2014


%\cite{Aaltonen:2012ie}
\bibitem{Aaltonen:2012ie} 
  T.~Aaltonen {\it et al.}  [CDF Collaboration],
  %``Measurement of the Bottom-Strange Meson Mixing Phase in the Full CDF Data Set,''
  Phys.\ Rev.\ Lett.\  {\bf 109}, 171802 (2012)
  [arXiv:1208.2967 [hep-ex]].
  %%CITATION = ARXIV:1208.2967;%%
  %24 citations counted in INSPIRE as of 14 Feb 2014


%\cite{Abazov:2011ry}
\bibitem{Abazov:2011ry} 
  V.~M.~Abazov {\it et al.}  [D0 Collaboration],
  %``Measurement of the CP-violating phase $\phi_s^{J/\psi \phi}$ using the flavor-tagged decay $B_s^0 \rightarrow J/\psi \phi$ in 8 fb$^{-1}$ of $p \bar p$ collisions,''
  Phys.\ Rev.\ D {\bf 85}, 032006 (2012)
  [arXiv:1109.3166 [hep-ex]].
  %%CITATION = ARXIV:1109.3166;%%
  %87 citations counted in INSPIRE as of 14 Feb 2014


%\cite{Aad:2012kba}
\bibitem{Aad:2012kba} 
  G.~Aad {\it et al.}  [ATLAS Collaboration],
  %``Time-dependent angular analysis of the decay $B_{s}^{0} \to J/{\psi\phi}$ and extraction of $\Delta\Gamma_{s}$ and the CP-violating weak phase $\phi_s$ by ATLAS,''
  JHEP {\bf 1212}, 072 (2012)
  [arXiv:1208.0572 [hep-ex]].
  %%CITATION = ARXIV:1208.0572;%%
  %40 citations counted in INSPIRE as of 14 Feb 2014


%\cite{Aaij:2013oba}
\bibitem{Aaij:2013oba} 
  RAaij {\it et al.}  [LHCb Collaboration],
  %``Measurement of $CP$ violation and the $B_s^0$ meson decay width difference with $B_s^0 \to J/\psi K^+K^-$ and $B_s^0\to J/\psi\pi^+\pi^-$ decays,''
  Phys.\ Rev.\ D {\bf 87}, 112010 (2013)
  [arXiv:1304.2600 [hep-ex]].
  %%CITATION = ARXIV:1304.2600;%%
  %47 citations counted in INSPIRE as of 14 Feb 2014


%\cite{Stone:2008ak}
\bibitem{Stone:2008ak} 
  S.~Stone and L.~Zhang,
  %``S-waves and the Measurement of CP Violating Phases in $B_s$ Decays,''
  Phys.\ Rev.\ D {\bf 79}, 074024 (2009)
  [arXiv:0812.2832 [hep-ph]].
  %%CITATION = ARXIV:0812.2832;%%
  %91 citations counted in INSPIRE as of 14 Feb 2014


%\cite{Stone:2009hd}
\bibitem{Stone:2009hd} 
  S.~Stone and L.~Zhang,
  %``Measuring the CP Violating Phase in B(s) Mixing Using B0(s) ---> J/psi f(0)(980),''
  arXiv:0909.5442 [hep-ex].
  %%CITATION = ARXIV:0909.5442;%%
  %26 citations counted in INSPIRE as of 14 Feb 2014


%\cite{Colangelo:2010bg}
\bibitem{Colangelo:2010bg} 
  P.~Colangelo, F.~De Fazio and W.~Wang,
  %``$B_s\to f_0(980)$ form factors and $B_s$ decays into $f_0(980)$,''
  Phys.\ Rev.\ D {\bf 81}, 074001 (2010)
  [arXiv:1002.2880 [hep-ph]].
  %%CITATION = ARXIV:1002.2880;%%
  %39 citations counted in INSPIRE as of 14 Feb 2014


%\cite{Colangelo:2010wg}
\bibitem{Colangelo:2010wg} 
  P.~Colangelo, F.~De Fazio and W.~Wang,
  %``Nonleptonic $B_s$ to charmonium decays: analyses in pursuit of determining the weak phase $\beta_s$,''
  Phys.\ Rev.\ D {\bf 83}, 094027 (2011)
  [arXiv:1009.4612 [hep-ph]].
  %%CITATION = ARXIV:1009.4612;%%
  %27 citations counted in INSPIRE as of 14 Feb 2014


%\cite{Leitner:2010fq}
\bibitem{Leitner:2010fq} 
  O.~Leitner, J.~-P.~Dedonder, B.~Loiseau and B.~El-Bennich,
  %``Scalar resonance effects on the B(s) - B-bar(s) mixing angle,''
  Phys.\ Rev.\ D {\bf 82}, 076006 (2010)
  [arXiv:1003.5980 [hep-ph]].
  %%CITATION = ARXIV:1003.5980;%%
  %17 citations counted in INSPIRE as of 14 Feb 2014


%\cite{Fleischer:2011au}
\bibitem{Fleischer:2011au} 
  R.~Fleischer, R.~Knegjens and G.~Ricciardi,
  %``Anatomy of $B^0_{s,d} \to J/\psi f_0(980)$,''
  Eur.\ Phys.\ J.\ C {\bf 71}, 1832 (2011)
  [arXiv:1109.1112 [hep-ph]].
  %%CITATION = ARXIV:1109.1112;%%
  %30 citations counted in INSPIRE as of 14 Feb 2014


%\cite{Aaij:2011fx}
\bibitem{Aaij:2011fx} 
  R.~Aaij {\it et al.}  [LHCb Collaboration],
  %``First observation of $B^0_s \to J/\psi f_0(980)$ decays,''
  Phys.\ Lett.\ B {\bf 698}, 115 (2011)
  [arXiv:1102.0206 [hep-ex]].
  %%CITATION = ARXIV:1102.0206;%%
  %58 citations counted in INSPIRE as of 14 Feb 2014


%\cite{Li:2011pg}
\bibitem{Li:2011pg} 
  J.~Li {\it et al.}  [Belle Collaboration],
  %``Observation of $B_s^0\to J/\psi f_0(980)$ and Evidence for $B_s^0\to J/\psi f_0(1370)$,''
  Phys.\ Rev.\ Lett.\  {\bf 106}, 121802 (2011)
  [arXiv:1102.2759 [hep-ex]].
  %%CITATION = ARXIV:1102.2759;%%
  %46 citations counted in INSPIRE as of 14 Feb 2014


%\cite{Aaltonen:2011nk}
\bibitem{Aaltonen:2011nk} 
  T.~Aaltonen {\it et al.}  [CDF Collaboration],
  %``Measurement of branching ratio and $B_s^0$ lifetime in the decay $B_s^0 \rightarrow J/\psi f_0(980)$ at CDF,''
  Phys.\ Rev.\ D {\bf 84}, 052012 (2011)
  [arXiv:1106.3682 [hep-ex]].
  %%CITATION = ARXIV:1106.3682;%%
  %41 citations counted in INSPIRE as of 14 Feb 2014


%\cite{Xie:2009fs}
\bibitem{Xie:2009fs}
  Y.~Xie, P.~Clarke, G.~Cowan and F.~Muheim,
  %``Determination of 2beta(s) in B0(s) ---> J/psi K+ K- Decays in the Presence of a K+ K- S-Wave Contribution,''
  JHEP {\bf 0909}, 074 (2009)
  [arXiv:0908.3627 [hep-ph]].
  %%CITATION = ARXIV:0908.3627;%%
  %31 citations counted in INSPIRE as of 11 Jul 2013



%\cite{Zhang:2012zk}
\bibitem{Zhang:2012zk}
  L.~Zhang and S.~Stone,
  %``Time-dependent Dalitz-plot formalism for B_q -> J/\psi\ h+ h-,''
  Phys.\ Lett.\ B {\bf 719}, 383 (2013)
  [arXiv:1212.6434 [hep-ph]].
  %%CITATION = ARXIV:1212.6434;%%
  %1 citations counted in INSPIRE as of 11 Jul 2013


%\cite{Dighe:1998vk}
\bibitem{Dighe:1998vk}
  A.~S.~Dighe, I.~Dunietz and R.~Fleischer,
  %``Extracting CKM phases and $B_s - \bar{B}_s$ mixing parameters from angular distributions of nonleptonic $B$ decays,''
  Eur.\ Phys.\ J.\ C {\bf 6}, 647 (1999)
  [hep-ph/9804253].
  %%CITATION = HEP-PH/9804253;%%
  %201 citations counted in INSPIRE as of 11 Jul 2013


%\cite{Fleischer:1999zi}
\bibitem{Fleischer:1999zi}
  R.~Fleischer,
  %``Extracting CKM phases from angular distributions of B(d,s) decays into admixtures of CP eigenstates,''
  Phys.\ Rev.\ D {\bf 60}, 073008 (1999)
  [hep-ph/9903540].
  %%CITATION = HEP-PH/9903540;%%
  %54 citations counted in INSPIRE as of 11 Jul 2013



  %\cite{Liu:2013nea}
\bibitem{Liu:2013nea} 
  X.~Liu, W.~Wang and Y.~Xie,
  %``Penguin Pollution in $B\to J/\psi V$ Decays and Impact on the Extraction of the $B_s-\bar B_s$ mixing phase,''
  arXiv:1309.0313 [hep-ph].
  %%CITATION = ARXIV:1309.0313;%%
  %4 citations counted in INSPIRE as of 24 Apr 2014


%\cite{Chen:2005ht}
\bibitem{Chen:2005ht} 
  C.~-H.~Chen and H.~-N.~Li,
  %``Nonfactorizable contributions to B meson decays into charmonia,''
  Phys.\ Rev.\ D {\bf 71}, 114008 (2005)
  [hep-ph/0504020].
  %%CITATION = HEP-PH/0504020;%%
  %30 citations counted in INSPIRE as of 14 Feb 2014


%\cite{Liu:2010zh}
\bibitem{Liu:2010zh} 
  X.~Liu, Z.~-Q.~Zhang and Z.~-J.~Xiao,
  %``B ---> (J/Psi, eta(c)) K decays in the perturbative QCD approach,''
  Chin.\ Phys.\ C {\bf 34}, 937 (2010).
  %%CITATION = CHPHD,C34,937;%%
  %3 citations counted in INSPIRE as of 14 Feb 2014


%\cite{Liu:2012ib}
\bibitem{Liu:2012ib} 
  X.~Liu, H.~-n.~Li and Z.~-J.~Xiao,
  %``Implications on $\eta$-$\eta'$-glueball mixing from $B_{d/s} \to J/\Psi \eta^{(')}$ Decays,''
  Phys.\ Rev.\ D {\bf 86}, 011501 (2012)
  [arXiv:1205.1214 [hep-ph]].
  %%CITATION = ARXIV:1205.1214;%%
  %12 citations counted in INSPIRE as of 14 Feb 2014


%\cite{Bhattacharya:2012ph}
\bibitem{Bhattacharya:2012ph} 
  B.~Bhattacharya, A.~Datta and D.~London,
  %``Reducing Penguin Pollution,''
  Int.\ J.\ Mod.\ Phys.\ A {\bf 28}, 1350063 (2013)
  [arXiv:1209.1413 [hep-ph]].
  %%CITATION = ARXIV:1209.1413;%%
  %6 citations counted in INSPIRE as of 14 Feb 2014


%\cite{Faller:2008gt}
\bibitem{Faller:2008gt} 
  S.~Faller, R.~Fleischer and T.~Mannel,
  %``Precision Physics with $B^0_s \to J/\psi \phi$ at the LHC: The Quest for New Physics,''
  Phys.\ Rev.\ D {\bf 79}, 014005 (2009)
  [arXiv:0810.4248 [hep-ph]].
  %%CITATION = ARXIV:0810.4248;%%
  %61 citations counted in INSPIRE as of 14 Feb 2014



%\cite{Aad:2012tfa}
\bibitem{Aad:2012tfa} 
  G.~Aad {\it et al.}  [ATLAS Collaboration],
  %``Observation of a new particle in the search for the Standard Model Higgs boson with the ATLAS detector at the LHC,''
  Phys.\ Lett.\ B {\bf 716}, 1 (2012)
  [arXiv:1207.7214 [hep-ex]].
  %%CITATION = ARXIV:1207.7214;%%
  %2534 citations counted in INSPIRE as of 30 Apr 2014
  

  %\cite{Chatrchyan:2012ufa}
\bibitem{Chatrchyan:2012ufa} 
  S.~Chatrchyan {\it et al.}  [CMS Collaboration],
  %``Observation of a new boson at a mass of 125 GeV with the CMS experiment at the LHC,''
  Phys.\ Lett.\ B {\bf 716}, 30 (2012)
  [arXiv:1207.7235 [hep-ex]].
  %%CITATION = ARXIV:1207.7235;%%
  %2507 citations counted in INSPIRE as of 30 Apr 2014  


  

\end{thebibliography}

\end{document}